\documentclass[aps,prd,preprintnumbers,showpacs,showkeys,nofootinbib,
superscriptaddress,fleqn,floatfix,tightenlines,10pt]{revtex4-1}
\usepackage{amsmath,amsfonts,amssymb,amscd,amsxtra,amsthm}
\usepackage{graphicx}  
\usepackage{epstopdf}
\usepackage{dcolumn}  
\usepackage{bm}        
\usepackage{slashed}
\usepackage{cancel} 
\usepackage{float}
\usepackage{mathtools}
\usepackage{amsbsy}
\usepackage{amstext}

\providecommand{\tabularnewline}{\\}
\usepackage[utf8]{inputenc} 
\usepackage{booktabs}
\usepackage[normalem]{ulem} 
\usepackage[dvipsnames]{xcolor} 
\usepackage{tabularx}
\usepackage{enumitem}  
\usepackage{array} 
\usepackage{slashed}
\usepackage{tikz}
\usepackage{float}
\usepackage{multirow}
\renewcommand\sout{\bgroup \color{red} \ULdepth=-.5ex \ULset}

\makeatletter

\begin{document}  
\preprint{INHA-NTG-01/2021}
\title{Electromagnetic transitions of the singly charmed baryons with
  spin 3/2} 
\author{June-Young Kim}
\email[E-mail: ]{Jun-Young.Kim@ruhr-uni-bochum.de}
\affiliation{Institut f\"ur Theoretische Physik II, Ruhr-Universit\"at
  Bochum, D-44780 Bochum, Germany}
\affiliation{Department of Physics, Inha University, Incheon 22212,
Republic of Korea}

\author{Hyun-Chul Kim}
\email[E-mail: ]{hchkim@inha.ac.kr}
\affiliation{Department of Physics, Inha University, Incheon 22212,
Republic of Korea}
\affiliation{School of Physics, Korea Institute for Advanced Study 
  (KIAS), Seoul 02455, Republic of Korea}

\author{Ghil-Seok Yang}
\email[E-mail: ]{ghsyang@ssu.ac.kr}
\affiliation{Department of Physics, Soongsil University, Seoul 06978,
  Republic of Korea}

\author{Makoto Oka}
\email[E-mail: ]{oka@post.j-parc.jp}
\affiliation{Advanced Science Research Center, Japan Atomic 
Energy Agency, Shirakata, Tokai, Ibaraki, 319-1195, Japan}
\date{\today}
\begin{abstract}
We investigate the electromagnetic transitions of the singly charmed
baryons with spin 3/2, based on a pion mean-field approach, also known
as the chiral quark-soliton model, taking into account the rotational
$1/N_c$ corrections and the effects of flavor SU(3) symmetry breaking.  
We examine the valence- and sea-quark contributions to the
electromagnetic transition form factors and find that the
quadrupole form factors of the sea-quark contributions dominate over
those of the valence-quark ones in the smaller $Q^2$
region, whereas the sea quarks only provide marginal contributions to
the magnetic dipole transition form factors of the baryon sextet with
spin 3/2. The effects of the flavor SU(3) symmetry 
breaking are in general very small except for the forbidden transition 
$\Xi_c^0\gamma\to \Xi_c^{*0}$ by $U$-spin symmetry. We also discuss
the widths of the radiative decays for the baryon sextet with spin
3/2, comparing the present results with those from other works. 
\end{abstract}
\maketitle
\section{Introduction}
It is of great importance to understand the electromagnetic (EM)
structure of a baryon, since it reveals how the baryon is shaped by
its constituents. A baryon with spin 3/2 has a finite value of the
electric quadrupole (E2) moment, which indicates that its charge
distribution is shown to be deformed to be either a cushion-like form
(oblate spheroid) or a rugby-ball-like one (prolate spheroid),
depending on the signature of its charge. This implies that a singly
heavy baryon with spin 3/2 may reveal a similar structure. It is also
known that the effects of the vacuum polarization or those of the pion
clouds are known to contribute significantly to the E2 moment of the
baryon decuplet~\cite{Kumano:1988wu}. This leads to an interpretation
that the E2 moment of a low-lying baryon with spin 3/2 is 
governed by long-distance pion clouds~\cite{Butler:1993ht}. 
While experimental information on the EM
transitions of the singly heavy baryons is still
inconclusive~\cite{Jessop:1998wt, Aubert:2006je, Solovieva:2008fw,
  Yelton:2016fqw}, there has been a great deal of theoretical
works within many different approaches such as chiral
perturbation theory~\cite{Cheng:1992xi, Savage:1994wa, Zhu:1998ih,
Banuls:1999br, Wang:2018cre},  the quark models~\cite{Dey:1994qi,
Tawfiq:1999cf, Ivanov:1999bk, Ivanov:1999bk}, QCD sum
rules~\cite{Aliev:2009jt, Aliev:2014bma, Aliev:2016xvq}, and so on
(see also a recent review~\cite{Cheng:2015iom}). 
In lattice QCD, the EM transition form factors for $\Omega_c^0 \gamma
\to \Omega_c^{*0}$ were calculated~\cite{Bahtiyar:2015sga,
  Bahtiyar:2019ykq}. Thus, anticipating that the experimental data on
the EM transitions of the singly heavy baryons will be available in
near future, it is of great interest to investigate the structure of
the EM transition form factors in a different theoretical framework. 

In the present work, we investigate the EM transition form factors of
the low-lying singly heavy baryons with spin 3/2 within the framework
of the chiral quark-soliton model ($\chi$QSM). The model is based on a
pion mean-field approach. As was proposed first by
Witten~\cite{Witten:1979kh}, in the large $N_c$ limit, the light
baryon can be viewed as a state of $N_c$ (the number of colors)
valence quarks bound by the pion mean 
fields that have been produced self-consistently by the $N_c$ valence 
quarks~\cite{Diakonov:1987ty, Blotz:1992pw}. The model was extended
to the description of the singly heavy baryons~\cite{Yang:2016qdz,
  Kim:2018xlc, Kim:2019rcx}, being motivated by
Ref.~\cite{Diakonov:2010tf}. In the limit of the infinitely heavy
quark mass ($m_Q\to\infty$), the spin of the heavy quark can not be
flipped, which makes the heavy-quark spin conserved. This causes also
the total spin of the light quarks inside a singly heavy baryon
conserved. In this limit of $m_Q\to\infty$, the flavor of the heavy
quark does not come into play. This is known as the heavy-quark
spin-flavor symmetry~\cite{Isgur:1989vq, Isgur:1991wq,
  Georgi:1990um}. Thus, the singly heavy baryons can be expressed
within the SU(3) representations. That is, two light valence quarks
($\bm{3}\otimes \bm{3}$) will allow one to have the baryon antitriplet
($\bar{\bm{3}}$) and sextet ($\bm{6}$). The spins of the two
light valence quarks can be aligned either in the spin-singlet state
($\bm{0}$) or in the spin-triplet one ($\bm{1}$). Hence, by combining
them with the spin of the heavy quark, one can have two degenerate
baryon sextets. This degeneracy can be removed by the color hyperfine
interaction in order $1/m_Q$~\cite{Yang:2016qdz}.  
Note that the infinitely heavy quark can be regarded as the mere static
color source. This indicates that the light quarks govern the
dynamics inside a singly heavy baryon. Based on this heavy-quark
spin-flavor symmetry, the pion mean-field approach was developed also
for the singly heavy baryons that can be regarded as the bound state
of $N_c-1$ valence quarks. The heavy quark inside a singly heavy
baryon is required only for the construction of a color-singlet state
of the singly heavy baryon. 

This pion mean-field approach or the $\chi$QSM described various
properties of the singly heavy baryons quantitatively well, compared
with the experimental data, without any free
parameters~\cite{Kim:2017jpx, Kim:2017khv, Yang:2016qdz, 
  Kim:2018xlc, Kim:2019rcx, Yang:2020klp, Yang:2019tst} (See also a
recent review~\cite{Kim:2018cxv}). The EM form factors of the
low-lying singly heavy baryons have been studied in the 
$\chi$QSM~\cite{Kim:2018nqf, Kim:2020uqo}. Since the heavy-quark mass
is taken to be infinitely heavy, the heavy quark gives a constant
contribution to the electric monopole form factor constrained by the
gauge invariance, whereas contributions to the magnetic dipole from
factor from the heavy quark is negligible. The numerical results were
in good agreement with the lattice data~\cite{Kim:2019wbg}. In the
present work, we want to investigate 
the EM transition form factors of the baryon sextet with spin 3/2. 
While the $\Omega_c^0\gamma\to \Omega_c^{*0}$ radiative decay was
computed in lattice QCD, there is no work on the EM
transition form factors for all possible radiative decays for the
baryon sextet with spin 3/2.  Thus, we will consider for the first
time the magnetic dipole (M1), and electric quadrupole
(E2) and Coulomb quadrupole (C2) transition form factors for the
baryon sextet ($\bm{6}$) with spin 3/2. We will compare the results
for the $\Omega_c^0\gamma\to \Omega_c^{*0}$ radiative decay with those
from the lattice calculation. We will compare the present numerical
results for the decay rates of the radiative decays for the singly
heavy baryons with those from other theoretical works. 

The present work is organized as follows: In Section II, we define the
M1, E2 and C2 transition form factors of the singly heavy
baryons. In Section III, we explain explicitly how the singly heavy baryon
state can be consistently constructed based on the heavy-quark
spin-flavor symmetry in the limit of the infinitely heavy-quark
mass. We show that the heavy-quark field can be decoupled from the singly
heavy baryon and its mass contributes to the classical mass of the
singly heavy baryon in a simple manner. In Section IV, we show briefly
how to compute them within the framework of the $\chi$QSM. In Section
V, we first compare the numerical results for 
$\Omega_c+\gamma \to 
\Omega_c^*$ with those of the corresponding lattice data. We also
examine the dependence of the EM transition form factors of
$\Omega_c+\gamma \to \Omega_c^*$ on the pion mass. We then scrutinize
the valence- and sea-quark contributions separately and show that the
sea quarks or the Dirac continuum play a crucial role in describing
the E2 and C2 transition form factors of the baryon sextet with spin 3/2,
which can be interpreted as the pion clouds. We also study the effects
of the explicit breaking of flavor SU(3) symmetry breaking on the EM
transition form factors of the baryon sextet with spin 3/2. 
We compare the present results for the decay rates of the radiative
decays for the singly heavy baryons with spin 3/2 with those from
other works. Finally, we summarize the results from
the present work and draw conclusions. 

\section{EM transition form factors of the baryon sextet with spin
  3/2} 
To describe the EM transition from a singly heavy baryon with spin
1/2 to that with spin 3/2, $B \gamma^*\to B^{*}$, we assume that the
baryon with spin 3/2 is at rest. In this rest frame, we define the
four-momenta for the baryon with spin 3/2, the baryon with spin 1/2
and the photon respectively as $p_{B^{*}}$, $p_{B}$, and $q$, which
are explicitly written as 
\begin{align}
p_{B^{*}} = ( M_{B^{*}}, \bm{0}), \ \  p = ( E_{B},
  -\bm{q}), \ \  {q} = ( \omega_q, \bm{q}) , 
\end{align}
where ${\bm q}$ and $\omega_q$ denote the three-momentum and energy of
the virtual photon. The energy-momentum relation is given 
by $E_{B}^{2} =  M_{B}^{2} +{ |\bm{q}|}^{2}$ and $E_{B^{*}}^{2} =
M_{B^{*}}^{2}$. Using this relation, we can express the momentum and
the energy of the virtual photon as follows
\begin{align}
|\bm{q}|^{2} =\left (\frac{M_{B^{*}}^{2} + M_{B}^{2}+ Q^{2}}{2
  M_{B^{*}}}\right )^{2} - M_{B}^{2}, \ \ \ 
\omega_q =\left (\frac{M_{B^{*}}^{2} - M_{B}^{2} - Q^{2} }{2 M_{B^{*}}}\right ),
\label{eq:kinematics}
\end{align}
where $Q^{2} = -q^{2} > 0$.

We start with the EM current defined by 
\begin{align}
  \label{eq:EMCurrent1}
V^\mu(x) = \bar{\psi}(x) \gamma^\mu \hat{\mathcal{Q}} \psi(x) +
\bar{\Psi}_h(x) \gamma^\mu Q_h \Psi_h(x),
\end{align}
where $\psi(x)$ and $\Psi_h(x)$ denote respectively the light and
heavy quarks. The first term in Eq.~\eqref{eq:EMCurrent1} is the EM
current for the light quarks with the charge operator defined by the
charges of the light quarks $\hat{\mathcal{Q}} =
\mathrm{diag}(2/3,-1/3,-1/3)$. The second term represents the EM
current for the heavy quark with a heavy quark charge $Q_h$. If one 
considers the charm quark, then $Q_h=2/3$. In the case of the bottom
baryon, we have $Q_h=-1/3$. In the present work, we will consider only
the charmed baryons. The transition EM matrix element
between $B^*$ and $B$ is then parametrized in terms of the three real
EM transition form factors 
\begin{align}
\langle B^{*}(p',\lambda') | V^{\mu}(0) | B(p,\lambda)
  \rangle = i \sqrt{\frac{2}{3}} \overline{u}_{\beta}(p',\lambda')
  \Gamma^{\beta   \mu} {u}(p,\lambda). 
\label{eq:matel1}
\end{align}
$\lambda$ and $\lambda'$ denote the helicities of the baryons with
spin 1/2 and 3/2, respectively.  
 $u_{\beta}(p,\lambda')$ and $u(p,\lambda)$ stand for the
 Rarita-Schwinger and Dirac spinors, respectively. $\Gamma^{\beta
   \mu}$ in Eq.~\eqref{eq:matel1} 
 denote the three real EM transition form factors:
\begin{align}
\Gamma^{\beta \mu} = G^{*}_{M1}(Q^{2}) \mathcal{K}_{M1}^{ \beta \mu} +
  G _{E2}^{*}(Q^{2}) \mathcal{K}_{E2}^{ \beta \mu} + G _{C2}^{*}(Q^{2})
  \mathcal{K}_{C2}^{ \beta \mu}, 
\end{align}
where $G^{*}_{M1}$, $G^{*}_{E2}$, and $G^{*}_{C2}$ are known
respectively as the magnetic dipole transition form factor, the
electric quadrupole one, and the Coulomb quadrupole 
one. The corresponding Lorentz tensors $\mathcal{K}^{ \beta \mu}_{M1}$
are written as 
\begin{align}
  {\mathcal{K}}_{M1}^{\beta \mu} &=
   \frac{-3(M_{B^{*}}+M_{B})}{2M_{B}[(M_{B^{*}}+M_{B})^{2}+Q^{2}]}
\varepsilon^{\beta \mu \sigma \tau} P_{\sigma} q_{\tau}, \cr 
{\mathcal{K}}_{E2}^{\beta \mu} &= -{\mathcal{K}}_{M}^{\beta \mu}  -
   \frac{6}{4M^{2}_{B^{*}} |\bm{q}|^{2}}
   \frac{M_{B^{*}}+M_{B}}{M_{B}}
   \varepsilon^{\beta \sigma \nu \gamma}
   P_{\nu}q_{\gamma}
   {{\varepsilon^{\mu}}_{\sigma}}^{\alpha  \delta}
   p_{B^{*}\alpha}q_{\delta} i\gamma^{5}, \cr
 {\mathcal{K}}_{C2}^{\beta \mu} &=-\frac{3}{4M^{2}_{B^{*}} |\bm{q}|^{2}}
   \frac{M_{B^{*}}+M_{B}}{M_{B}} q^{\beta} [q^{2} P^{\mu} -
 q\cdot P q^{\mu}]i\gamma^{5}.
\end{align}
The Lorentz tensors are required to satisfy the gauge-invariant
identities $q_{\mu} {\mathcal{K}}_{M1,E2,C2}^{\beta \mu} = 0$, which
arises from the conservation of the EM current.  

The EM transition form factors can be extracted experimentally by
using the helicity amplitudes. The transverse and Coulomb helicity
amplitudes are defined respectively in terms of the spatial and
temporal components of the EM current 
\begin{align}
  A_{\lambda} &= -\frac{e}{\sqrt{2\omega_q}}
  \int d^{3} r e^{i {\bm{q}} \cdot {\bm{r}}} \bm{\epsilon}_{+1}
  \cdot \langle B^{*} (3/2 , \lambda) | \bar{\psi} (\bm{r})
  \hat{{Q}} \bm{\gamma} \psi (\bm{r}) |
  B(1/2,\lambda-1) \rangle, \cr
 S_{1/2} &= -\frac{e}{\sqrt{2\omega_q}} \frac{1}{\sqrt{2}}
   \int d^{3} r e^{i {\bm{q}} \cdot {\bm{r}}}  \langle B^{*} (3/2,1/2)
           | \bar{\psi} (\bm{r}) \hat{{Q}} \gamma^{0}
           \psi (\bm{r}) | B(1/2,1/2) \rangle,
\label{eq:MatrixEl1}
\end{align}
where $\lambda$ is the corresponding value of the helicity of the
baryon $B^{*}$ with spin 3/2, i.e. $\lambda=3/2$ or $1/2$.  
Note that the transverse photon polarization vector is defined as 
$\hat{\bm{\epsilon}} =  -1/\sqrt{2} (1,i,0)$. The helicity amplitudes
are expressed in terms of the EM transition form
factors
\begin{align}
  A_{1/2} &= - \frac{e}{\sqrt{2\omega_{q}}}\frac{1}{4c_{\Delta}}
      ( G^{*}_{M1} - 3 G^{*}_{E2}), \ \ 
       A_{3/2} = - \frac{e}{\sqrt{2\omega_{q}}}\frac{\sqrt{3}}{4c_{\Delta}}
       ( G^{*}_{M1} + G^{*}_{E2}), \ \ S_{1/2}=  \frac{e}{\sqrt{2\omega_{q}}}
      \frac{|\bm{q}|}{4c_{\Delta}M_{B^{*}}}G^{*}_{C2},
\label{eq:Amplitude_decom}
\end{align}
where $c_{\Delta} = \sqrt{\frac{M^{3}_{B} }{2 M_{B^{*}} |\bm{q}|^{2}}}
\sqrt{1 + \frac{Q^{2}}{(M_{B^{*}}+M_{B})^{2}}}$.
Then, we are able to express the EM transition form factors inversely
by the transition amplitudes
\begin{align}
G^{*}_{M1}(Q^{2})& =  - 2c_{\Delta}  \int d^{3} r 3
 j_{1}(|\bm{q}||\bm{r}|)
 \langle B^{*} (3/2 , 1/2) |   [\hat{\bm r} \times
 {\bm V}]_{11}
 | B(1/2,-1/2) \rangle, \cr
G^{*}_{E2}(Q^{2})& \simeq -2c_{\Delta}  \int d^{3} r
  \sqrt{\frac{20\pi}{27}}  \frac{\omega_{q}}{|{\bm
  q}|} \left( \frac{\partial}{\partial r} r
  j_{2}(|{\bm q}||{\bm r}|)\right)  \langle B^{*}
  (3/2 , 1/2) | Y_{21} (\hat{\bm r}) V_{0} |
  B(1/2,-1/2) \rangle, \cr 
G^{*}_{C2}(Q^{2})& = 4c_{\Delta}\frac{M_{B^{*}}}{|\bm{q}|}  \int d^{3}
    r \sqrt{10\pi} j_{2}(|\bm{q}||\bm{r}|)   \langle
    B^{*} (3/2 , 1/2) | Y_{20} (\hat{\bm r}) V_{0} |
    B(1/2,1/2) \rangle. 
 \end{align}
Note that we neglect a term that provides a tiny correction to the
E2 transition form factor at low-energy regions, which implements the
current conservation.

From the form factors, the well-known quantities $R_{EM}$ and
$R_{SM}$, which are defined respectively as 
\begin{align}
R_{EM}(Q^{2})  = -\frac{G^{*}_{E2}(Q^{2})}{G^{*}_{M1}(Q^{2})}, \;\;\;
R_{SM}(Q^{2})  =
  -\frac{|\bm{q}|}{2M_{B^*}}\frac{G^{*}_{C2}(Q^{2})}{G^{*}_{M1}(Q^{2})}, 
\label{eq:rem}
\end{align}
can be obtained. The decay width is expressed in terms of the helicity
amplitudes~\cite{Tanabashi:2018oca}:  
\begin{align}
\Gamma(B^{*} \to B \gamma) &= \frac{\omega^{2}_{q}}{\pi}
          \frac{M_{B}}{2M_{B^{*}}} \left(
          |A_{1/2} |^{2}+ |A_{3/2}
          |^{2}\right) \cr
  &= \frac{\alpha_{\mathrm{EM}}}{16}
    \frac{(M^{2}_{B^{*}}-M^{2}_{B})^{3}}{M^{3}_{B^{*}}M^{2}_{B}} 
  \left(|G^{*}_{M1}(0)|^{2}+ 
  3 |G^{*}_{E2}(0)|^{2}\right) 
\label{eq:decay_width}
\end{align}
with the EM fine structure constant $\alpha_{\mathrm{EM}}$.
\section{A singly heavy baryon in the chiral quark-soliton model}
The pion mean-field approach or the $\chi$QSM has one great
virtue. The model allows one to describe both light baryons and singly
heavy baryons on an equal footing. While various properties of singly
heavy baryons were investigated in the previous works based on the
$\chi$QSM, it was not discussed formally how a singly heavy baryon can
be explicitly constructed in the $\chi$QSM. Thus, before we compute
the EM transition form factors of singly heavy baryons, we first want
to show how a singly heavy baryon can be formulated in the present
approach. Let us first define the normalization of the baryon
state $\langle B(p',J_3') | B (p,J_3)\rangle = 2 p_0
\delta_{J_3'J_3} (2\pi)^{3}\delta^{(3)}(\bm{p}'-\bm{p})$. In the large
$N_c$ limit, this normalization can be expressed as $\langle
B(p',J_3') | B (p,J_3)\rangle = 2 M_B \delta_{J_3'J_3}
(2\pi)^{3}\delta^{(3)}(\bm{p}'-\bm{p})$, 
where $M_B$ is a baryon mass. 
Since a singly heavy baryon consists of the $N_c-1$ valence quarks and
a heavy quark, it can be expressed in terms of the Ioffe-type current
of the $N_c-1$ valence quarks  and a heavy-quark field in Euclidean
space as follows: 
\begin{align}
|B,p\rangle &= \lim_{x_4\to-\infty} \exp(ip_{4}x_{4})
              \mathcal{N}(\bm{p}) \int d^3 x
              \exp(i\bm{p}\cdot \bm{x}) (-i\Psi_h^\dagger(\bm{x}, x_4)
              \gamma_4) J_B^\dagger (\bm{x},x_4) 
              |0\rangle,\cr
\langle B,p| &= \lim_{y_4\to \infty} \exp(-ip'_4 y_4) 
               \mathcal{N}^*(\bm{p}') \int d^3 y
              \exp(-i\bm{p}'\cdot \bm{y}) \langle 0| J_{B}
                 (\bm{y},y_4) \Psi_h (\bm{y},y_4), 
\end{align}
where $\mathcal{N}(\bm{p}) (\mathcal{N}^*(\bm{p}'))$ stands for
the normalization factor depending on the initial (final) momentum. 
$J_B(x)$ and $J_B^\dagger(y)$ denote the Ioffe-type current of the
$N_c-1$ valence quarks~\cite{Diakonov:1987ty} defined by 
\begin{align}
J_B(x) &= \frac1{(N_c-1)!} \epsilon_{\alpha_1\cdots \alpha_{N_c-1}} 
\Gamma_{(TT_3Y)(JJ_3Y_R)}^{f_1\cdots  f_{N_c-1}}
\psi_{f_1 \alpha_1}(x)\cdots \psi_{f_{N_c-1} \alpha_{N_c-1}}(x),\cr  
J_{B}^\dagger(y) &= \frac1{(N_c-1)!} \epsilon_{\alpha_1\cdots \alpha_{N_c-1}}
\Gamma_{(TT_3Y)(JJ_3'Y_R)}^{f_1\cdots f_{N_c-1}}
(-i\psi^\dagger(y)\gamma_4)_{f_1\alpha_1} \cdots
  (-i\psi^\dagger(y)\gamma_4)_{f_{N_c-1}\alpha_{N_c-1}} ,
\end{align}
where $f_1\cdots f_{N_c-1}$ and $\alpha_1\cdots\alpha_{N_c-1}$ denote
respectively the spin-isospin and color
indices. $\Gamma_{(TT_3Y)(JJ_3Y_R)}$ are matrices with the quantum
numbers $(TT_3Y)(JJ_3Y_{R})$ for the corresponding baryon. 
For example, a singly heavy baryon $\Sigma_c^+$ can be identified as
the state with $J=1/2$, $T=1$, $T_3=0$, and $Y=2/3$. The right
hypercharge $Y_{R}$ for singly heavy baryons is constrained by the
number of the valence quarks. Note that $Y_{R}=N_c/3$ for a light
baryon whereas $Y_{R}=(N_c-1)/3$ for a singly heavy baryon. The right
hypercharge $Y_{R}=1$ with $N_c=3$ allows one to get the lowest-lying
representations for the SU(3) baryons, i.e., the baryon octet
($\bm{8}$) and decuplet ($\bm{10}$) for the light baryons. On the
other hand, we find the 
baryon antitriplet ($\bar{\bm{3}}$), sextet ($\bm{6}$) and so
on~\cite{Yang:2016qdz, Kim:2018cxv}. $\psi_{f_k \alpha_k}(x)$ denotes the
light-quark field and $\Psi_h(x)$ stands for the heavy-quark field. 
In the limit of $m_Q\to \infty$, a singly heavy baryon satisfies the
heavy-quark flavor symmetry. Then the heavy-quark field can be written
as 
\begin{align}
\Psi_h(x) = \exp(-im_Q v\cdot  x) \tilde{\Psi}_h(x), 
\end{align}
where $\tilde{\Psi}_h(x)$ is a rescaled heavy-quark field almost on
mass-shell. It carries no information on the heavy-quark mass in the
leading order approximation in the heavy-quark expansion. $v$ denotes
the velocity of the heavy quark~\cite{Isgur:1989vq, Isgur:1991wq,
  Georgi:1990um,Shifman:1995dn}.  

We can show explicitly that the normalization factor
$\mathcal{N}^*(\bm{p}')\mathcal{N}(\bm{p})$ correctly turns out to be
$2 M_B$. The normalization of the baryon state can be computed as
follows:       
\begin{align}
 \label{eq:normal}
\langle B(p',J_3') | B (p,J_3)\rangle &=
\frac{1}{\mathcal{Z}_{\mathrm{eff}}}
  \mathcal{N}^*(p')\mathcal{N}(p)
\lim_{x_4\to -\infty}  \lim_{y_4\to \infty}   
\exp\left(-iy_4p_4'+ix_4p_4\right)  \cr
& \times \int  d^3x d^3y 
  \exp(-i\bm{p}'\cdot \bm{y}+  i\bm{p}\cdot \bm{x})
 \int \mathcal{D} U  \mathcal{D} \psi
  \mathcal{D}\psi^\dagger \mathcal{D} \tilde{\Psi}_h \mathcal{D}
  \tilde{\Psi}_h^\dagger    
J_B (y)\Psi_h(y) (-i\Psi_h^\dagger(x) \gamma_4) J_B^\dagger(x) \cr
&\times \exp\left[\int d^4z\left\{
  (\psi^\dagger(z))_{\alpha}^{f} \left( i\rlap{/}{\partial}  + i
  MU^{\gamma_5} + i  \hat{m} \right)_{fg} \psi^{g\alpha}(z) +
  \Psi_h^\dagger(z) v\cdot \partial 
  \Psi_h(z) \right\} \right] \cr
&=
  \frac{1}{\mathcal{Z}_{\mathrm{eff}}}\mathcal{N}^*(p')\mathcal{N}(p)
  \lim_{x_4\to -\infty} \lim_{y_4\to \infty} 
 \exp\left(-iy_4p_4'+ix_4p_4\right) \cr
& \times \int  d^3x d^3y 
  \exp(-i\bm{p}'\cdot y+  i\bm{p}\cdot \bm{x}) \langle 
J_B (y) \Psi_h(y) (-i\Psi_h^\dagger(x) \gamma_4) J_B^\dagger(x)\rangle_0,  
\end{align}
where $\mathcal{Z}_{\mathrm{eff}}$ represents the low-energy effective
QCD partition function defined as
\begin{align}
\mathcal{Z}_{\mathrm{eff}} = \int   \mathcal{D} U \exp(-S_{\mathrm{eff}}).    
\end{align}
$S_{\mathrm{eff}}$ is called the effective chiral action 
(E$\chi$A) defined by  
\begin{align}
S_{\mathrm{eff}} = -N_c \mathrm{Tr} \ln \left[ i\rlap{/}{\partial}  +
  iMU^{\gamma_5} + i   \hat{m}\right]. 
\end{align} 
$\langle ... \rangle_{0}$ in Eq.~\eqref{eq:normal} denotes the vacuum
expectation value of the baryon correlation function. 
$M$ represents the dynamical quark mass that arises from the
spontaneous breakdown of chiral symmetry.
The $U^{\gamma_5}$ denotes the chiral field that is defined by
\begin{align}
  \label{eq:1}
U^{\gamma_5} (z) = \frac{1-\gamma_5}{2}   U(z) + U^\dagger(z)
  \frac{1+\gamma_5}{2} 
\end{align}
with
\begin{align}
  \label{eq:2}
U(z) = \exp[{i\pi^a(z) \lambda^a}],  
\end{align}
where $\pi^a(z)$ represents the pseudo-Nambu-Goldstone (pNG) fields
and $\lambda^a$ are the flavor Gall-Mann matrices. 
$\hat{m}$ designates the mass
matrix of current quarks $\hat{m} = \mathrm{diag}(m_{\mathrm{u}},\,
m_{\mathrm{d}},\,m_{\mathrm{s}})$.
Note that we deal with the strange current quark mass $m_{\mathrm{s}}$
perturbatively. Thus, we will consider it when we make a zero-mode
quantization for a collective baryon state. The propagators of a light
quark in the $\chi$QSM~\cite{Diakonov:1987ty} is given by   
\begin{align}
G(y,x) &= \left \langle y \left |
                      \frac1{i\rlap{/}{\partial} + i 
  MU^{\gamma_5}  +i\overline{m}} (i\gamma_4)\right | x\right 
\rangle \cr
&= \Theta(y_4   -x_4) 
\sum_{E_n>0} e^{-E_n(y_4-x_4)} \psi_n(\bm{y})
  \psi_n^\dagger(\bm{x}) - \Theta(x_4  -y_4) 
\sum_{E_n<0} e^{-E_n(y_4-x_4)}
  \psi_n(\bm{y})\psi_n^{\dagger}(\bm{x}), 
\end{align}
where $\Theta(y_4   -x_4)$ stands for the Heaviside step
function. Here, $\overline{m}$ represents the average mass of the
  up and down current quarks: $\overline{m}=(m_{\mathrm{u}}+
  m_{\mathrm{d}})/2$. $E_n$ is the energy eigenvalues of the
  single-quark state  
given by
\begin{align}
H \psi_n(\bm{x}) = E_n \psi_n(\bm{x}),  
\end{align}
where $H$ denotes the one-body Dirac Hamiltonian in the presence of
the pNG boson fields, which is defined by
\begin{align}
  H = \gamma_4 \gamma_i \partial_i + \gamma_4 MU^{\gamma^5}
    + \gamma_4 \bar{m} \mathbf{1}.
\end{align}
The heavy-quark propagator in the limit of $m_Q\to\infty$ is expressed
as 
\begin{align}
G_h (y,x) = \left \langle y \left |   \frac1{\partial_4 }\right |
  x\right \rangle= \Theta(y_4-x_4) \delta^{(3)}(\bm{y} - \bm{x}) .  
\end{align}
Using these quark propagators and taking the limit of
$y_4-x_4=T\to\infty$, we can derive the baryon correlation 
function $\langle J_B (y)\Psi_h(y) (-i\Psi^{\dagger}_h(x)\gamma_4)
J_B^\dagger(x)\rangle_0$ as follows~\cite{Diakonov:1987ty,
  Christov:1995vm}:  
\begin{align}
\langle J_B (y)\Psi_h(y) (-i\Psi_h^\dagger(x) \gamma_4)
J_B^\dagger(x)\rangle_0 &  \sim \exp\left[-\{(N_c-1)E_{\mathrm{val}} +
  E_{\mathrm{sea}}+m_Q\}T\right] = \exp[-M_B T].  
\label{eq:baryon_corr}
\end{align}
Since
the result for the correlation function given in
Eq.~\eqref{eq:baryon_corr} is canceled with the term
$\exp\left(-iy_4p_4'+ix_4p_4\right)= \exp[M_BT] $ in the large $N_c$ 
limit, i.e., $-ip'_{4}=-ip_{4}=M_{B}=\mathcal{O}(N_{c})$. Thus, the
normalization factor becomes
$\mathcal{N}^*(\bm{p}')\mathcal{N}(\bm{p})=2M_{B}$.  
Using this normalization and Eq.~\eqref{eq:baryon_corr}, we are able
to produce the classical mass of the singly heavy baryon correctly to
be   
\begin{align}
M_B = (N_c-1) E_{\mathrm{val}} + E_{\mathrm{sea}} + m_Q,  
\end{align}
which was already defined in a previous work~\cite{Kim:2018xlc}.
\section{EM transition form factors from the chiral quark-soliton
  model}   
In the present Section, we will present here only the final
expressions of the EM transition form factors, since detailed
formalisms of how to derive the form factors of the SU(3) baryons can
be found in previous works. For a detailed calculation, we refer to 
Refs.~\cite{Kim:1995mr, Ledwig:2008es, Kim:2019gka} (see also a
review~\cite{Christov:1995vm}). The EM current for the heavy quark
given in the second term of Eq.~\eqref{eq:EMCurrent1} can be expressed
in terms of the effective heavy quark field~\cite{Cho:1992nt} 
\begin{align}
-i \Psi_h^\dagger (x) \gamma^\mu Q_h \Psi_h(x) &= -i\exp(-im_Qv\cdot x)
  \tilde{\Psi}_h^\dagger (x) 
  \left[  v_\mu + \frac{i}{2m_Q} (\overleftarrow{\partial}_\mu -
  \overrightarrow{\partial}_\mu) + \frac1{2m_Q} \sigma_{\mu\nu} 
  (\overleftarrow{\partial}_\mu + \overrightarrow{\partial}_\mu)   
\right]Q_h \tilde{\Psi}_h(x) \cr 
&\approx -i\exp(-im_Qv\cdot x)\tilde{\Psi}_h^\dagger (x)
  v_\mu   Q_h \tilde{\Psi}_h(x). 
\end{align}
This indicates that the heavy quark does not contribute to the EM
transition form factors of the singly heavy baryons. It only gives a
constant contribution to their electric
form factors, which yields the correct charges corresponding to the
singly heavy baryon. Thus, we can simply 
consider the light-quark current to compute the matrix element
of the EM current~\cite{Kim:2019gka}
\begin{align}
\langle B^*,p' |V^\mu(0) |B,p\rangle &= 
\frac1{\mathcal{Z}_{\mathrm{eff}}} \lim_{T\to\infty} 
\exp\left[-i(p_4' + p_4) \frac{T}{2}\right] \int d^3 x d^3 y
 \exp(-i\bm{p}'\cdot \bm{y} + i \bm{p}\cdot \bm{x}) 
\int \mathcal{D} U  \mathcal{D} \psi 
  \mathcal{D} \psi^\dagger  \cr
&\hspace{-3cm} \times  J_{B^*} (\bm{y},T/2) 
(-i\psi^\dagger(0))\gamma_\mu \hat{\mathcal{Q}} \psi(0)
J_B^\dagger(\bm{x},-T/2) \exp\left[\int d^4z
  (\psi^{\dagger}(z))_\alpha^f  \left(
  i\rlap{/}{\partial}  + i MU^{\gamma_5} + i 
  \hat{m} \right)_{fg} \psi^{g\alpha}(z)\right]
\end{align} 
Since we ignore the meson fluctuation, the integration over the pNG
fields can be carried out easily. However, there are the rotational
and translational zero modes that are not at all small, so that we
need to integrate exactly over these zero modes. This is known as the
collective zero-mode quantization. For details about the zero-mode
quantization in the SU(3) $\chi$QSM, we refer to
Refs.~\cite{Blotz:1992pw, Christov:1995vm}. 

Having taking into account the rotational $1/N_{c}$ and linear $m_{s}$
corrections, we obtain the magnetic dipole form factors $G_{M1}^{B\to
  B^{*} *}$ as 
\begin{align}
  G_{M1}^{B\to B^{*} *}(Q^{2}) &=-c_{\Delta}  \int d^{3} r \frac{6}{\sqrt{2}}
                      j_{1}(|\bm{q}||\bm{r}|)\mathcal{G}^{B\to
                      B^{*}}_{M1}(\bm{r}), 
\label{eq:M1}
\end{align}
where the corresponding magnetic dipole densities
$\mathcal{G}^{B\to B^{*}}_{M1}(\bm{r})$ are defined as  
\begin{align}
  \mathcal{G}^{B\to B^{*}}_{M1}(\bm{r}) &=
\left(  \mathcal{Q}_{0} (\bm{r})  +
\frac{1}{I_{1}}  \mathcal{Q}_{1} (\bm{r}) \right)
\langle B^{*} | D^{(8)}_{Q3 } | B \rangle
-  \frac{1}{\sqrt{3}} \frac{1}{I_{1}}  \mathcal{X}_{1} (\bm{r})
\langle B^{*} | D^{(8)}_{Q 8}J_{3} | B \rangle \cr
&-   \frac{1}{I_{2}}  \mathcal{X}_{2} (\bm{r}) \langle B^{*} |
 d_{pq3} D^{(8)}_{Qp} J_{q} | B \rangle   
 - \frac{2}{3} m_{s} \left(\frac{K_{1}}{I_{1}} \mathcal{X}_{1}
  (\bm{r}) -   \mathcal{M}_{1} (\bm{r})\right)  \langle B^{*} | D^{(8)}_{83}
 D^{(8)}_{Q8} |B \rangle \cr
&-\frac{2}{\sqrt{3}} m_{s} \left(\frac{K_{2}}{I_{2}} \mathcal{X}_{2} (\bm{r})
     -    \mathcal{M}_{2} (\bm{r})\right) \langle B^{*} | d_{pq3}  D^{(8)}_{8p}
  D^{(8)}_{Qq} | B \rangle  \cr 
  & - \frac{2}{3}   m_{s} \mathcal{M}_{0} (\bm{r})
    \langle B^{*} | D^{(8)}_{Q3} | B \rangle +
    \frac{2}{3} m_{s} \mathcal{M}_{0} (\bm{r})
    \langle B^{*} | D^{(8)}_{88} D^{(8)}_{Q3} | B\rangle. 
\label{eq:M1den}
\end{align}
The explicit expressions for the densities $\mathcal{Q}_i$,
$\mathcal{X}$, and $\mathcal{M}_i$ can be found in Appendix~\ref{app:a}.
The $\langle B^{*} |...| B  \rangle$  stands for the matrix elements
of collective operators~\cite{Yang:2019tst}, of which the explicit
values are found in Appendix~\ref{app:b}. $I_i$ and $K_i$ stand for
the moments and anomalous moments of inertia~\cite{Christov:1995vm}.  
The expression for the electric quadrupole form factors is given as
\begin{align}
G_{E2}^{B\to B^{*} *} (Q^{2}) &=c_{\Delta} \int d^{3} r \sqrt{\frac{10}{9}}  
\frac{\omega_{q}}{|\bm{q}|} \left( \frac{\partial}{\partial r} r  
j_{2}(|\bm{q}||\bm{r}|)\right)     \mathcal{G}^{B\to B^{*}}_{E2} (\bm{r}), 
\label{eq:E2}
\end{align}
with the electric quadrupole densities $\mathcal{G}^{B\to B^{*}}_{E2} (\bm{r})$
\begin{align}
\mathcal{G}^{B\to B^{*}}_{E2}(\bm{r}) =& - \frac{2}{I_{1}}
\mathcal{I}_{1E2} (\bm{r})\left(3
    \langle B^{*} | D^{(8)}_{Q 3} J_{3} | B \rangle
    - \langle B^{*} | D^{(8)}_{Q i} J_{i} |B \rangle  \right) \cr
 & - \frac{4}{\sqrt{3}} m_{s}\left( \frac{K_{1}}{I_{1}} \mathcal{I}_{1E2}(\bm{r}) -
    \mathcal{K}_{1E2}(\bm{r})\right) \left( 3\langle B^{*} | D^{(8)}_{8 3}
    D^{(8)}_{Q 3} | B\rangle -\langle B^{*} | D^{(8)}_{8 i} D^{(8)}_{Q i}
   |B \rangle \right). 
\label{eq:E2den}
\end{align}
The explicit expressions for $\mathcal{I}_{1E2}(\bm{r})$ and $
\mathcal{K}_{1E2}(\bm{r})$ can be found in Appendix~\ref{app:a}. 
The Coulomb quadrupole form factor $G_{C2}^{B\to B^{*}*}$ is
written as  
\begin{align}
  G_{C2}^{B\to B^{*}*}(Q^{2}) &=c_{\Delta} \sqrt{40}\int d^{3} r  
                      \, \frac{M_{B^{*}}}{|\bm{q}|} j_{2}(|{\bm q}||{\bm r}|)
                      \mathcal{G}^{B\to B^{*}}_{C2} (\bm{r}), 
\label{eq:C2}
\end{align}
where  ${\cal G}^{B\to B^{*} }_{C2} (\bm{r})$ is simply the same as
${\cal G}^{B\to B^{*} }_{E2} (\bm{r})$. Note that for the E2 and
C2 form factors, the leading contributions in the large $N_c$
expansion vanish, so that the rotational $1/N_c$ corrections take over
the role of the leading-order contributions. 

In order to scrutinize each contribution, it is more convenient to
decompose the densities into three different terms 
\begin{align}
  \mathcal{G}^{B\to B^{*} }_{(M1,E2,C2)}(\bm{r}) =
     \mathcal{G}^{B\to B^{*}(0)}_{(M1,E2,C2)}(\bm{r})
  + \mathcal{G}^{B\to B^{*} (\mathrm{op})}_{(M1,E2,C2)}(\bm{r})
  + \mathcal{G}^{B\to B^{*} (\mathrm{wf})}_{(M1,E2,C2)}(\bm{r}).
\end{align}
The first term represents the SU(3) symmetric terms including both the
leading and rotational $1/N_c$ contributions, the second one denotes the
linear $m_{\mathrm{s}}$ corrections arising from the current-quark
mass term of the effective chiral action, and the last terms 
come from the collective baryon wave functions. When the effects of
the flavor SU(3) symmetry breaking are considered, a collective baryon
wave function is not any longer in a pure state but becomes a state
mixed with higher representations. Thus,
there are two different terms that provide the effects of flavor SU(3)
symmetry breaking. The explicit expressions of these three terms for
the M1 form factors are then given as follows
\begin{align}
  \mathcal{G}^{\overline{\bm{3}}_{1/2}\to \bm{6}_{3/2} (0)}_{M1}(\bm{r}) & =
   \frac{1}{4\sqrt{3}} 
 \mathcal{Q}_{B\to B^{*}}\left(  \mathcal{Q}_{0}  (\bm{r}) + \frac{1}{I_{1}}
 \mathcal{Q}_{1} (\bm{r}) + \frac{1}{2}
 \frac{1}{I_{2}}  \mathcal{X}_{2} (\bm{r}) \right)  ,
\label{eq:M1leadingcon} \\
\mathcal{G}^{\overline{\bm{3}}_{1/2}\to \bm{6}_{3/2}
  (\mathrm{op})}_{M1}(\bm{r}) &= 
-\frac{m_{s}}{60\sqrt{3}}\begin{pmatrix} \mathcal{Q}_{\Lambda_{c} \to
    \Sigma_{c}^*} \\-4\mathcal{Q}_{\Xi_{c} \to \Xi_{c}^*} + 3
  \\  \end{pmatrix} 
  \left(\frac{K_{1}}{I_{1}}\mathcal{X}_{1}
  (\bm{r}) -   \mathcal{M}_{1} (\bm{r})\right) \cr 
  &-\frac{m_{s}}{10\sqrt{3}}\begin{pmatrix} \mathcal{Q}_{\Lambda_{c}
      \to \Sigma_{c}^*} \\-\mathcal{Q}_{\Xi_{c} \to \Xi_{c}^*} + 2
    \\  \end{pmatrix} 
  \left(\frac{K_{2}}{I_{2}}\mathcal{X}_{2}
  (\bm{r}) -   \mathcal{M}_{2} (\bm{r})\right)
  +\frac{m_{s}}{60\sqrt{3}}\begin{pmatrix} -7\mathcal{Q}_{\Lambda_{c}
      \to \Sigma_{c}^*} \\ \mathcal{Q}_{\Xi_{c} \to \Xi_{c}^*}+1
    \\  \end{pmatrix} 
  \mathcal{M}_{0}(\bm{r}),
\label{eq:M1op} \\
\mathcal{G}^{\overline{\bm{3}}_{1/2}\to \bm{6}_{3/2}
  (\mathrm{wf})}_{M1}(\bm{r}) &= 
\frac{q_{\overline{15}}}{40\sqrt{6}}
\begin{pmatrix}
  4\mathcal{Q}_{\Lambda_{c} \to \Sigma_{c}^*} \\-\mathcal{Q}_{\Xi_{c}
    \to \Xi_{c}^*} + 2 \\  
\end{pmatrix} 
  \left(\mathcal{Q}_{0}  (\bm{r}) + \frac{1}{I_{1}}
 \mathcal{Q}_{1} (\bm{r}) -\frac{1}{2} \frac{1}{I_{2}}\mathcal{X}_{2}
  (\bm{r})\right) \cr 
&-\frac{p_{\overline{15}}}{240}
\begin{pmatrix} 2\mathcal{Q}_{\Lambda_{c} \to \Sigma_{c}^*}
  \\-5\mathcal{Q}_{\Xi_{c} \to \Xi_{c}^*} + 4 \\  
\end{pmatrix} 
  \left(\mathcal{Q}_{0}  (\bm{r}) + \frac{1}{I_{1}} 
 \mathcal{Q}_{1} (\bm{r}) +\frac{3}{2} \frac{1}{I_{2}}\mathcal{X}_{2}
  (\bm{r})\right), 
\label{eq:M1wf} 
\end{align}
in the basis of the $[\Lambda_{c} \to \Sigma_{c}^*, \Xi_{c} \to
\Xi_{c}^*]$ for $\overline{\bm{3}}_{1/2}\to \bm{6}_{3/2}$ 
\begin{align}
  \mathcal{G}^{\bm{6}_{1/2}\to \bm{6}_{3/2} (0)}_{M1}(\bm{r}) & =
   \frac{1}{30\sqrt{2}} 
 (3\mathcal{Q}_{B\to B^{*}}-2)\left(  \mathcal{Q}_{0}  (\bm{r}) + \frac{1}{I_{1}}
 \mathcal{Q}_{1} (\bm{r}) + \frac{1}{3}
 \frac{1}{I_{1}}  \mathcal{X}_{1} (\bm{r})+ \frac{1}{2}
 \frac{1}{I_{2}}  \mathcal{X}_{2} (\bm{r}) \right)  ,
\label{eq:M1leadingcon1} \\
\mathcal{G}^{\bm{6}_{1/2}\to \bm{6}_{3/2} (\mathrm{op})}_{M1}(\bm{r})
&= -\frac{m_{s}}{270\sqrt{2}}\begin{pmatrix} 4\mathcal{Q}_{\Sigma_{c}
    \to \Sigma_{c}^*}-5 \\ 2\mathcal{Q}_{\Xi'_{c} \to \Xi_{c}^*}-1 \\
  2\mathcal{Q}_{\Omega_{c} \to \Omega_{c}^*}+3  \end{pmatrix} 
  \left(\frac{K_{1}}{I_{1}}\mathcal{X}_{1}
  (\bm{r}) -   \mathcal{M}_{1} (\bm{r})\right) \cr 
  &-\frac{m_{s}}{135\sqrt{2}}\begin{pmatrix} 5\mathcal{Q}_{\Sigma_{c}
      \to \Sigma_{c}^*}+7 \\ 7\mathcal{Q}_{\Xi'_{c} \to \Xi_{c}^*}-2
    \\ \mathcal{Q}_{\Omega_{c} \to \Omega_{c}^*}+3  \end{pmatrix} 
  \left(\frac{K_{2}}{I_{2}}\mathcal{X}_{2}
  (\bm{r}) -   \mathcal{M}_{2} (\bm{r})\right)
  +\frac{m_{s}}{270\sqrt{2}}\begin{pmatrix} -14\mathcal{Q}_{\Sigma_{c}
      \to \Sigma_{c}^*} +7 \\ -16\mathcal{Q}_{\Xi'_{c} \to
      \Xi_{c}^*}+11 \\ -16\mathcal{Q}_{\Omega_{c} \to \Omega_{c}^*}+15
    \\  \end{pmatrix} 
  \mathcal{M}_{0}(\bm{r}),
  \label{eq:M1op1}  \\
\mathcal{G}^{{\bm{6}}_{1/2}\to \bm{6}_{3/2}
  (\mathrm{wf})}_{M1}(\bm{r}) &= 
\frac{q_{\overline{15}}}{180}\begin{pmatrix} 4\mathcal{Q}_{\Sigma_{c}
    \to \Sigma_{c}^*} - 8 \\ \mathcal{Q}_{\Xi'_{c} \to \Xi_{c}^*} - 4
  \\ 0  \end{pmatrix} 
  \left(\mathcal{Q}_{0}  (\bm{r}) + \frac{1}{I_{1}}
 \mathcal{Q}_{1} (\bm{r}) - \frac{1}{I_{1}}\mathcal{X}_{1}(\bm{r})
  -\frac{1}{2} \frac{1}{I_{2}}\mathcal{X}_{2} (\bm{r})\right) \cr 
&-\frac{q_{\overline{24}}}{180\sqrt{5}}\begin{pmatrix}
  \mathcal{Q}_{\Sigma_{c} \to \Sigma_{c}^*} + 1 \\
  2\mathcal{Q}_{\Xi'_{c} \to \Xi_{c}^*} + 4 \\
  3\mathcal{Q}_{\Omega_{c} \to \Omega_{c}^*} + 6  \end{pmatrix} 
  \left(\mathcal{Q}_{0}  (\bm{r}) + \frac{1}{I_{1}}
 \mathcal{Q}_{1} (\bm{r})  - \frac{2}{I_{1}}\mathcal{X}_{1}(\bm{r})  -
  \frac{2}{I_{2}}\mathcal{X}_{2} (\bm{r})\right),  
\label{eq:M1wf1} 
\end{align}
in the basis of the $[\Sigma_{c} \to \Sigma_{c}^*, \Xi'_{c} \to
\Xi_{c}^*, \Omega_{c} \to \Omega_{c}^*]$ for $\bm{6}_{1/2}\to
\bm{6}_{3/2}$. Similarly, the densities for the E2 form factors are
written by  
\begin{align}
  \mathcal{G}^{\bm{6}_{1/2}\to \bm{6}_{3/2} (0)}_{E2}(\bm{r}) & =
  -\frac{1}{5\sqrt{2}} 
 (3\mathcal{Q}_{B\to B^{*}}-2) \frac{1}{I_{1}}\mathcal{I}_{1E2}  ,
\label{eq:E2leadingcon} \\
\mathcal{G}^{\bm{6}_{1/2}\to \bm{6}_{3/2} (\mathrm{op})}_{E2}(\bm{r}) &=
-\frac{8m_{s}}{135\sqrt{2}}\begin{pmatrix} -2\mathcal{Q}_{\Sigma_{c}
    \to \Sigma_{c}^*}+1 \\ 8\mathcal{Q}_{\Xi'_{c} \to \Xi_{c}^*}-1 \\
  -8\mathcal{Q}_{\Omega_{c} \to \Omega_{c}^*}+3  \end{pmatrix} 
  \left(\frac{K_{1}}{I_{1}}\mathcal{I}_{1E2}
  (\bm{r}) -   \mathcal{K}_{1E2} (\bm{r})\right),
\label{eq:E2op} \\
\mathcal{G}^{{\bm{6}}_{1/2}\to \bm{6}_{3/2} (\mathrm{wf})}_{E2}(\bm{r}) &=
-\frac{1}{30}\left[q_{\overline{15}} \begin{pmatrix}
    4\mathcal{Q}_{\Sigma_{c} \to \Sigma_{c}^*}-8 \\
    5\mathcal{Q}_{\Xi'_{c} \to \Xi_{c}^*}-4 \\ 0  \end{pmatrix}
  -\frac{q_{\overline{24}}}{\sqrt{5}} \begin{pmatrix}
    \mathcal{Q}_{\Sigma_{c} \to \Sigma_{c}^*}+1 \\
    2\mathcal{Q}_{\Xi'_{c} \to \Xi_{c}^*}+4 \\
    3\mathcal{Q}_{\Omega_{c} \to \Omega_{c}^*}+6  \end{pmatrix} 
 \right]\frac{1}{I_{1}}\mathcal{I}_{1E2}(\bm{r}),
\label{eq:E2wf} 
\end{align}
in the basis of the $[\Sigma_{c} \to \Sigma_{c}^*, \Xi'_{c} \to
\Xi_{c}^*, \Omega_{c} \to \Omega_{c}^*]$ for $\bm{6}_{1/2}\to
\bm{6}_{3/2}$.  $\mathcal{Q}_{B \to B^{*}}$ stand for the charges of
the corresponding heavy baryons. 
Note that the E2 and C2 transtion from factors
($\overline{\bm{3}}_{1/2}(J=0) \to \bm{6}_{3/2}(J=1)$) are forbidden
within this model. 

\section{Results and discussion}
\subsection{Comparison with the lattice data}
\begin{figure}[htp] 
\includegraphics[scale=0.25]{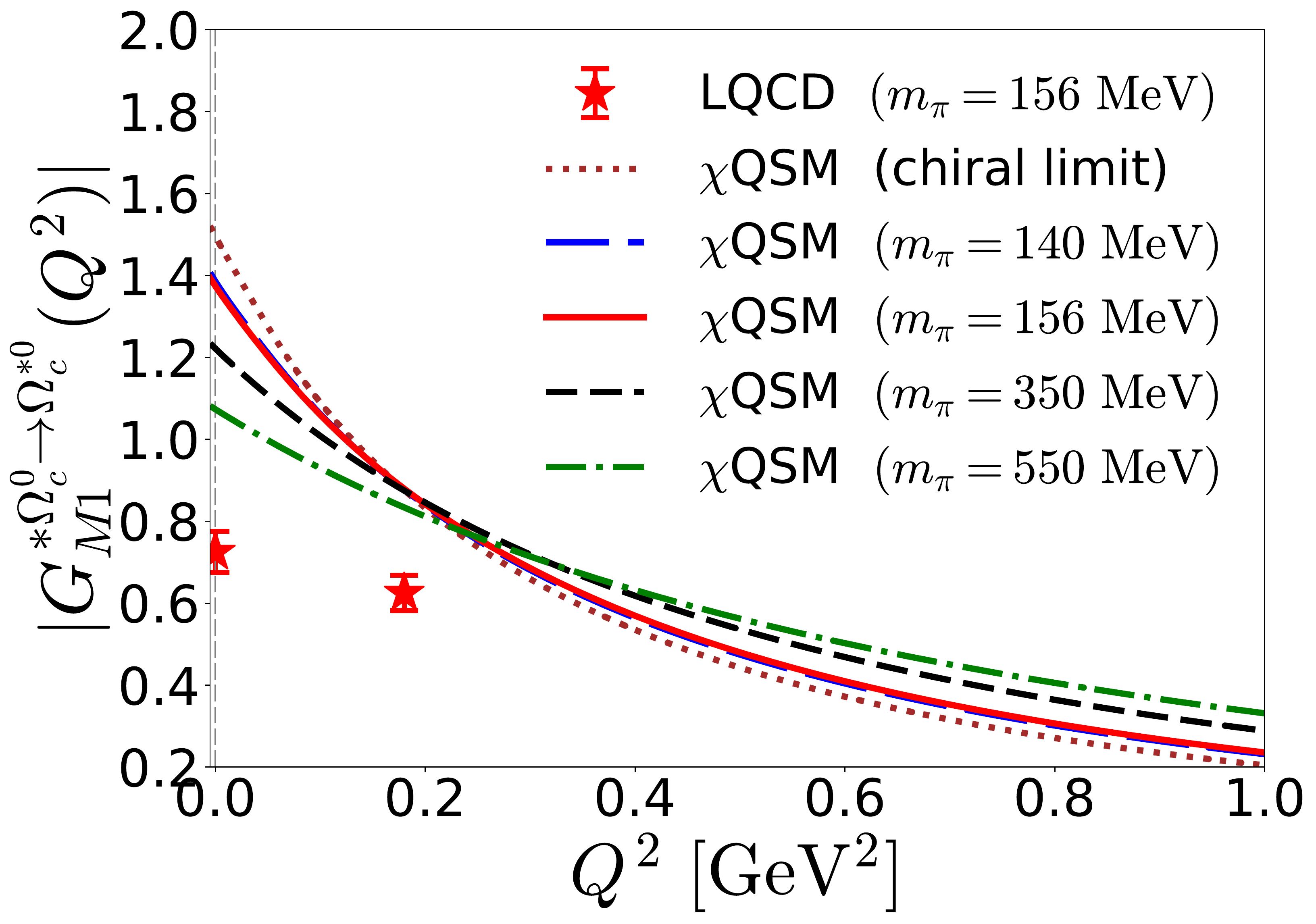}
\includegraphics[scale=0.25]{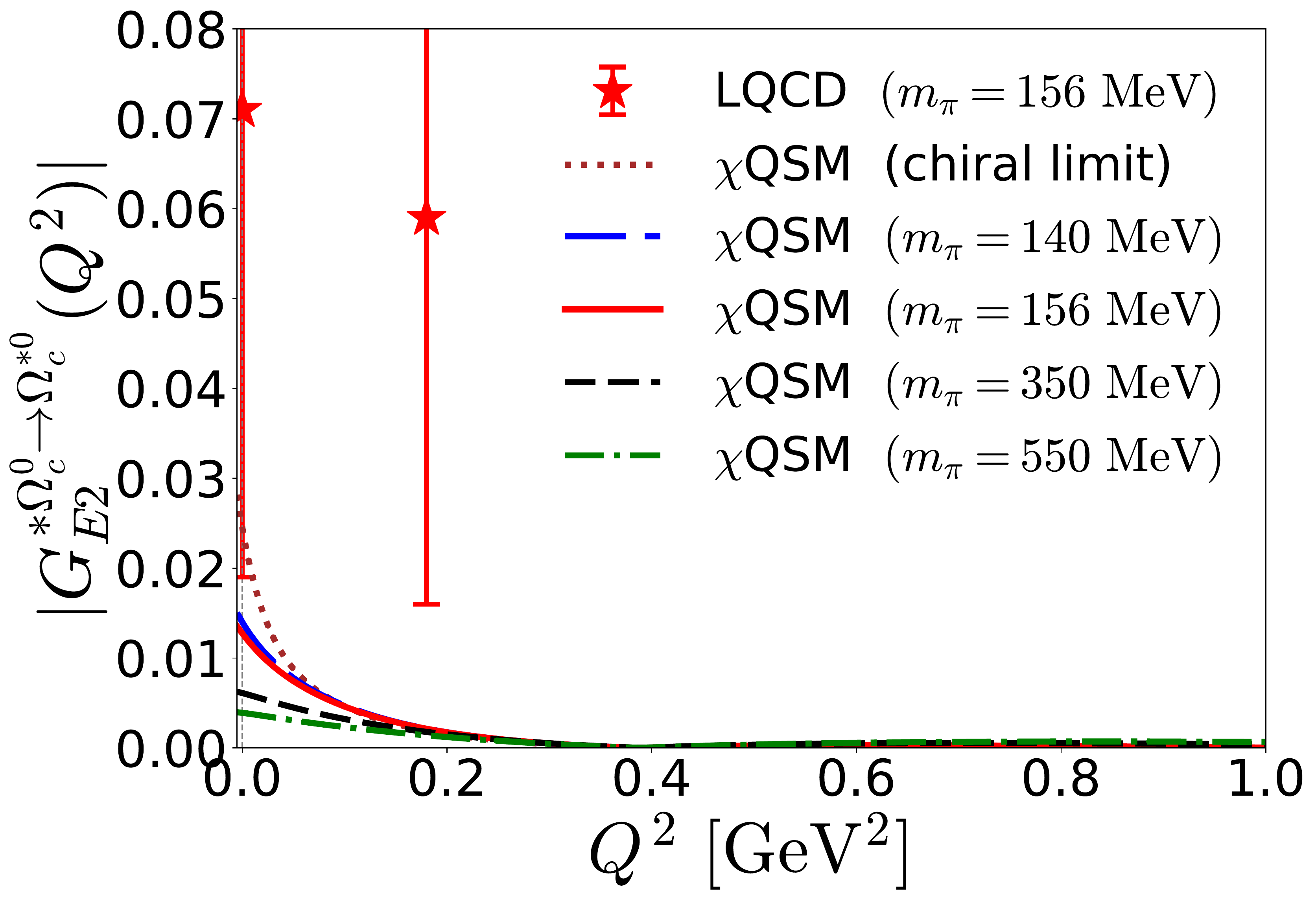}
\caption{Numerical results for the magnetic dipole and electric
  quadrupole transition form factors of the $\Omega_{c} \gamma \to 
  \Omega^{*}_{c}$ transition with the pion mass varied from 0 to
  550 MeV, drawn respectively in the left and right panels. 
  The results are compared with the lattice data taken
  from~Ref.\cite{Bahtiyar:2015sga}}  
\label{fig:1} 
\end{figure}

Since we use exactly the same set of the model parameters as in
Refs.~\cite{Kim:2018nqf, Kim:2020uqo, Kim:2018xlc}, we proceed to
present the numerical results and discuss them. To compare the present
results with those from lattice QCD~\cite{Bahtiyar:2015sga}, we need
to employ the values of the unphysical pion mass. We refer to
Ref.~\cite{Kim:2019wbg} for details.  
In the left and right panels of Fig.~\ref{fig:1}, we draw,
respectively, the results for the M1 and E2 transition form
factors with the pion mass varied from the chiral limit ($m_\pi=0$) to
$m_\pi=550$ MeV. We find that when the larger value of the pion mass
is used, the results for both the M1 and E2 form 
factors fall off more slowly, as $Q^2$ increases. This feature is
already known from the results for the EM form factors of both the
light and singly heavy baryons~\cite{Kim:2018nqf, Kim:2019wbg,
  Kim:2020uqo, Kim:2019gka}. There are only two lattice data at
$Q^2=0$ and $Q^2=0.2\,\mathrm{GeV}^2$ and they indicate that the
lattice data on the M1 transition form factor of the $\Omega_c^0
\gamma\to \Omega_c^{*0}$ process falls off rather slowly, compared to
the present results with the corresponding value of the pion mass,
i.e., $m_\pi=156$ MeV. As shown in the left panel of Fig.~\ref{fig:1},
the present results are overestimated approximately by 50~\%. 
Considering the fact that the lattice data on the $\Omega_c^0 \to
\Omega_c^{*0}$ E2 transition form factor show large
numerical uncertainties, we are not able to draw any definitive
conclusions from the comparison with the lattice data. Actually, the
lattice data on the E2 form factors of $\Omega_c^{*0}$ contain
similar uncertainties as shown in Ref.~\cite{Can:2015exa}. We
anticipate future experimental and lattice data, which will allow one
to make a quantitative comparison.

It is also interesting to see that the magnitude of the E2 form factor of
the $\Omega_c^{0*}\to \Omega_c$ transition increases drastically as
$Q^2$ gets closer to zero. This is in line with what was found in 
Ref.~~\cite{Savage:1994wa}, where the radiative decay $\Sigma_c^*\to
\Lambda_c \gamma$ was examined. This indicates that the effects of the
vacuum polarization or the sea quarks become dominant over those of
the valence quarks as $Q^2$ decreases. We will later discuss each 
contribution of the valence and sea quarks to the E2 form factors in
detail. 

It is of great importance to know the magnetic dipole transition form 
factors of the baryon sextet with spin 3/2, since they provide 
essential information on their radiative decays. As expressed in
Eq.~\eqref{eq:decay_width}, the values of the M1 and E2 transition
form factors at $Q^2=0$ will determine the decay rates of the
radiative decays of the baryon sextet with spin 3/2. However, since
the values of the E2 transition form factors are known to be rather
small in the case of the baryon decuplet, we expect that they would be
also small in the case of the baryon sextet with spin 3/2. As will be
shown later, the magnitudes of the E2 transition form factors are
indeed very small, compared with those of the M1 transition form
factors. 

\subsection{Valence- and sea-quark contributions}
\begin{figure}
\centering
\includegraphics[scale=0.25]{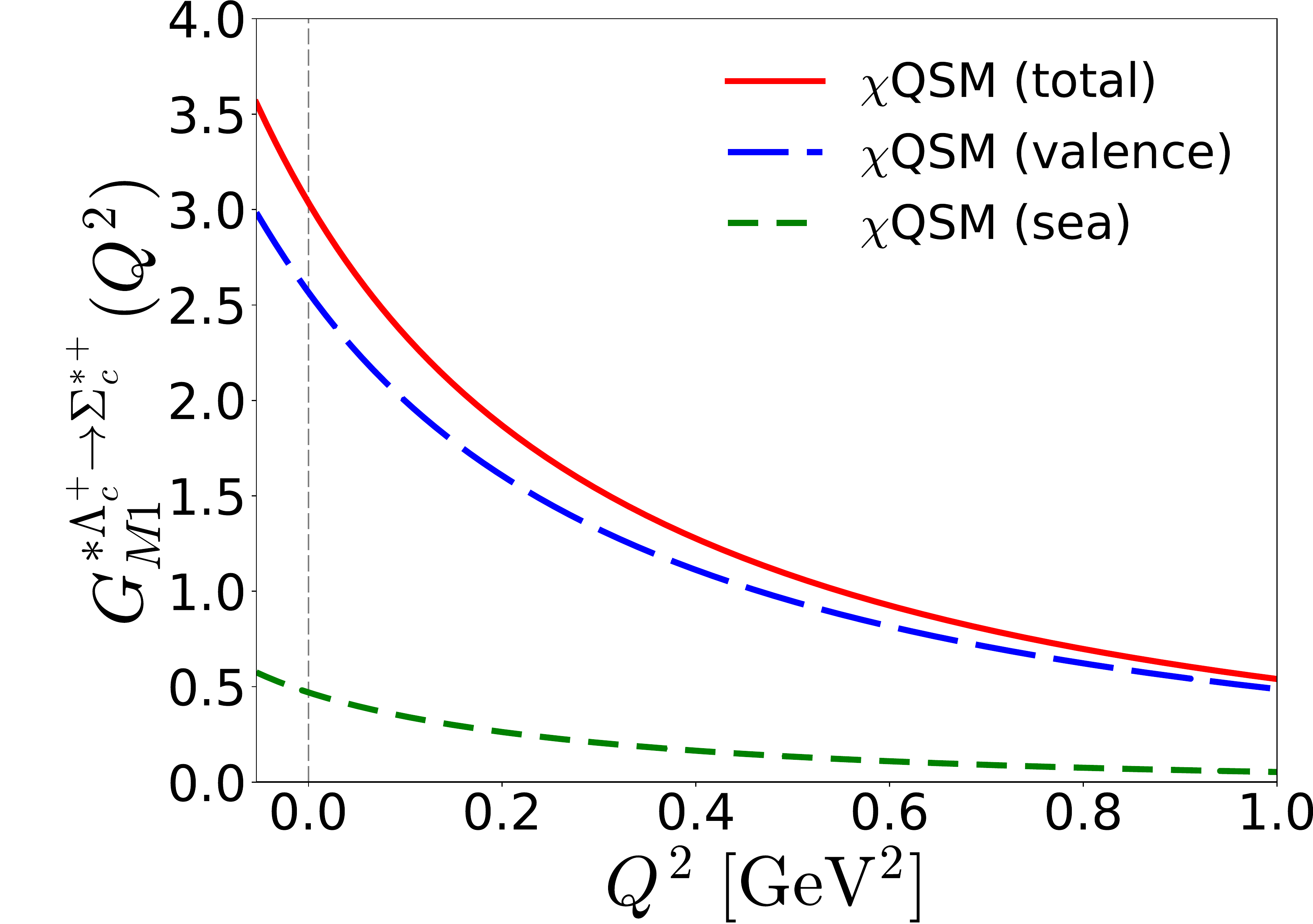}
\includegraphics[scale=0.25]{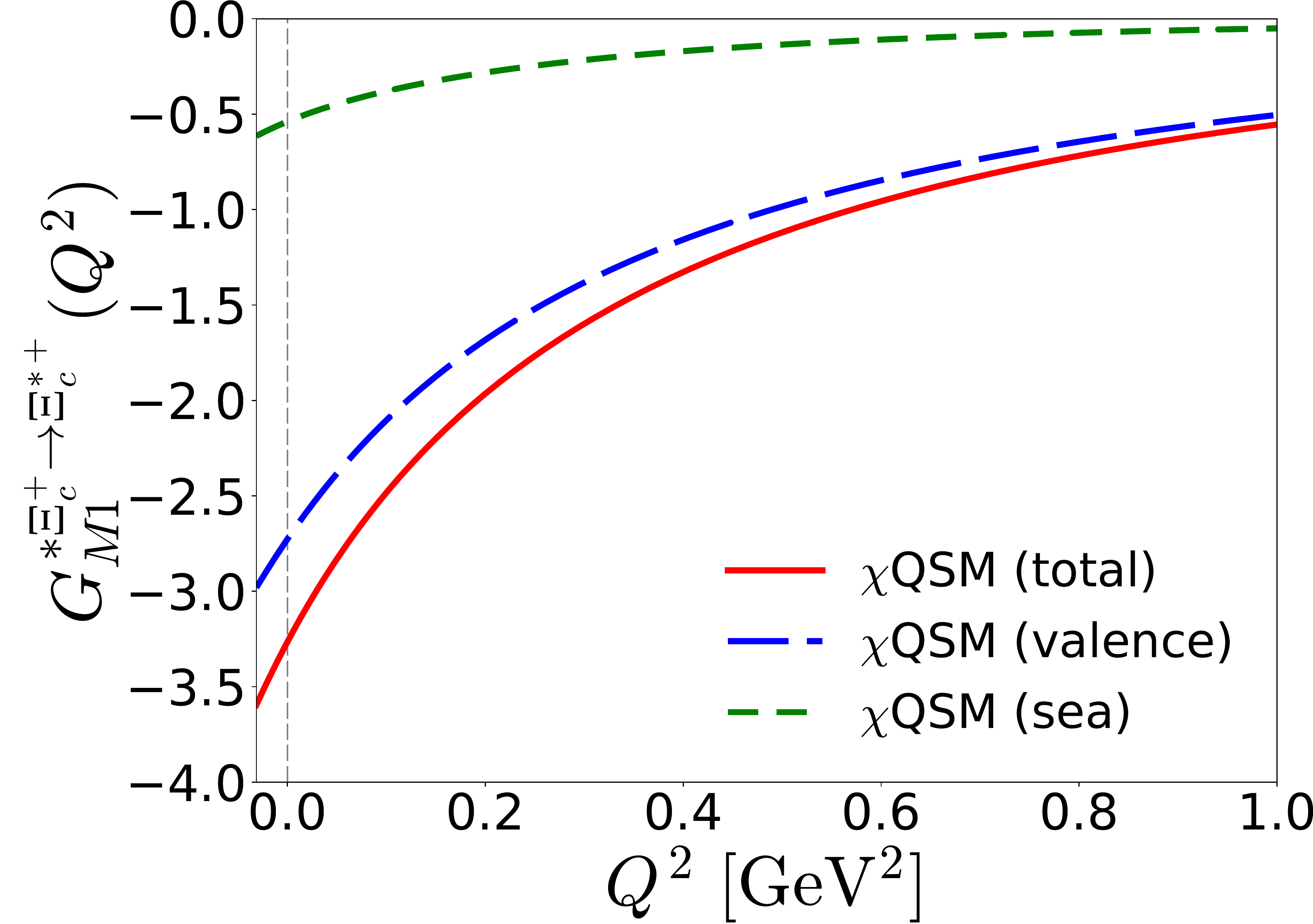}
\includegraphics[scale=0.25]{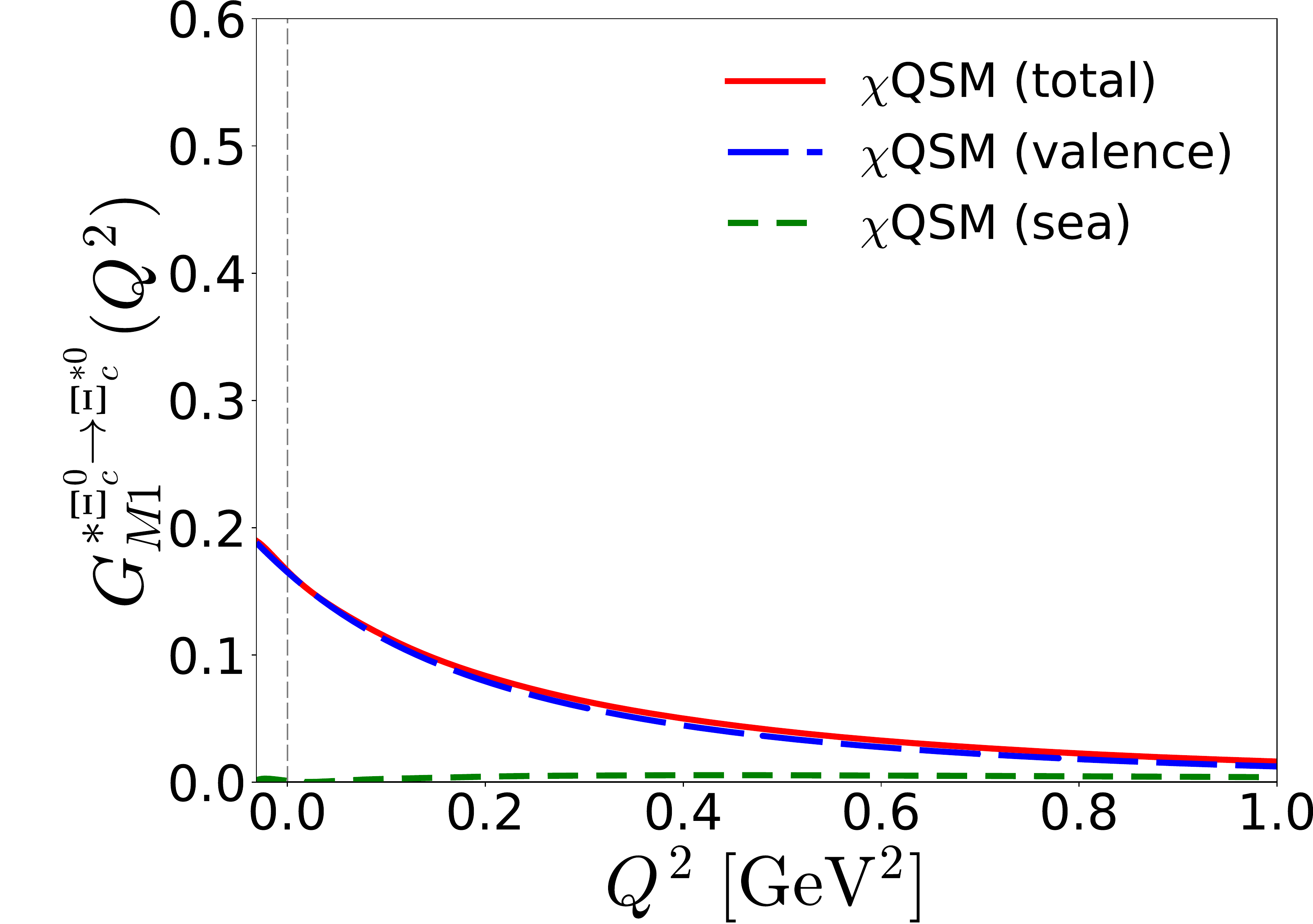}
\caption{Results for the magnetic dipole transition form
  factors from the baryon antitriplet to the baryon sextet with
  spin 3/2, with the valence- and sea-quark contributions 
  separated. The dashed and short-dashed curves draw the valence- and
  sea-quark contributions, respectively. The solid ones depict the
  total results.} 
\label{fig:2}
\end{figure}

\begin{figure}
\centering
\includegraphics[scale=0.25]{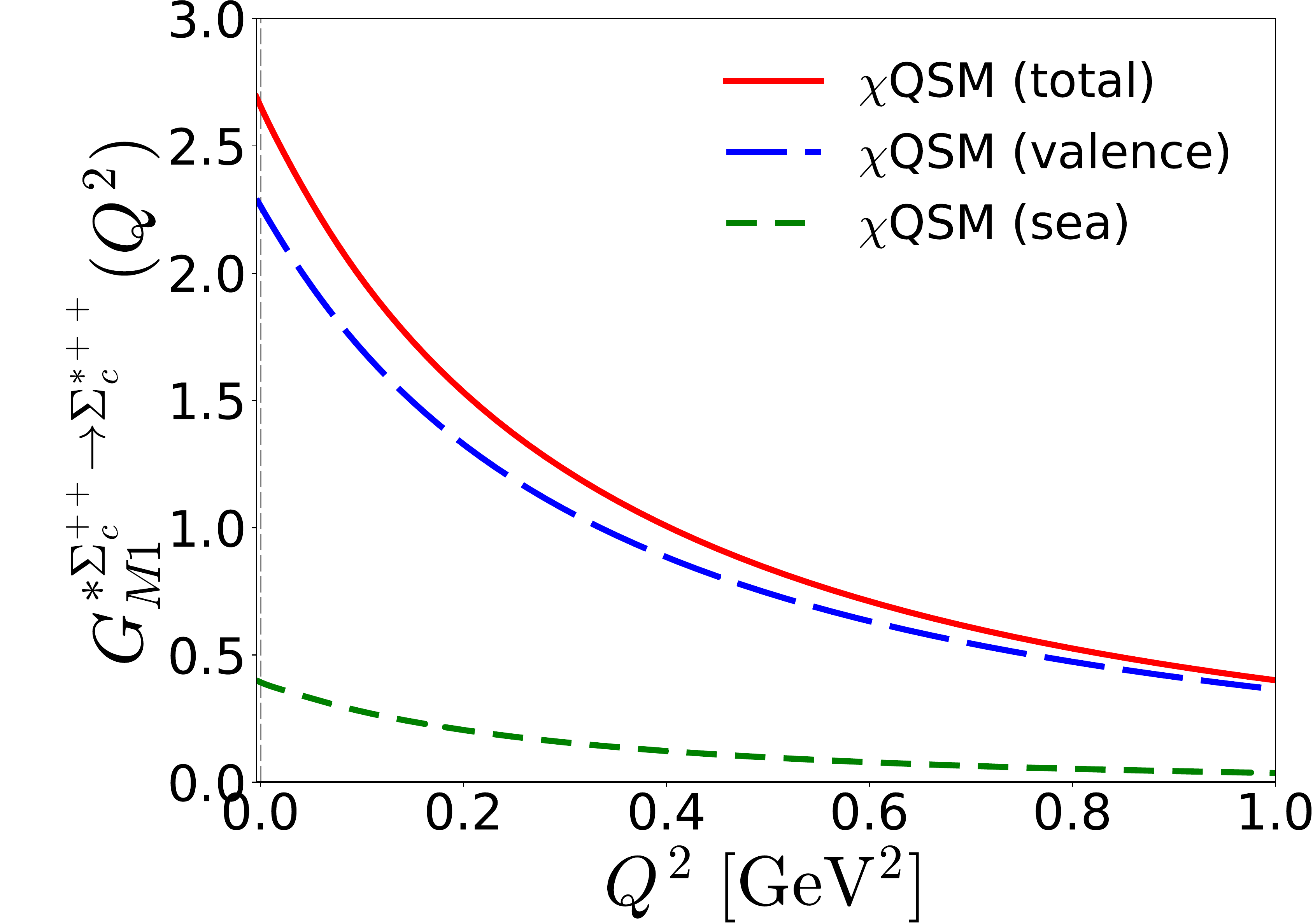}
\includegraphics[scale=0.25]{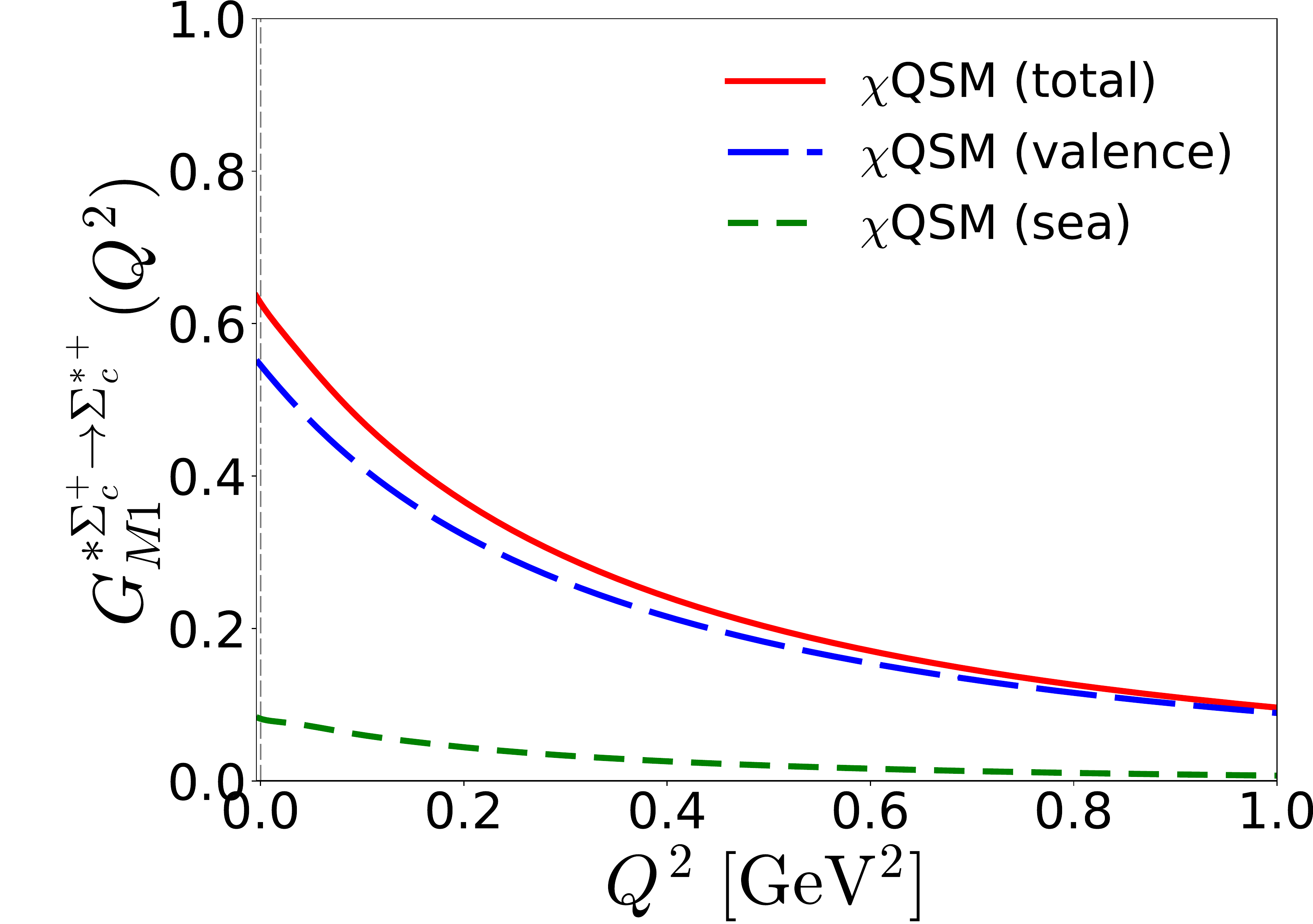}
\includegraphics[scale=0.25]{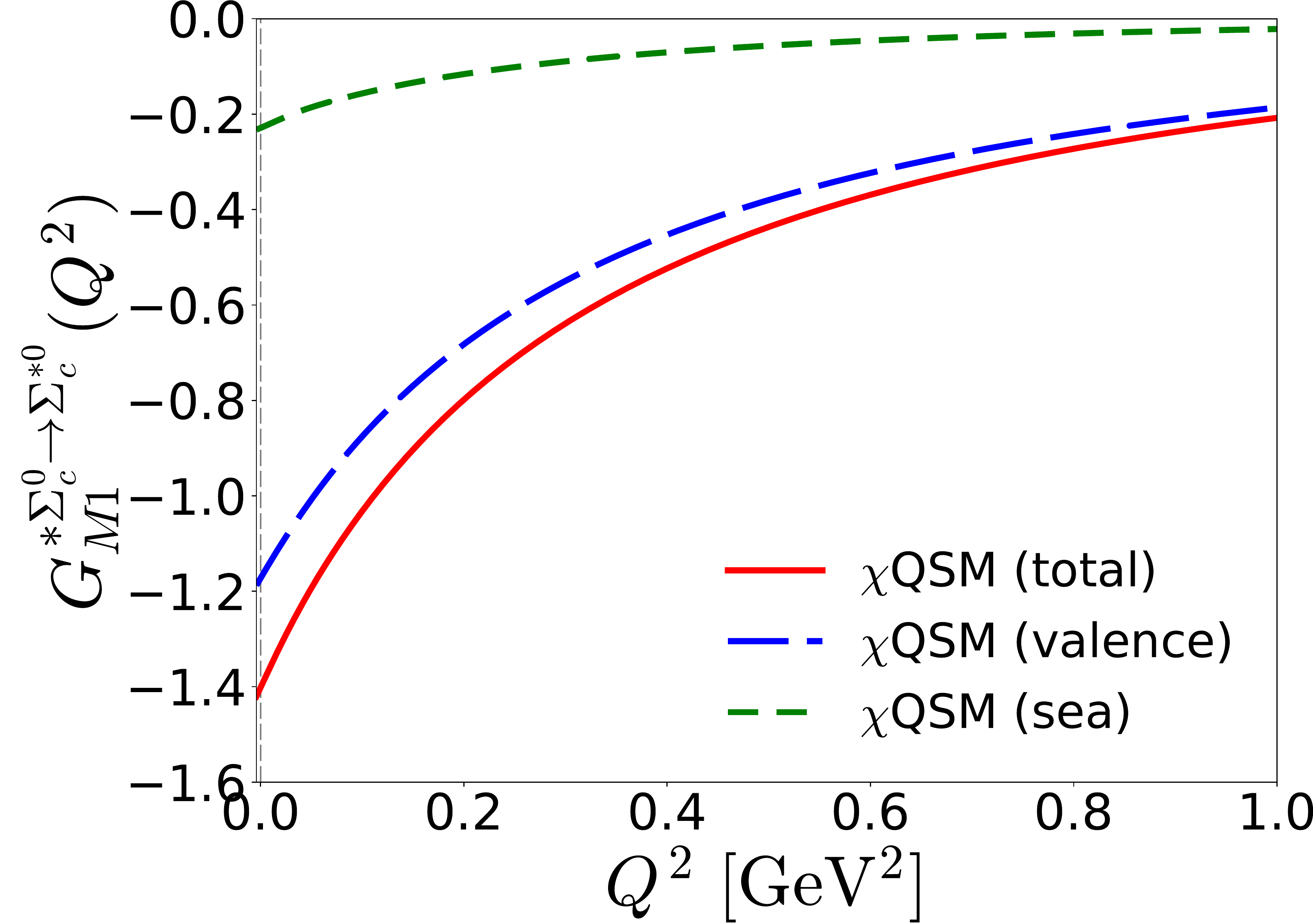}
\includegraphics[scale=0.25]{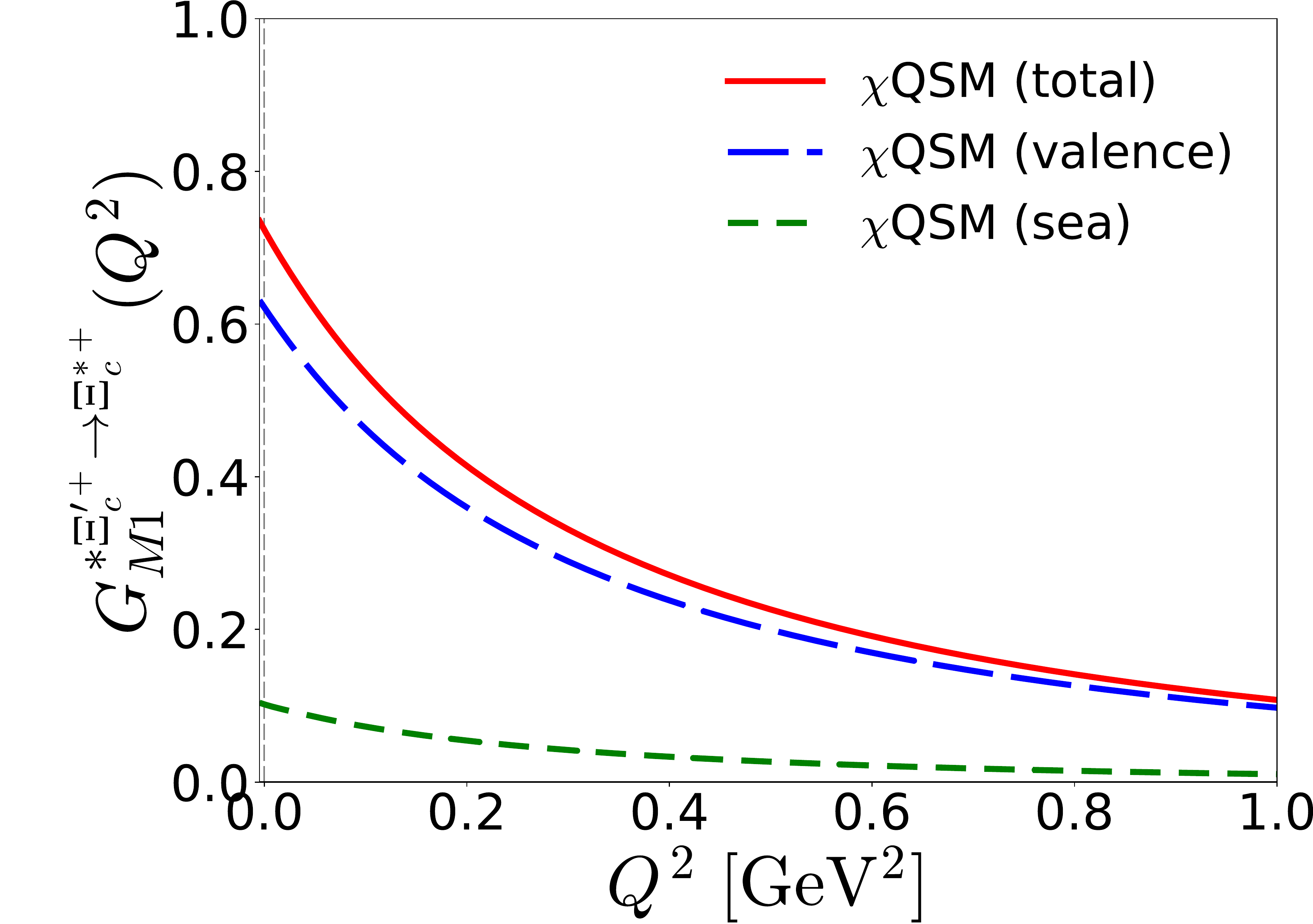}
\includegraphics[scale=0.25]{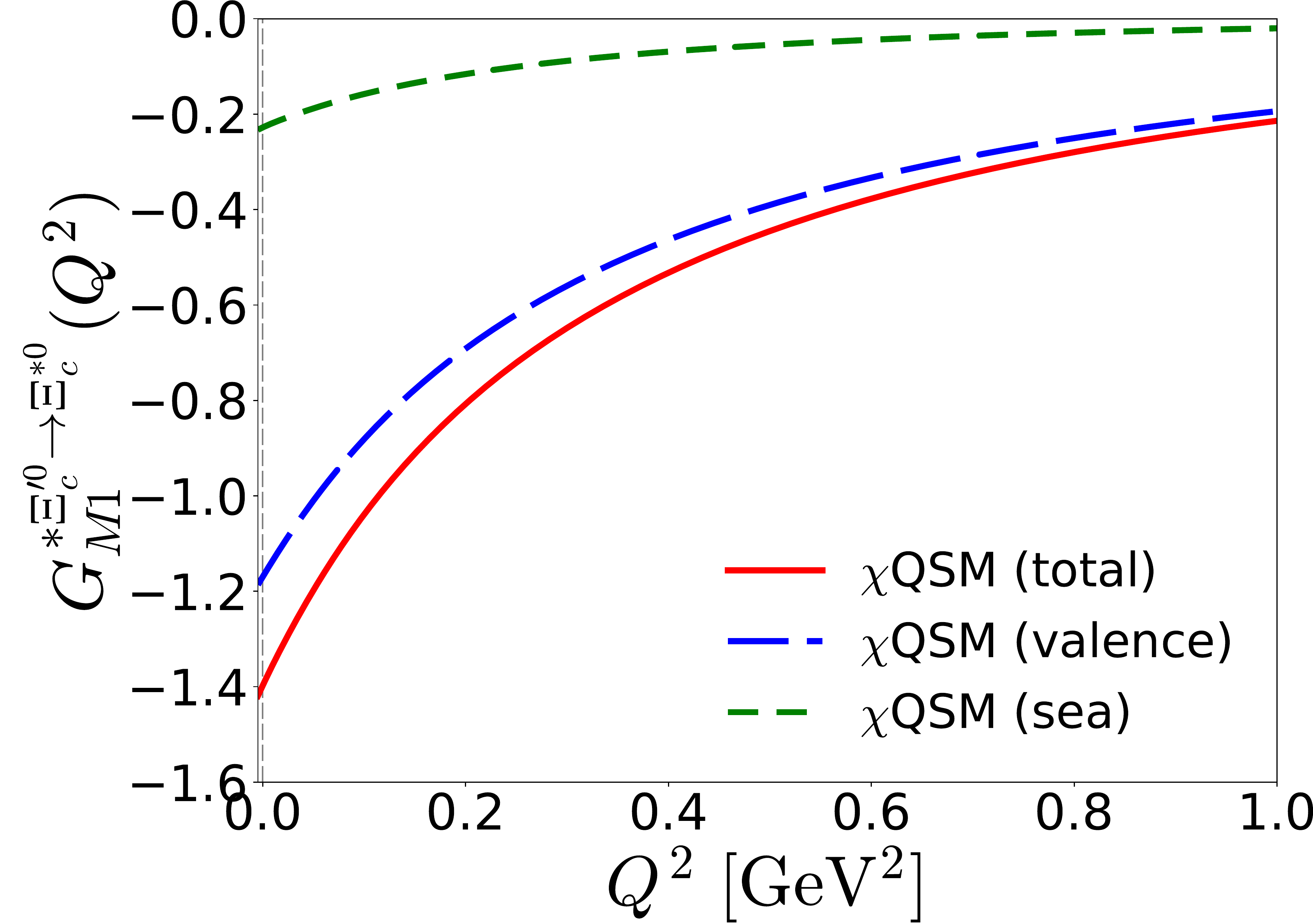}
\includegraphics[scale=0.25]{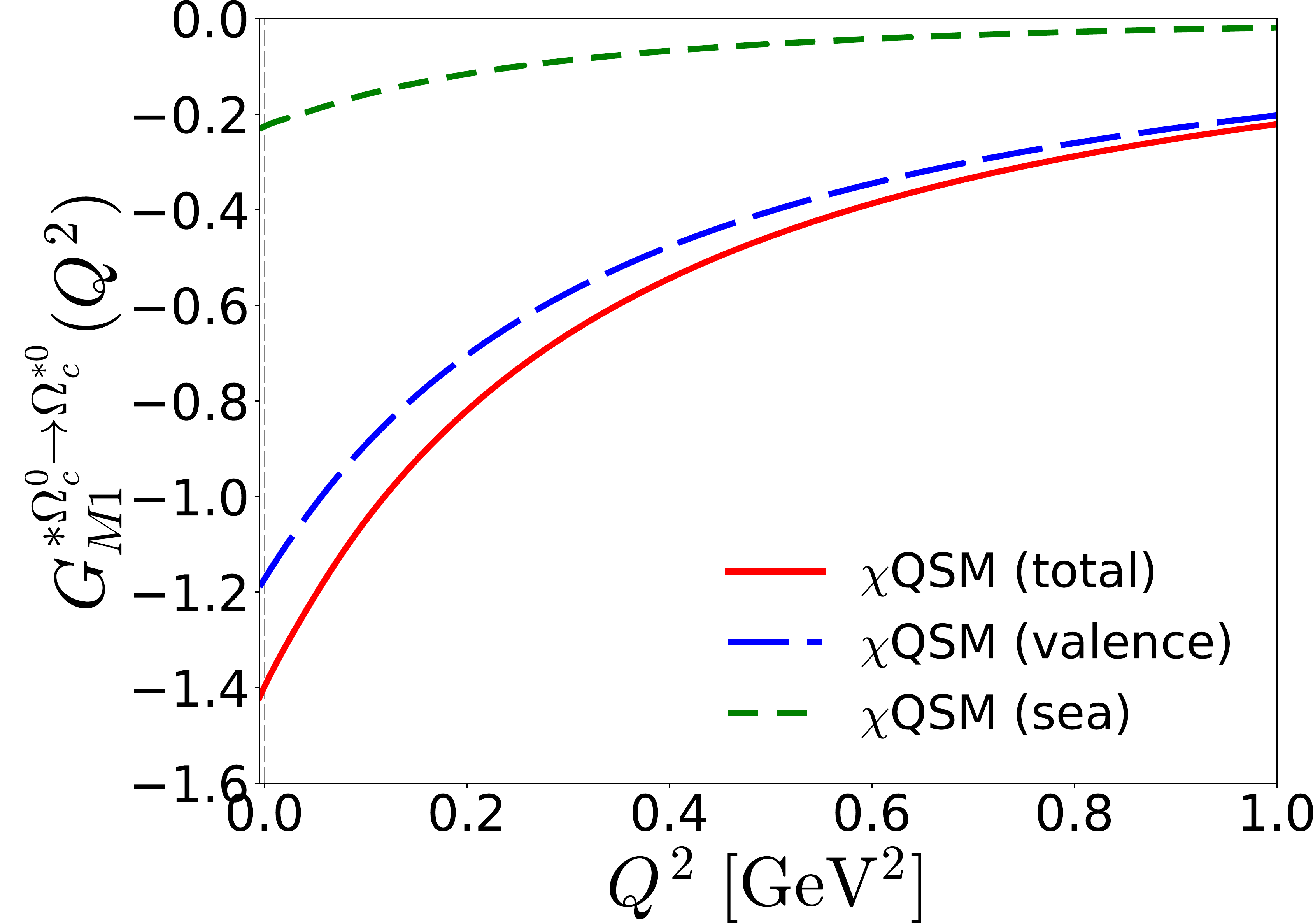}
\caption{Results for the magnetic dipole transition form
  factors from the baryon sextet with spin 1/2 to the baryon sextet
  with spin 3/2, with the valence- and sea-quark contributions 
  separated. The notations are the same as in Fig.~\ref{fig:2}.}
\label{fig:3}
\end{figure}
In Fig.~\ref{fig:2}, we show the results for the M1 form factors of
the EM transitions from the baryon antitriplet to the baryon sextet
with spin 3/2, drawing separately the valence- and sea-quark 
contributions. On the other hand, Figure~\ref{fig:3} depicts the
results for those from the baryon sextet with spin 1/2 to the baryon
sextet with spin 3/2, again with the valence- and sea-quark 
contributions separated. In general, the valence-quark contributions
dominate over those of the sea quarks. In the case of the radiative
excitation $\Xi_c^0 \gamma \to \Xi_c^{*0}$, the effect of the sea
quarks is negligibly small. Note that the magnitude of this M1 form
factor is approximately more than ten times smaller, compared with
those for the  $\Lambda_c^+ \gamma \to \Sigma_c^{*+}$ and  $\Xi_c^+
\gamma \to  \Xi_c^{*+}$ excitations. This is due to the $U$-spin
symmetry, which will be discussed later. Comparing the results in
Fig.~\ref{fig:2}  with those in Fig.~\ref{fig:3}, we find that the M1
excitations for the baryon antitriplet are larger than those for the
baryon sextet except for the $\Xi_c^0 \gamma \to \Xi_c^{*0}$
 excitation. While this can be understood by comparing 
 Eq.~\eqref{eq:M1leadingcon} with Eq.~\eqref{eq:M1leadingcon1}, the
 physical interpretation of these results is originated from the spin
 configuration of the light valence quarks inside heavy baryons. The
spins of the light valence quarks for the baryon antitriplet are in
the spin-singlet state ($S_L=0$), whereas those for the baryon sextet
are in the spin-triplet state ($S_L=1$). The M1 transitions occur
more likely due to the spin flip of the light valence quarks based on
the naive quark model. Thus, the M1 form factors for the radiative
excitations from the baryon antitriplet to the baryon sextet with spin
3/2 turn out naturally larger than those from the baryon sextet with
spin 1/2 to the baryon sextet with spin 3/2. 

\begin{figure}
\centering
\includegraphics[scale=0.25]{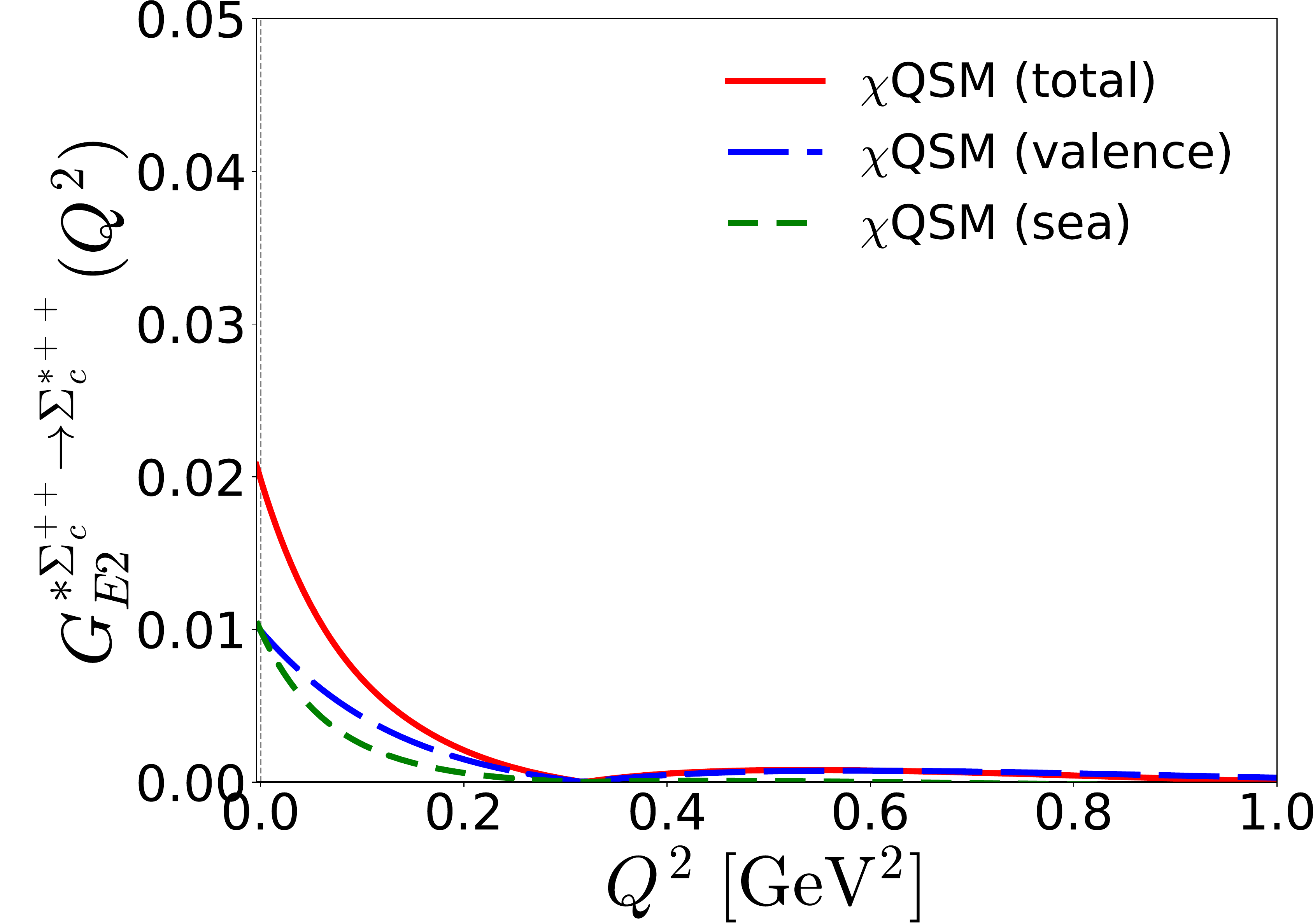}
\includegraphics[scale=0.25]{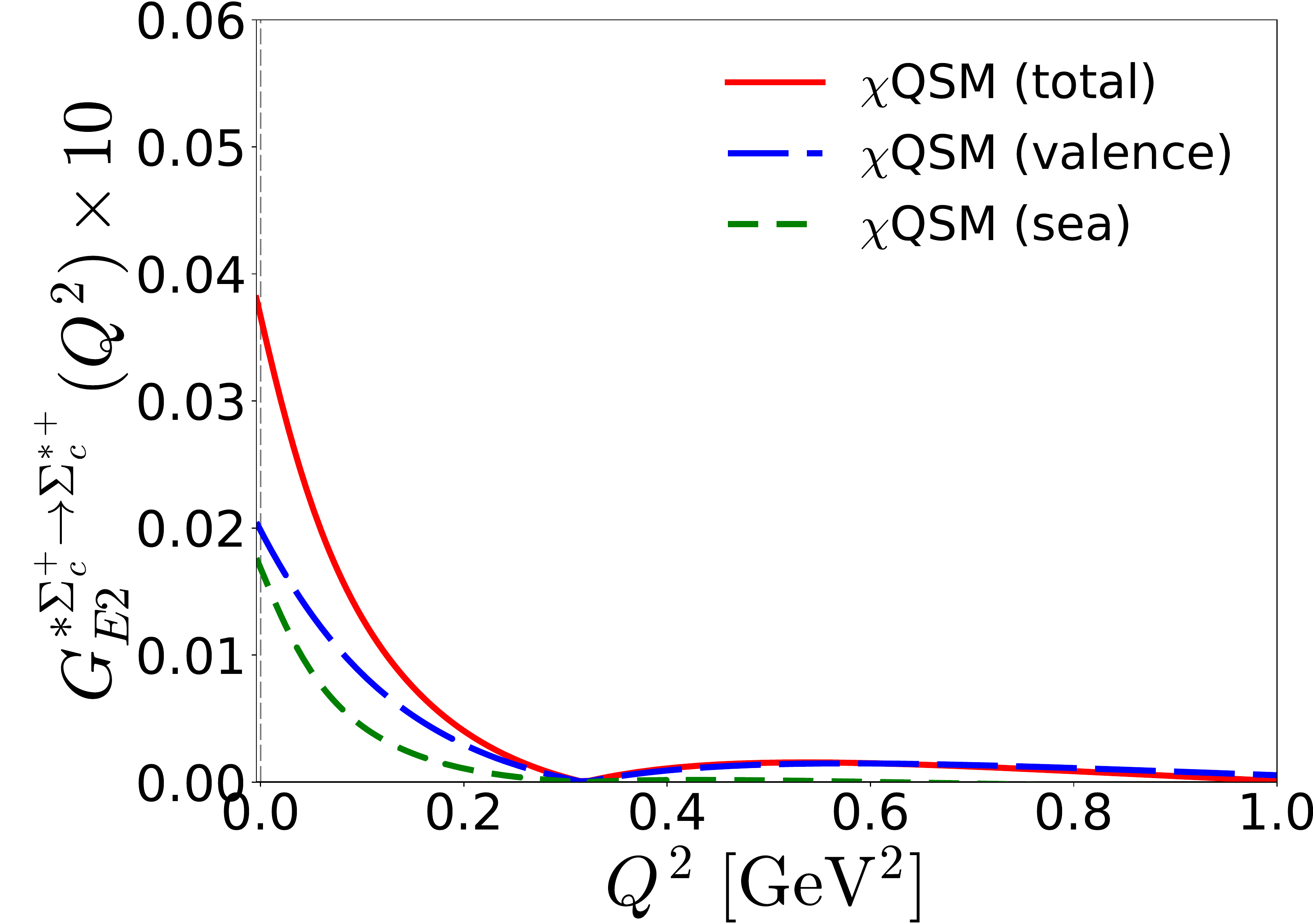}
\includegraphics[scale=0.25]{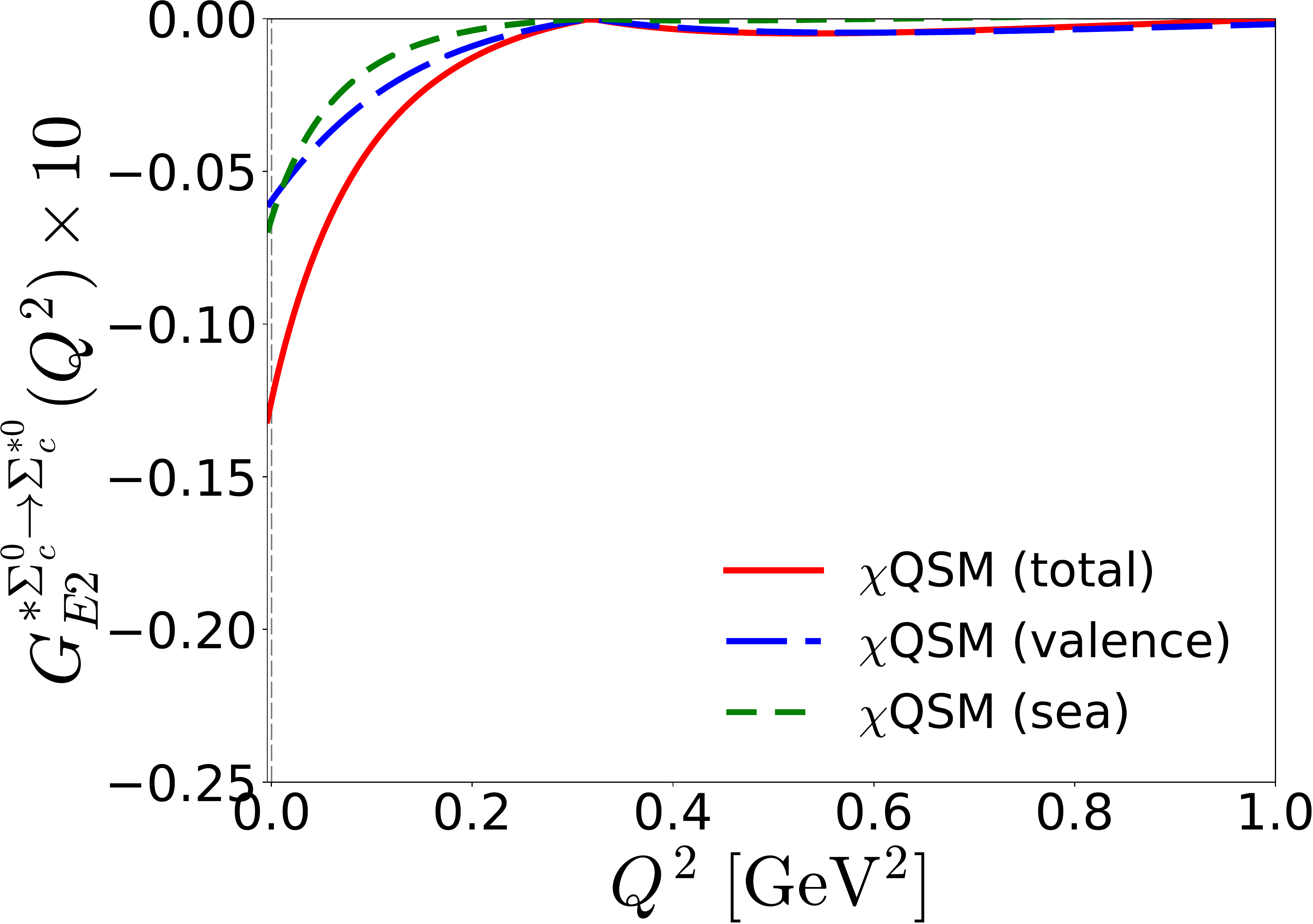}
\includegraphics[scale=0.25]{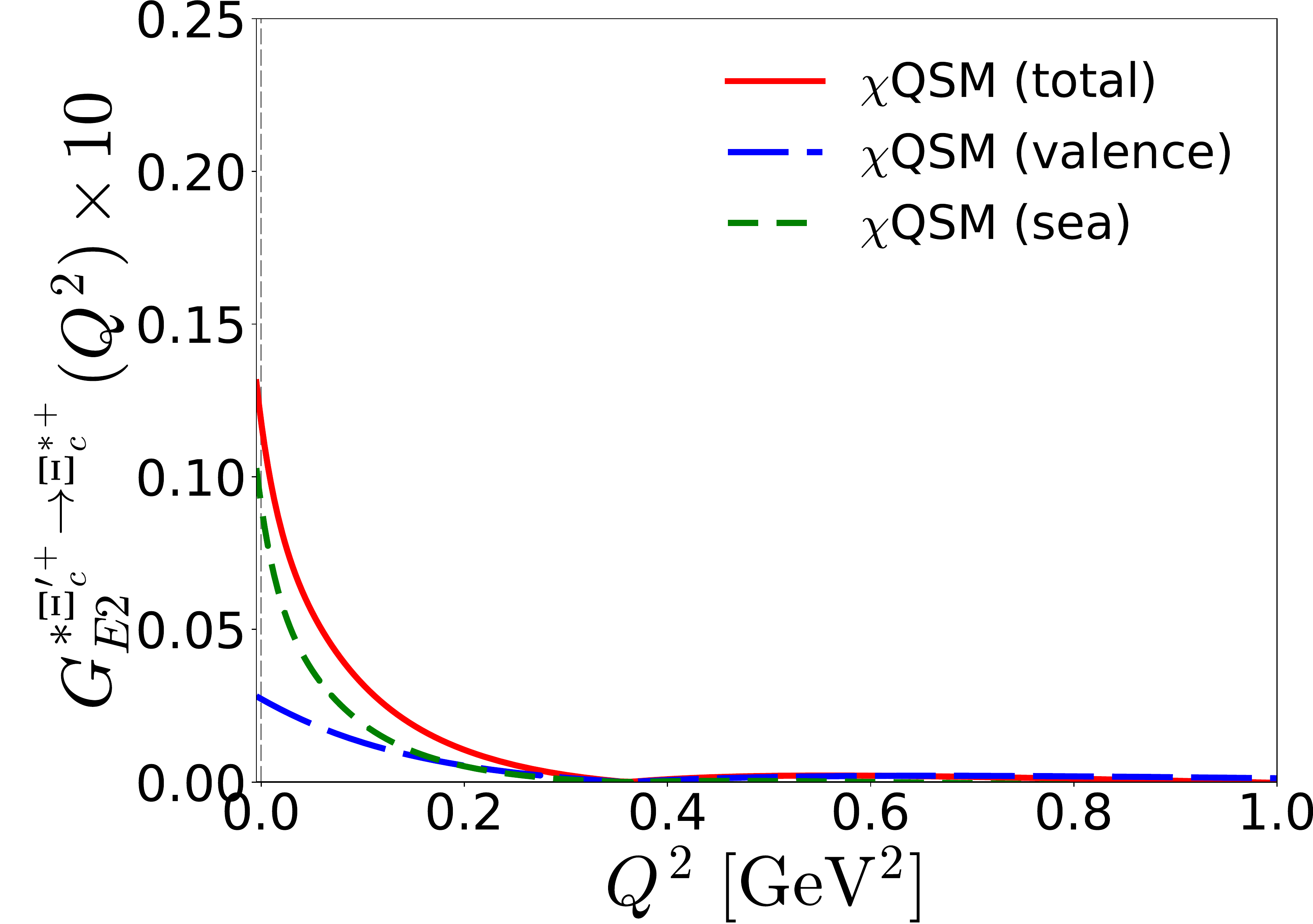}
\includegraphics[scale=0.25]{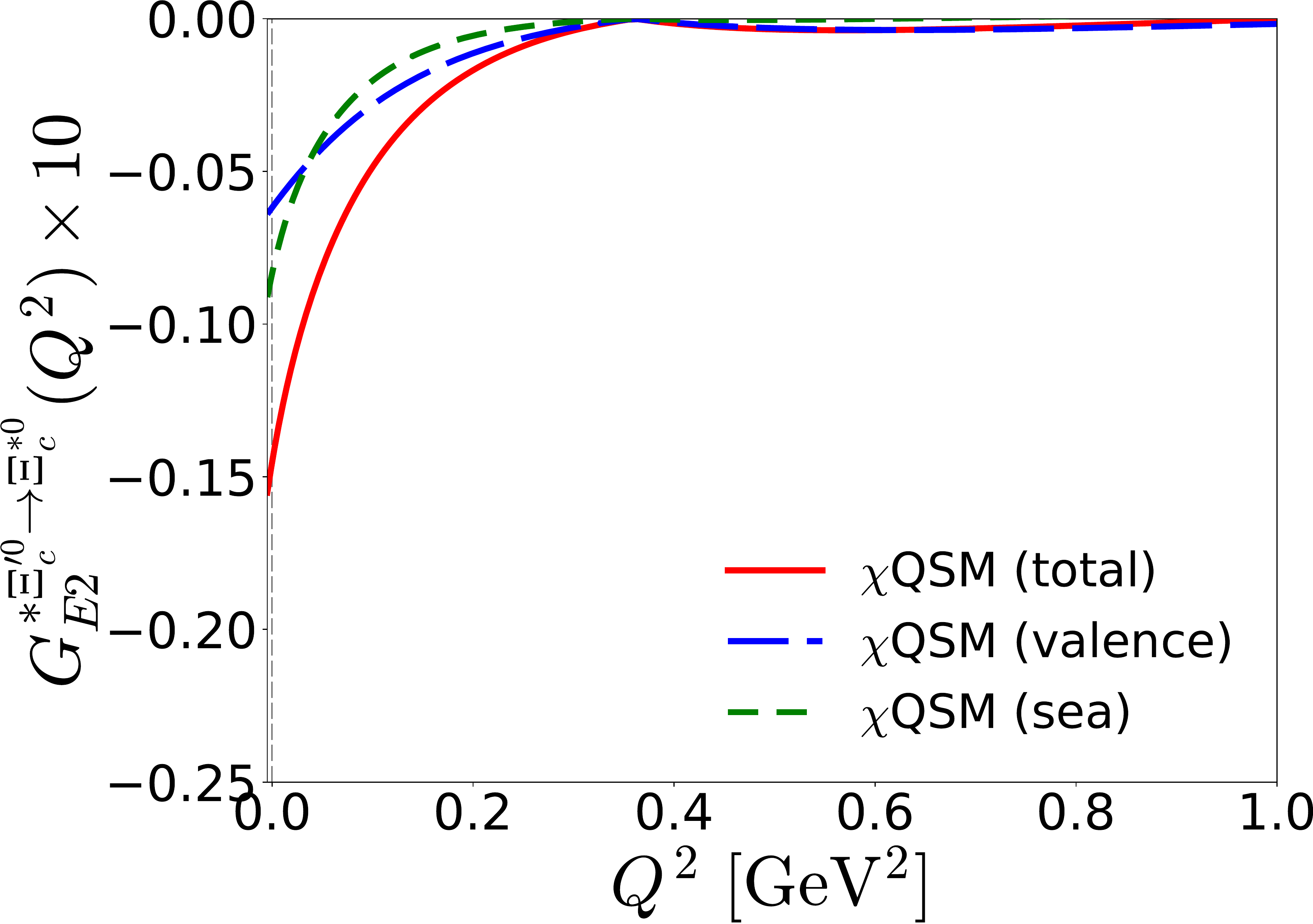}
\includegraphics[scale=0.25]{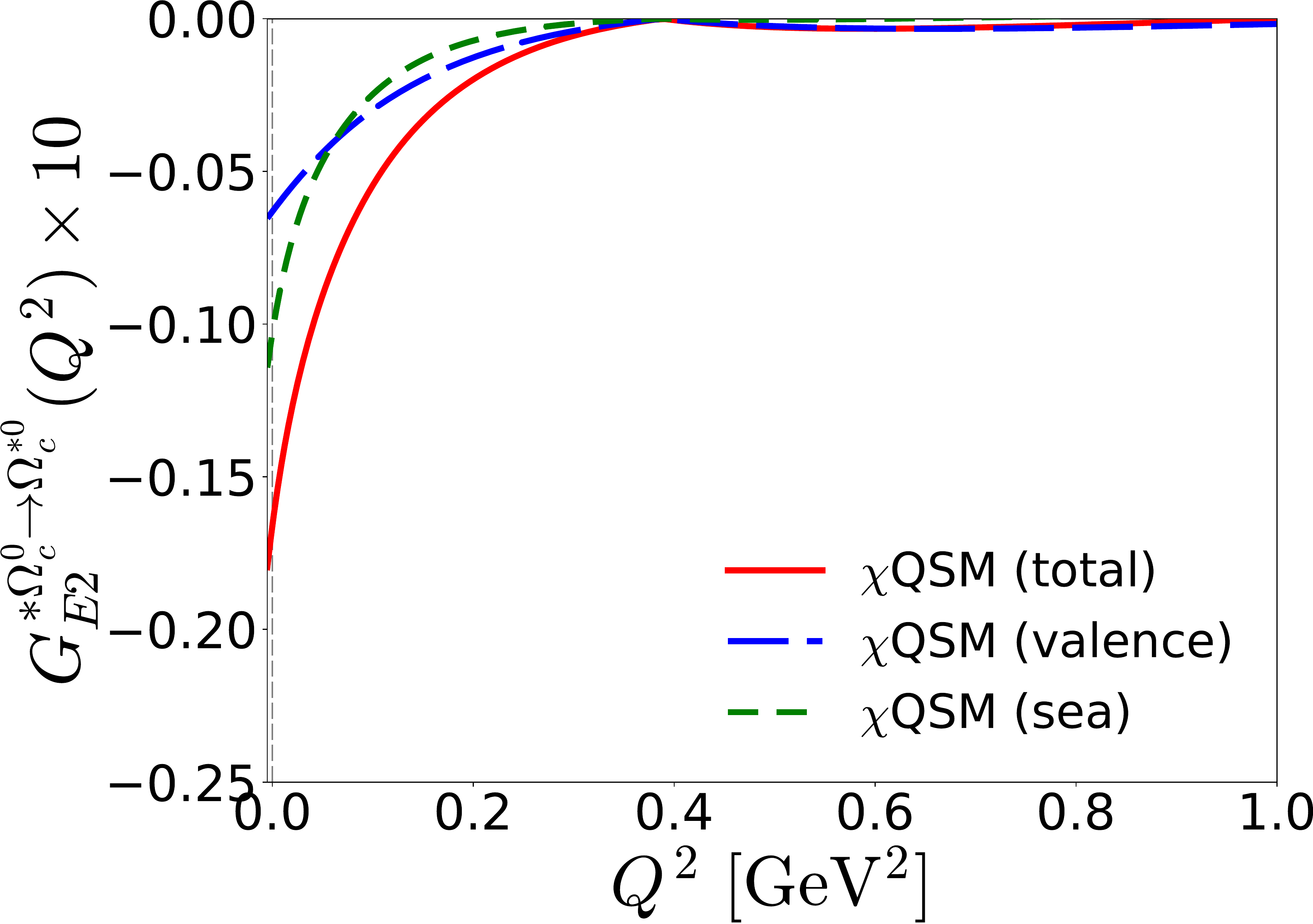}

\caption{Results for the electric quadrupole transition form
  factors from the baryon sextet with spin 1/2 to the baryon sextet
  with spin 3/2, with the valence- and sea-quark contributions 
  separated. The notations are the same as in Fig.~\ref{fig:2}.}
\label{fig:4}
\end{figure}

\begin{figure}
\centering
\includegraphics[scale=0.25]{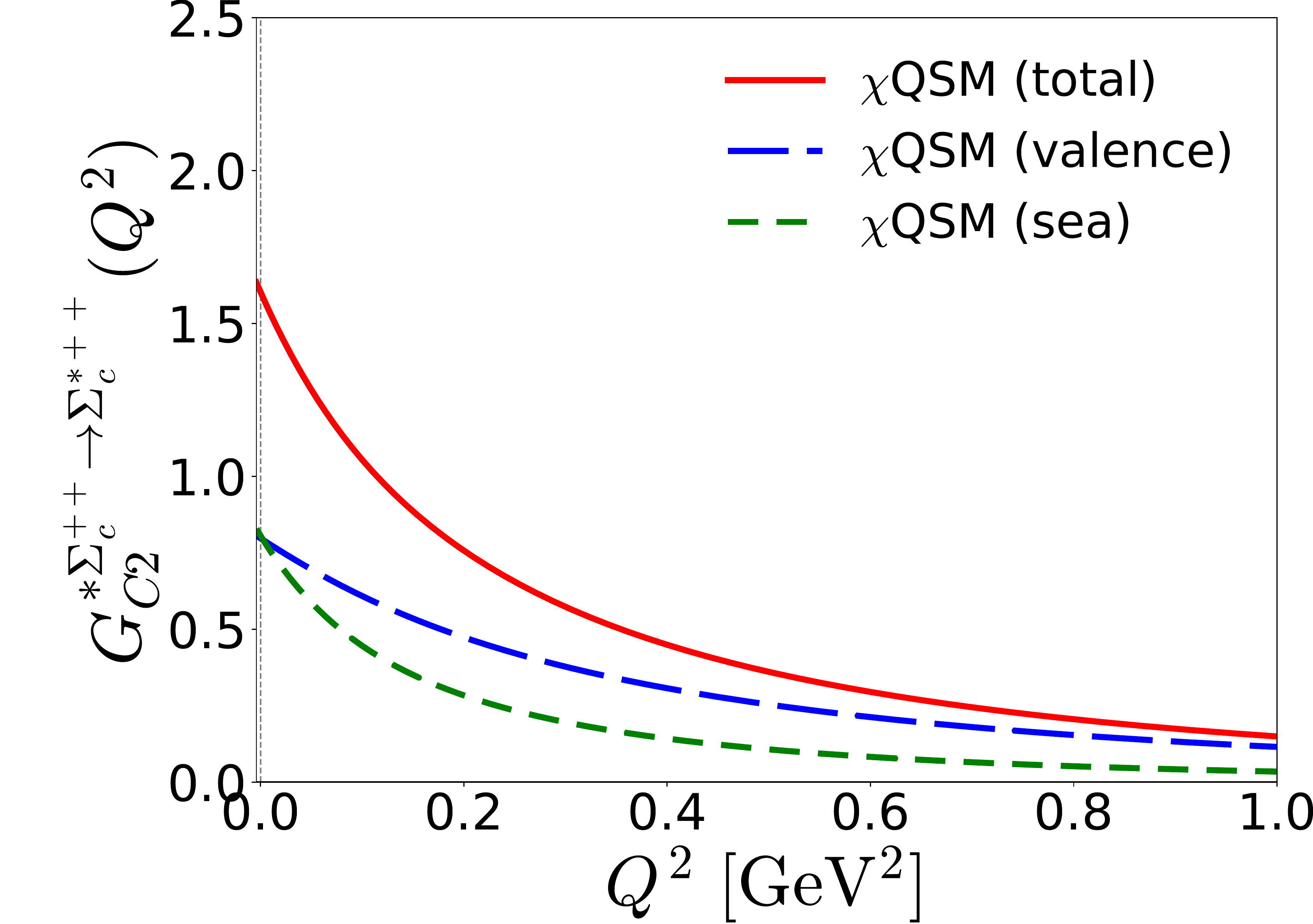}
\includegraphics[scale=0.25]{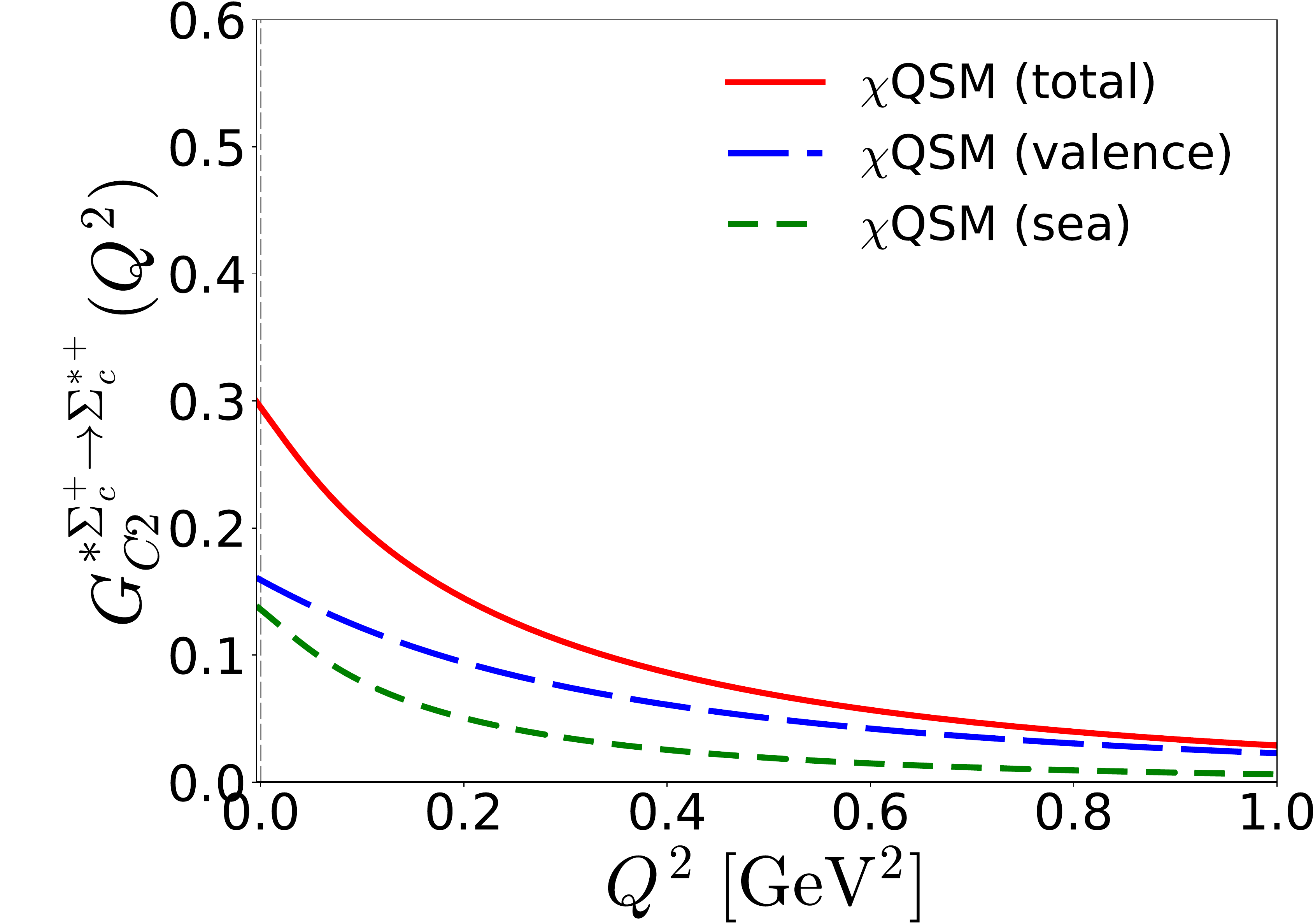}
\includegraphics[scale=0.25]{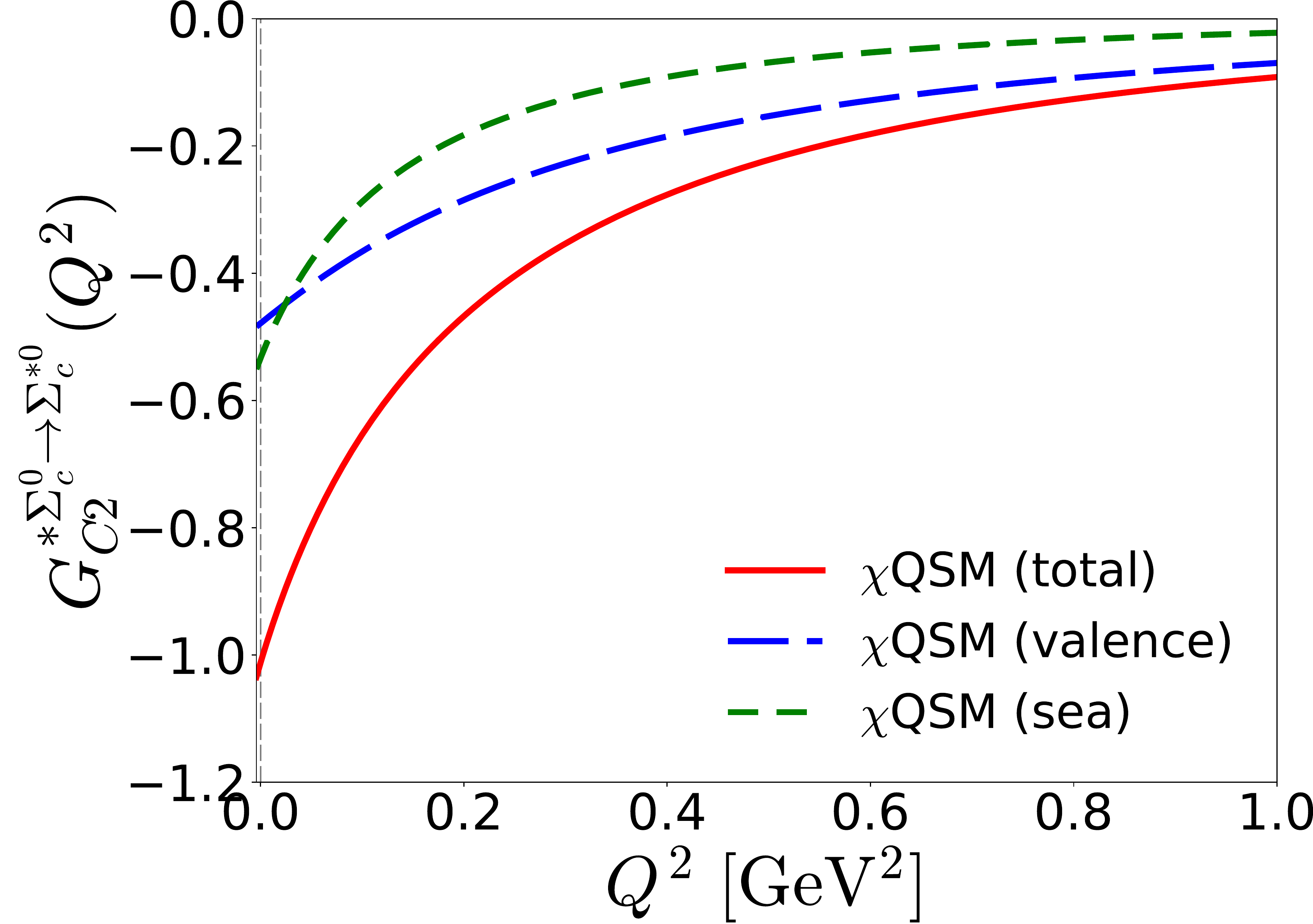}
\includegraphics[scale=0.25]{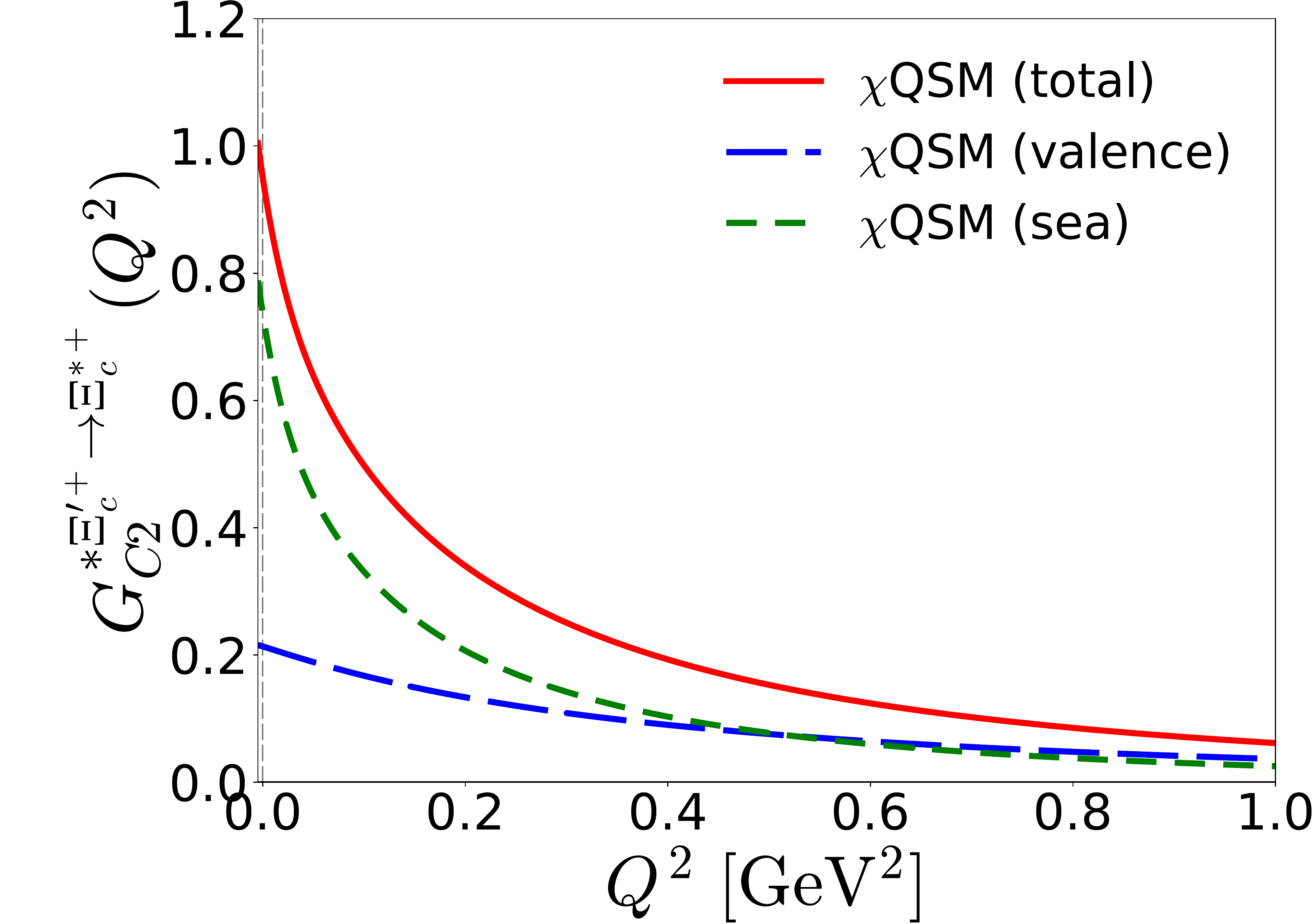}
\includegraphics[scale=0.25]{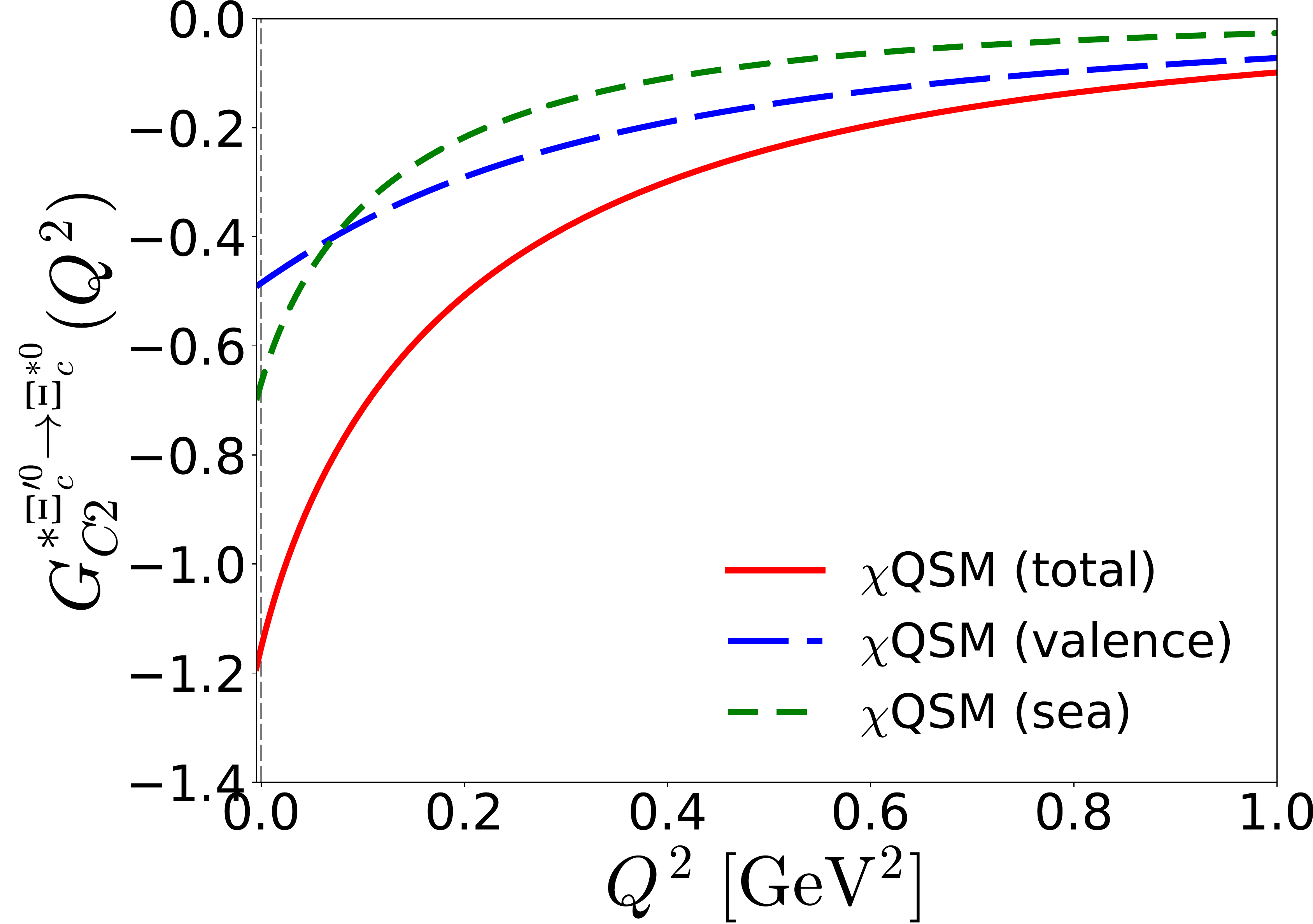}
\includegraphics[scale=0.25]{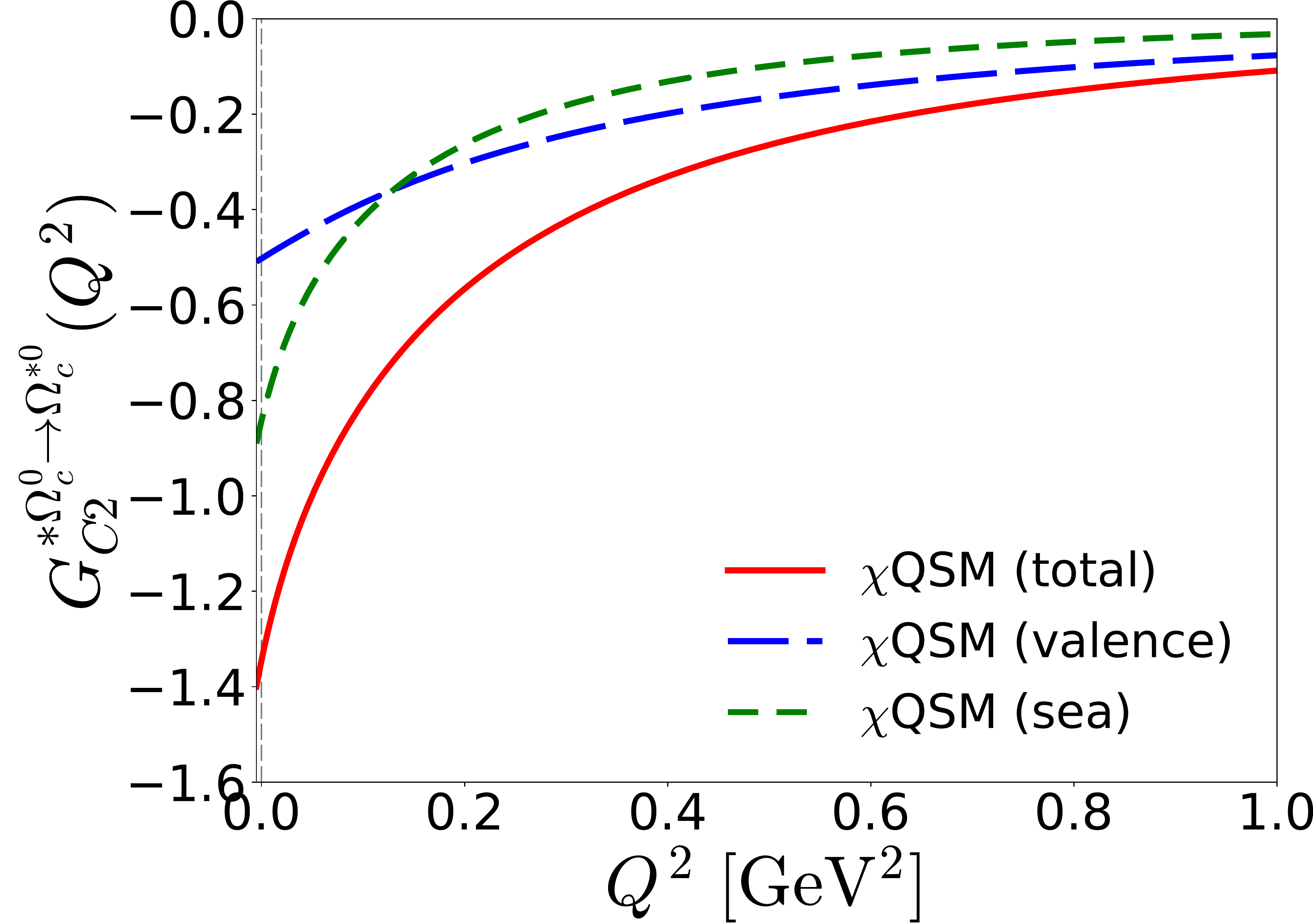}
\caption{Results for the Coulomb quadrupole transition form
  factors from the baryon sextet with spin 1/2 to the baryon sextet
  with spin 3/2, with the valence- and sea-quark contributions 
  separated. The notations are the same as in Fig.~\ref{fig:2}.}
\label{fig:5}
\end{figure}
When it comes to the E2 transitions, the situation is the other way
around. The E2 transitions are forbidden between the different spin
states. Thus, we have only the finite results for the E2 form factors
for the EM transitions between the baryon sextet with spin 1/2 and
with spin 3/2, as shown in Fig.~\ref{fig:4}. The sea-quark
contributions to the E2 form factors are remarkably sizable, which was
already shown in those of the baryon decuplet~\cite{Ledwig:2008es,
  Kim:2019gka}.  In the case of the $\Xi_c^{\prime 0}\to \Xi_c^{*0}$
and $\Omega_c^0\to \Omega_c^{*0}$ EM transitions, the sea-quark
contributions dominate over those of the valence quarks, in
particular, in the smaller $Q^2$ region. This implies that the effects
of the pion clouds play a significant role in describing the E2 form
factors. Knowing the fact that the E2 transition form factors reveal
how a baryon with spin 3/2 is deformed, the outer part of the charge
distribution governs the E2 form factors. As shown in
Refs.~\cite{Kim:1995mr, Goeke:2007fp}, the sea-quark part constructs
the outer place of the charge and mechanical distributions in a
baryon, whereas the valence-quark part is responsible for the inner
part of these distributions. In this sense, it is natural that the
sea-quark contributions contribute significantly to the E2 transition
form factors in the smaller $Q^2$ region. Note that as $Q^2$
increases, the sea-quark contributions fall off much faster than those
of the valence quarks, which is also understandable.

The magnitudes of the E2 transition form factors of a baryon are in
general much smaller than those of the M1 transition form factors. The
leading contribution to the E2 form factors vanishes within the
$\chi$QSM, so that the rotational $1/N_c$ correction takes the place
of the leading contribution as shown in
Eq.~\eqref{eq:E2leadingcon}. This indicates that the magnitudes of the
E2 form factors should be smaller than those of the M1 ones. Moreover,
the E2 form factors are suppressed by the mass of a decaying
baryon. Considering the fact that the mass of a singly heavy baryon is
much larger than those of the baryon decuplet, one can expect that the
E2 transition form factors of the baryon sextet would turn out to be
much smaller than those of the baryon decuplet. In addition, 
the matrix elements of the SU(3) Wigner $D$ functions for the baryon
sextet are smaller than those for the baryon decuplet. Thus, the
magnitudes of the E2 transition form factors for the baryon sextet
become approximately five to ten times smaller than those for the
baryon decuplet. Figure~\ref{fig:5} presents the numerical results for 
the Coulomb quadrupole form factors from the baryon sextet with spin
1/2 to that with spin 3/2. The main conclusion is the same as in the
case of the E2 transition form factors. The sea-quark contributions
are again dominant over those of the valence-quarks in the smaller
$Q^2$ region. 

\subsection{Effects of explicit breaking of flavor SU(3) symmetry}
\begin{figure}
\centering
\includegraphics[scale=0.25]{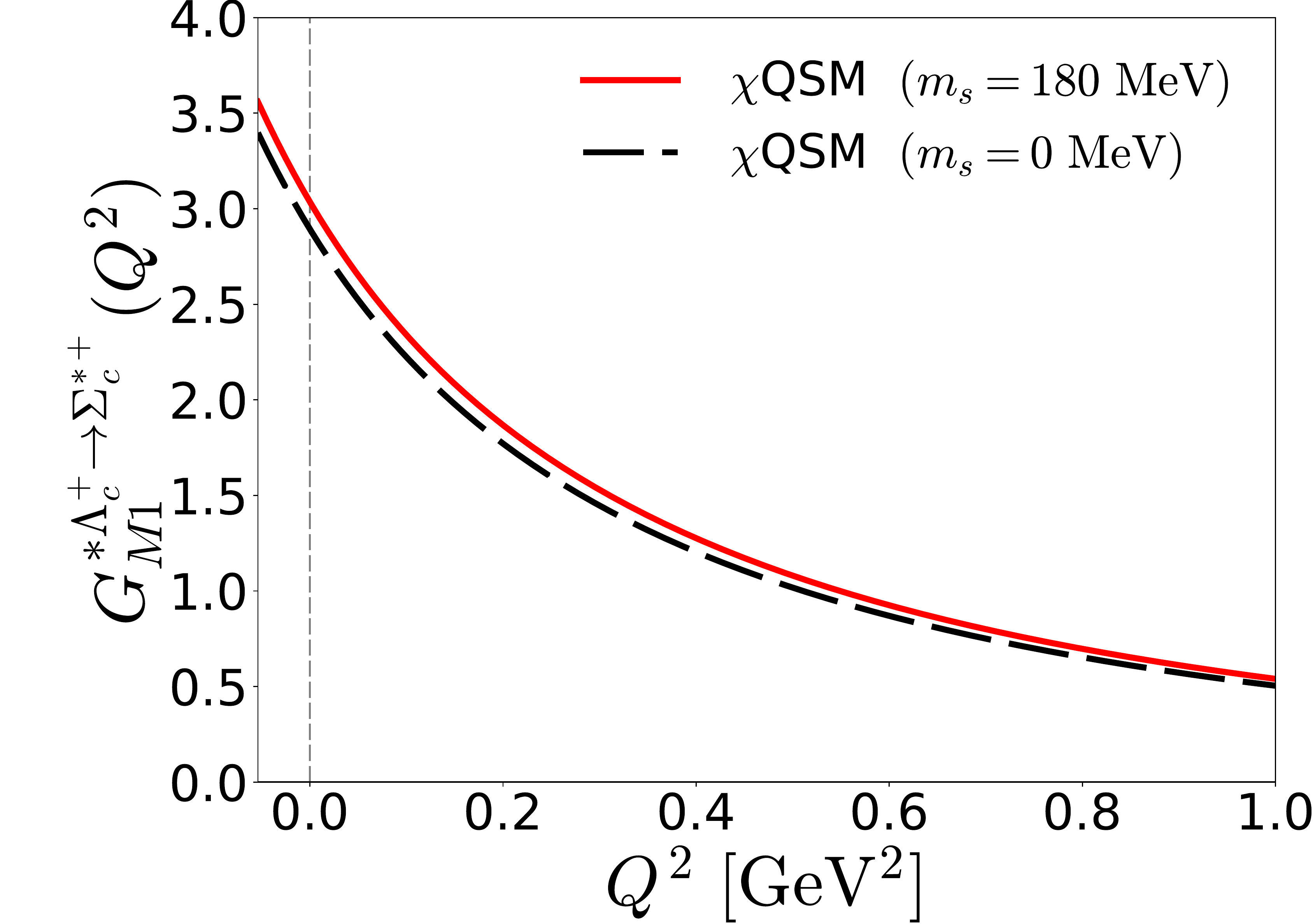}
\includegraphics[scale=0.25]{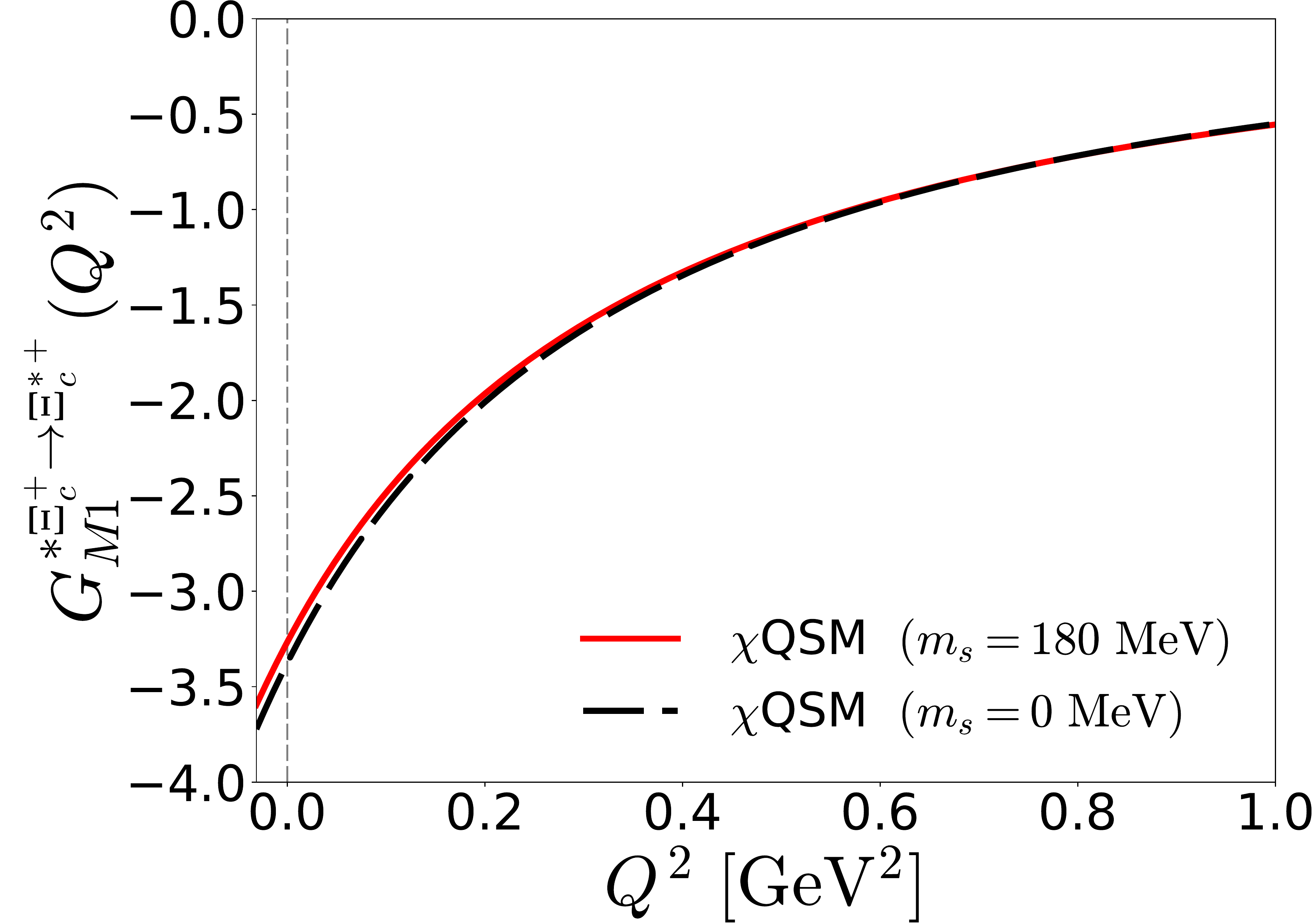}
\includegraphics[scale=0.25]{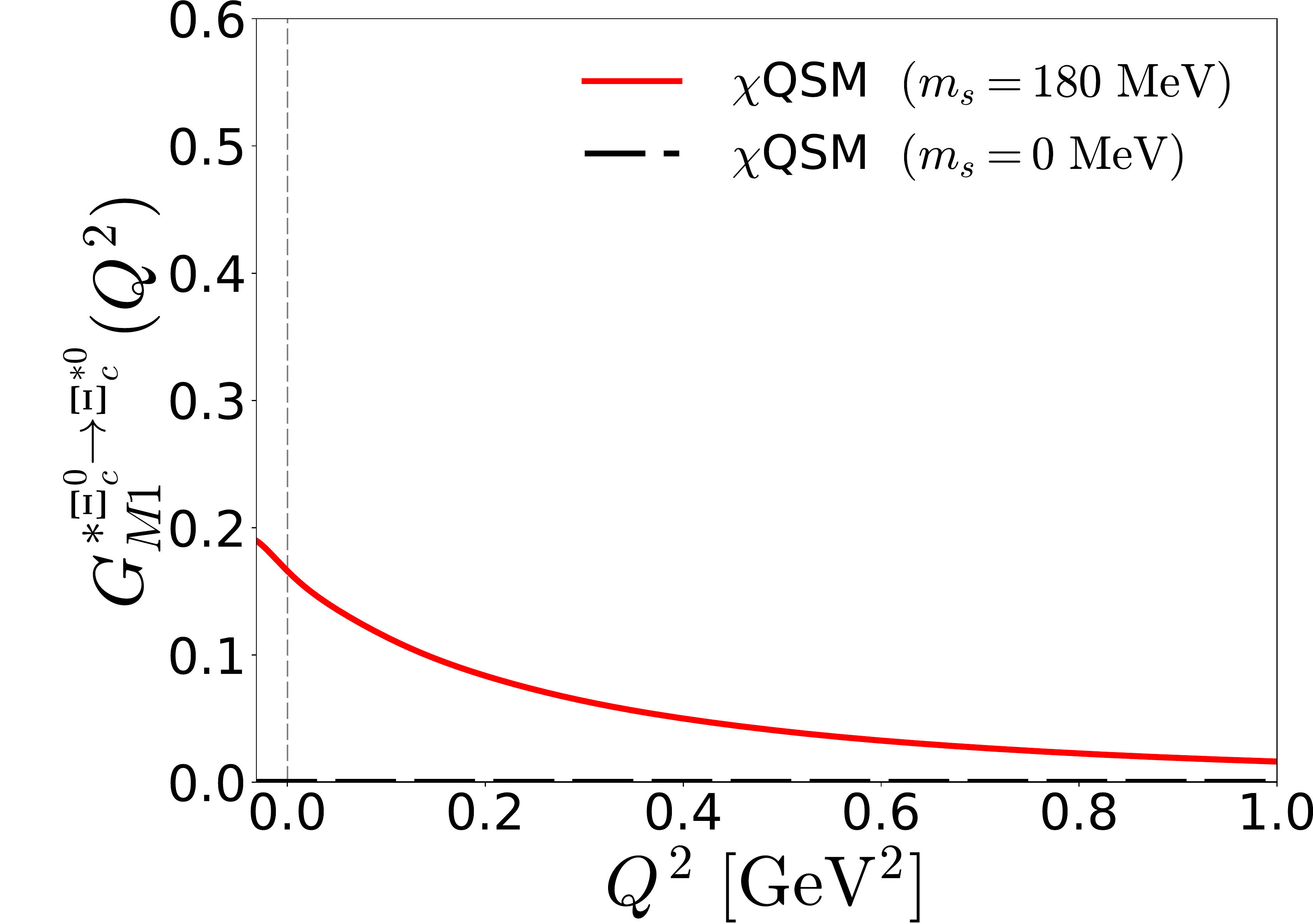}
\caption{Results for the magnetic dipole transition form
  factors from the baryon antitriplet to the baryon sextet
  with spin 3/2. The dashed curves draw the results in flavor SU(3)
  symmetry, whereas the solid ones depic those with flavor SU(3)
  symmetry breaking taken into account.}
\label{fig:6}
\end{figure}

\begin{figure}
\centering
\includegraphics[scale=0.25]{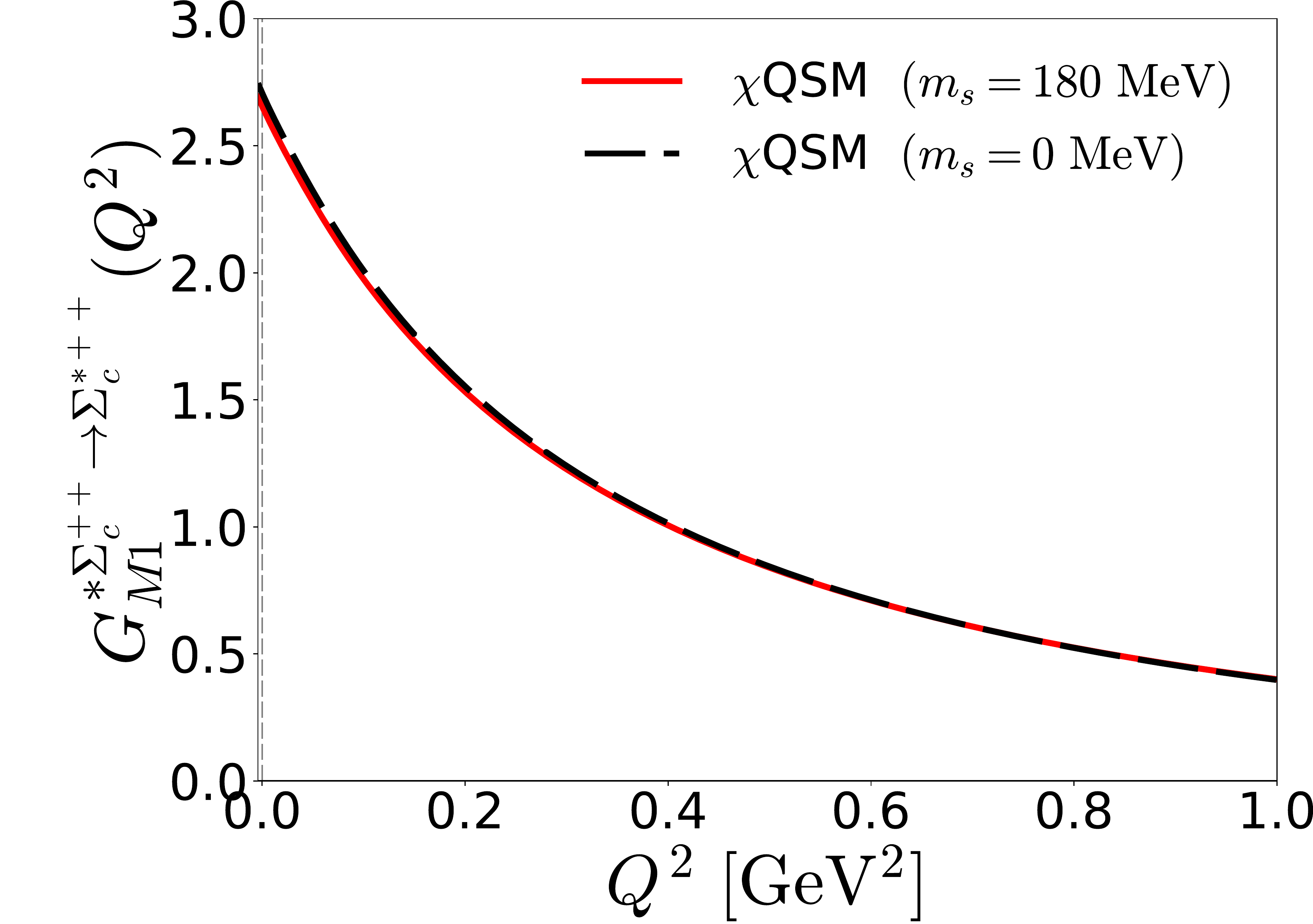}
\includegraphics[scale=0.25]{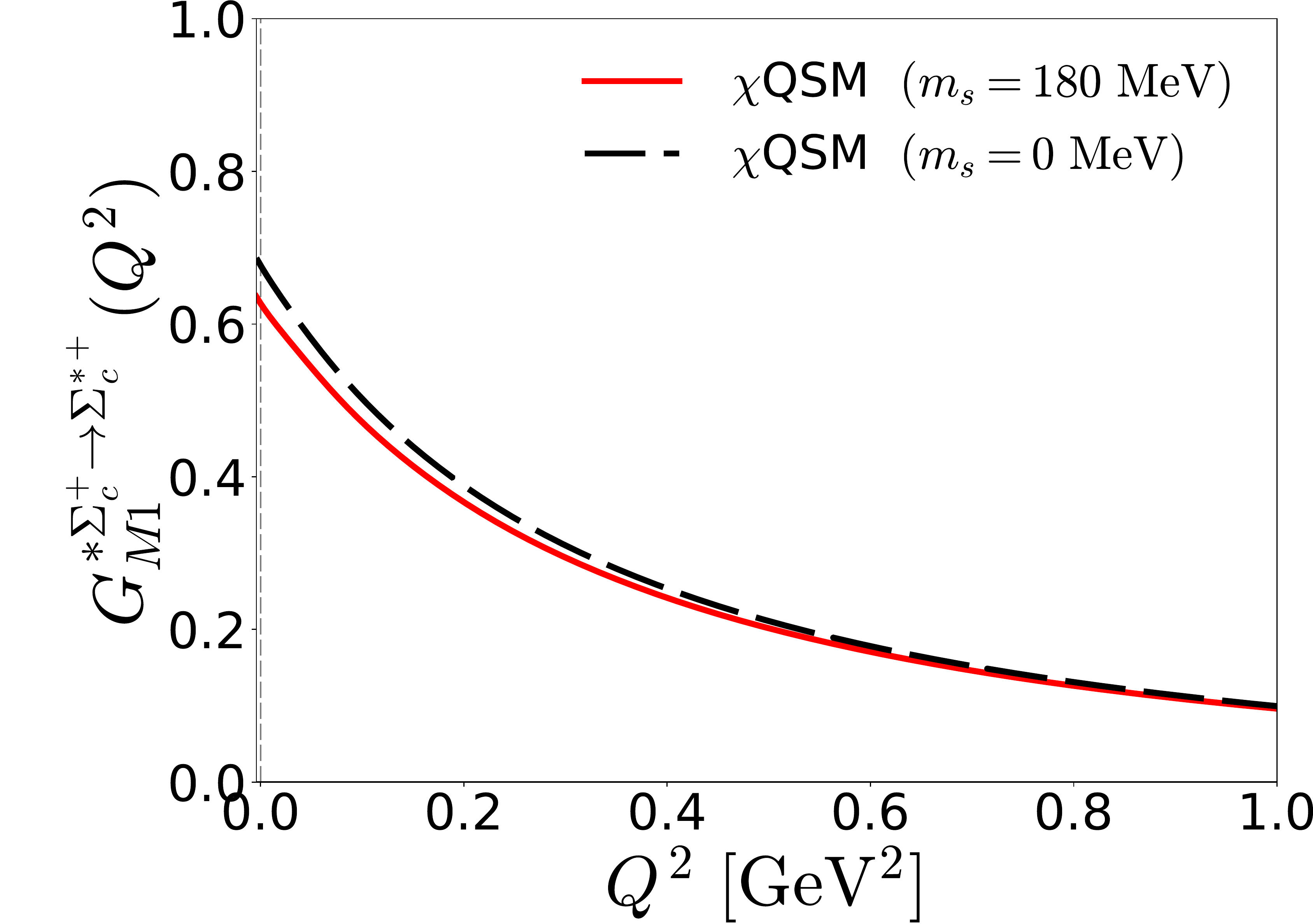}
\includegraphics[scale=0.25]{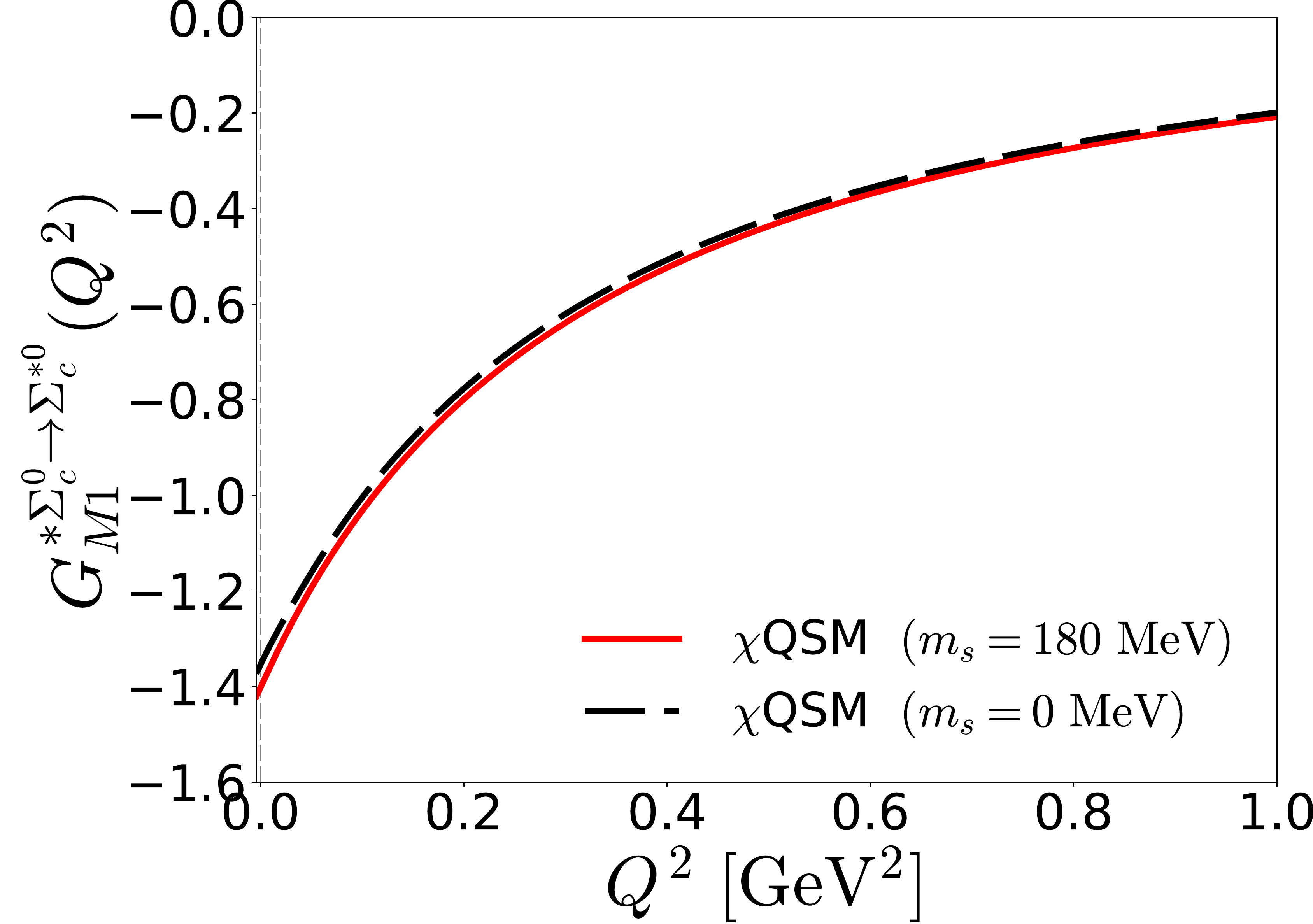}
\includegraphics[scale=0.25]{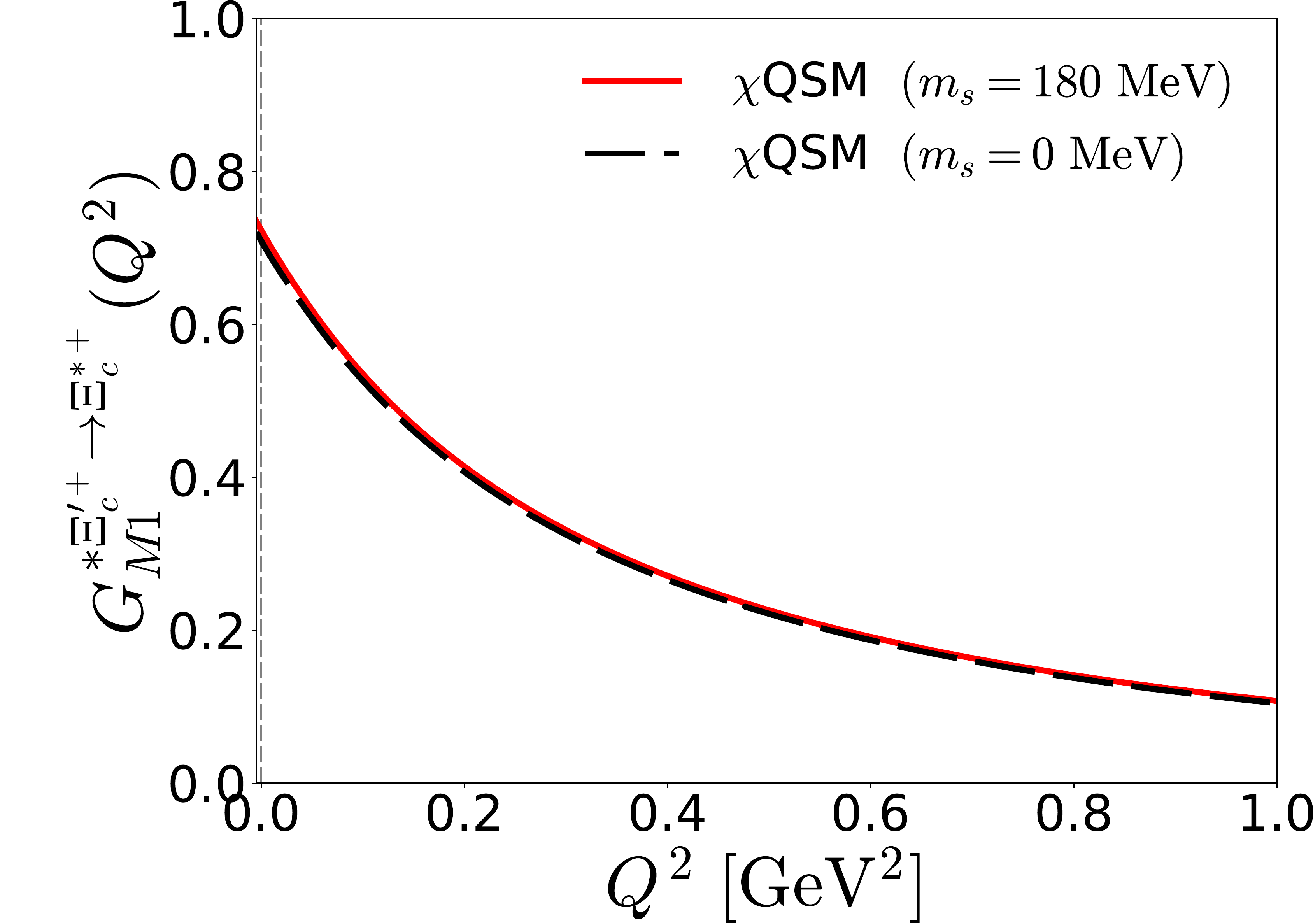}
\includegraphics[scale=0.25]{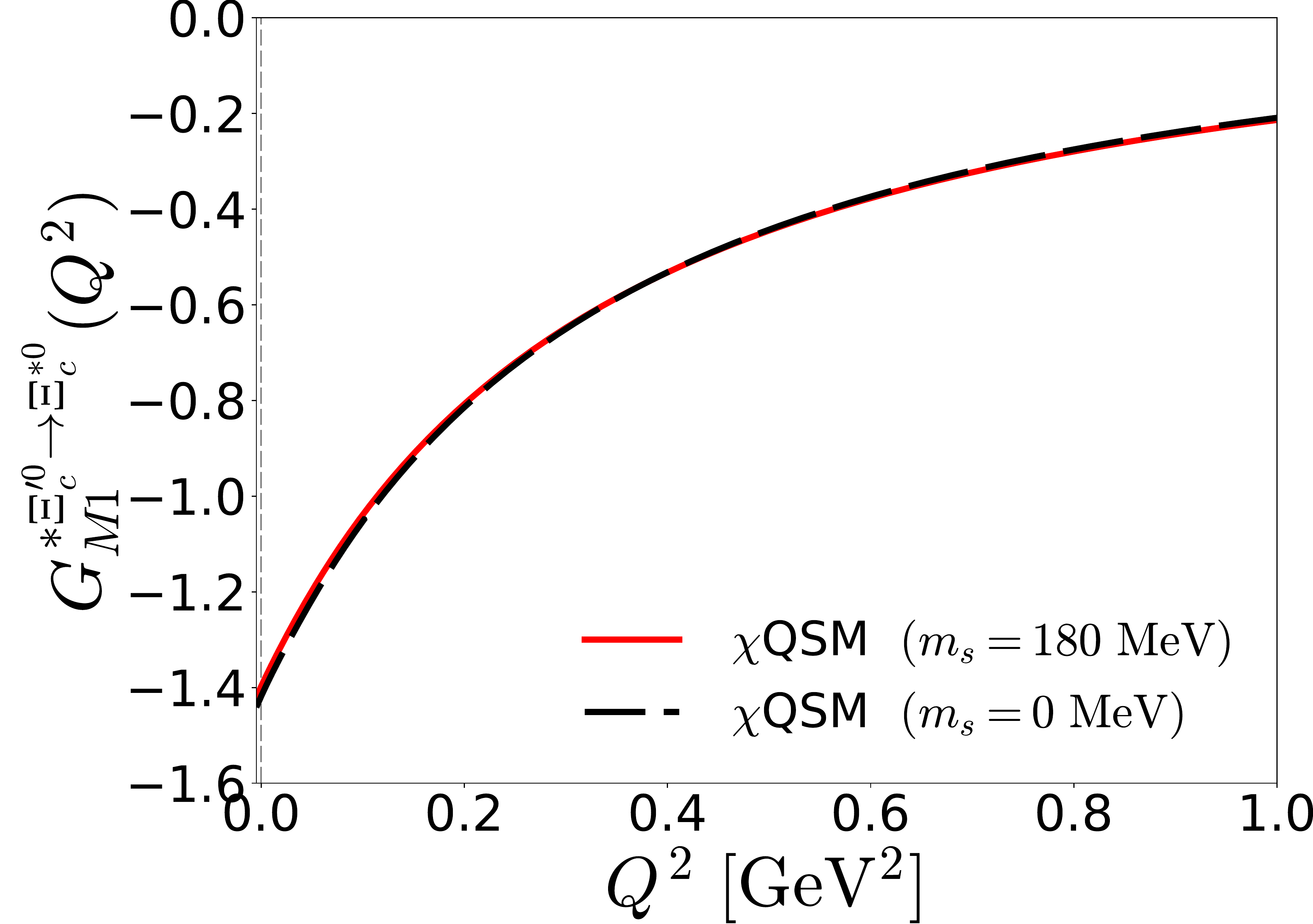}
\includegraphics[scale=0.25]{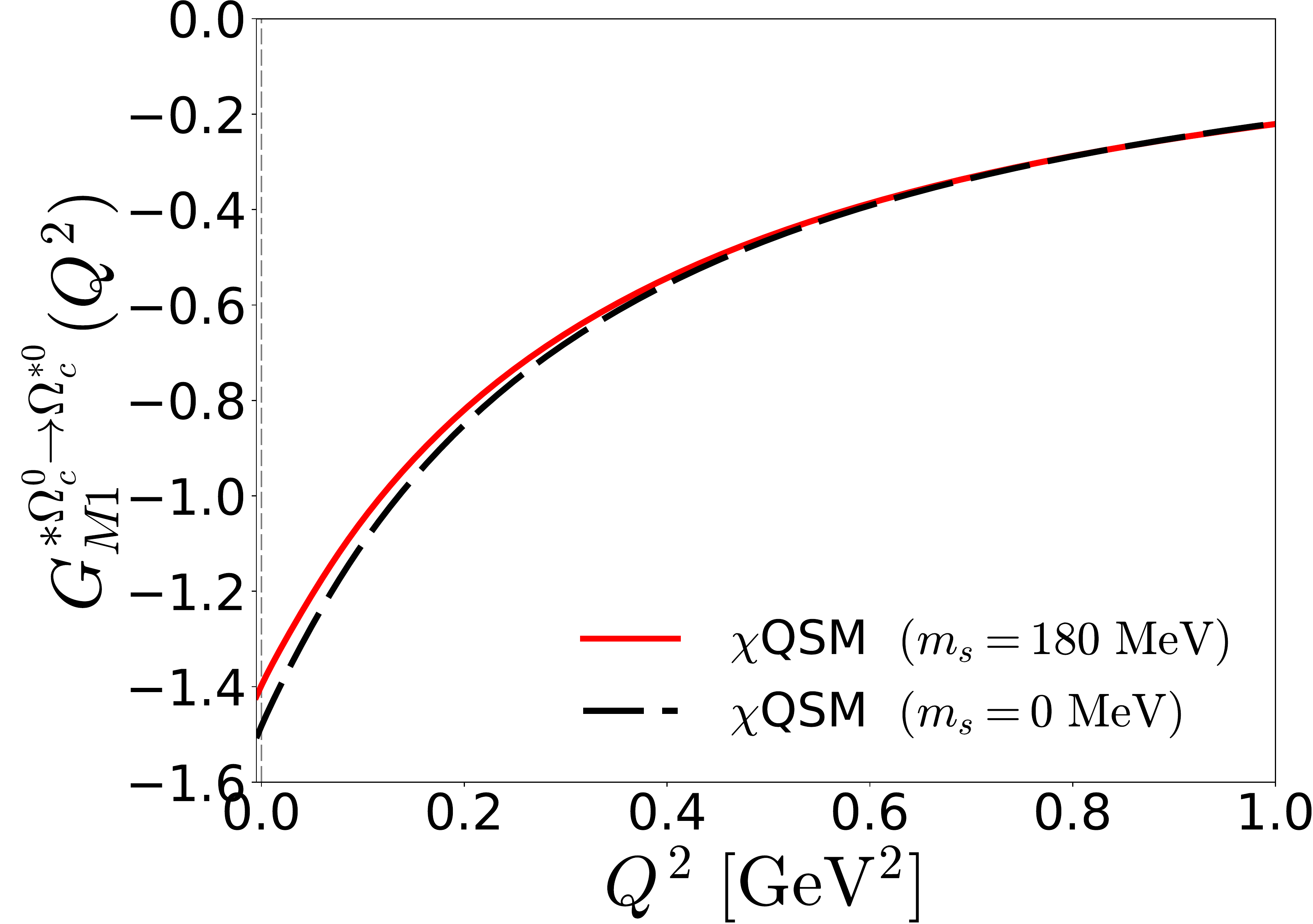}
\caption{Results for the magnetic dipole transition form
  factors from the baryon sextet with spin 1/2 to the baryon sextet
  with spin 3/2. The notations are the same as in Fig.~\ref{fig:6}.}
\label{fig:7}
\end{figure}

\begin{figure}
\centering
\includegraphics[scale=0.25]{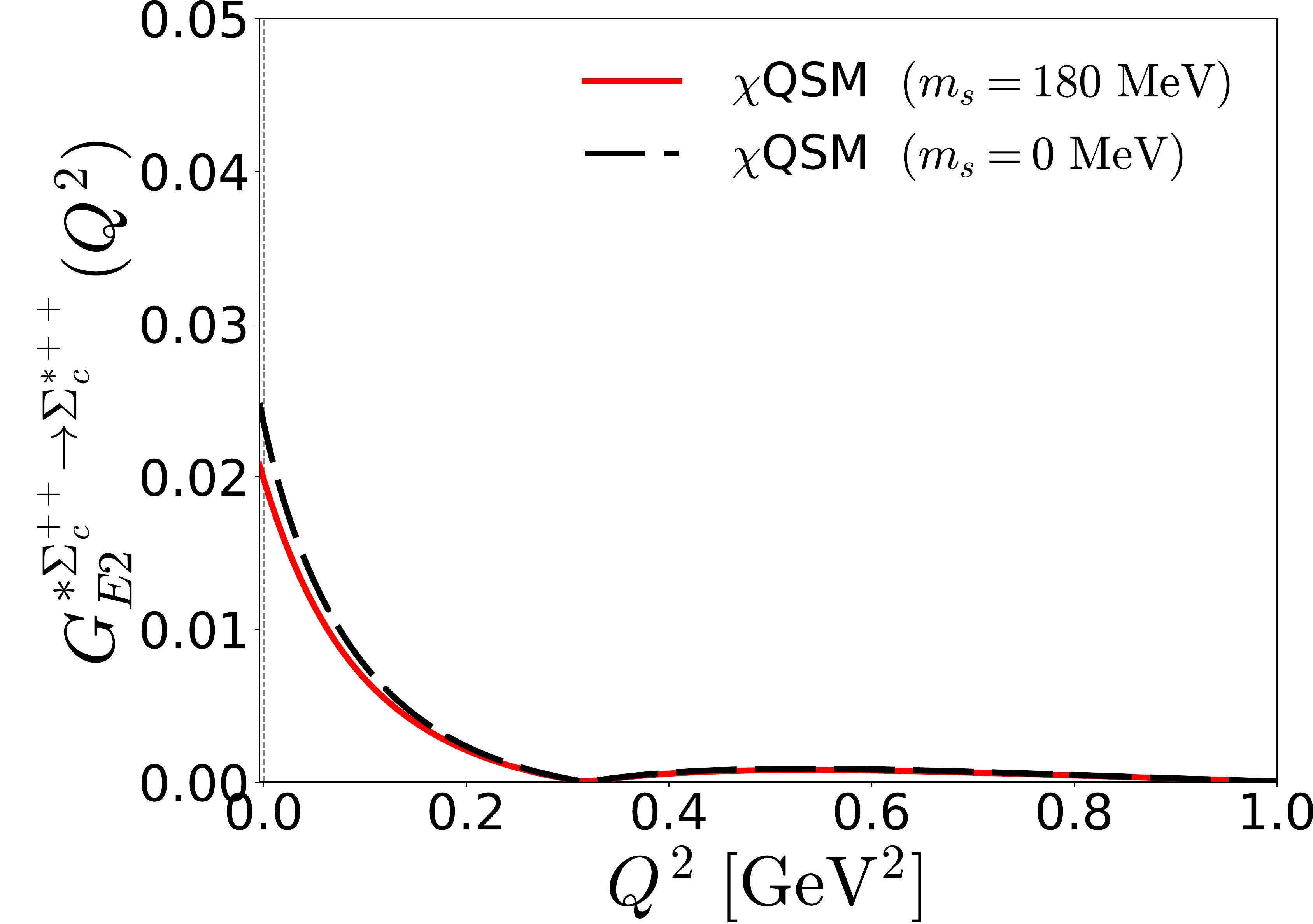}
\includegraphics[scale=0.25]{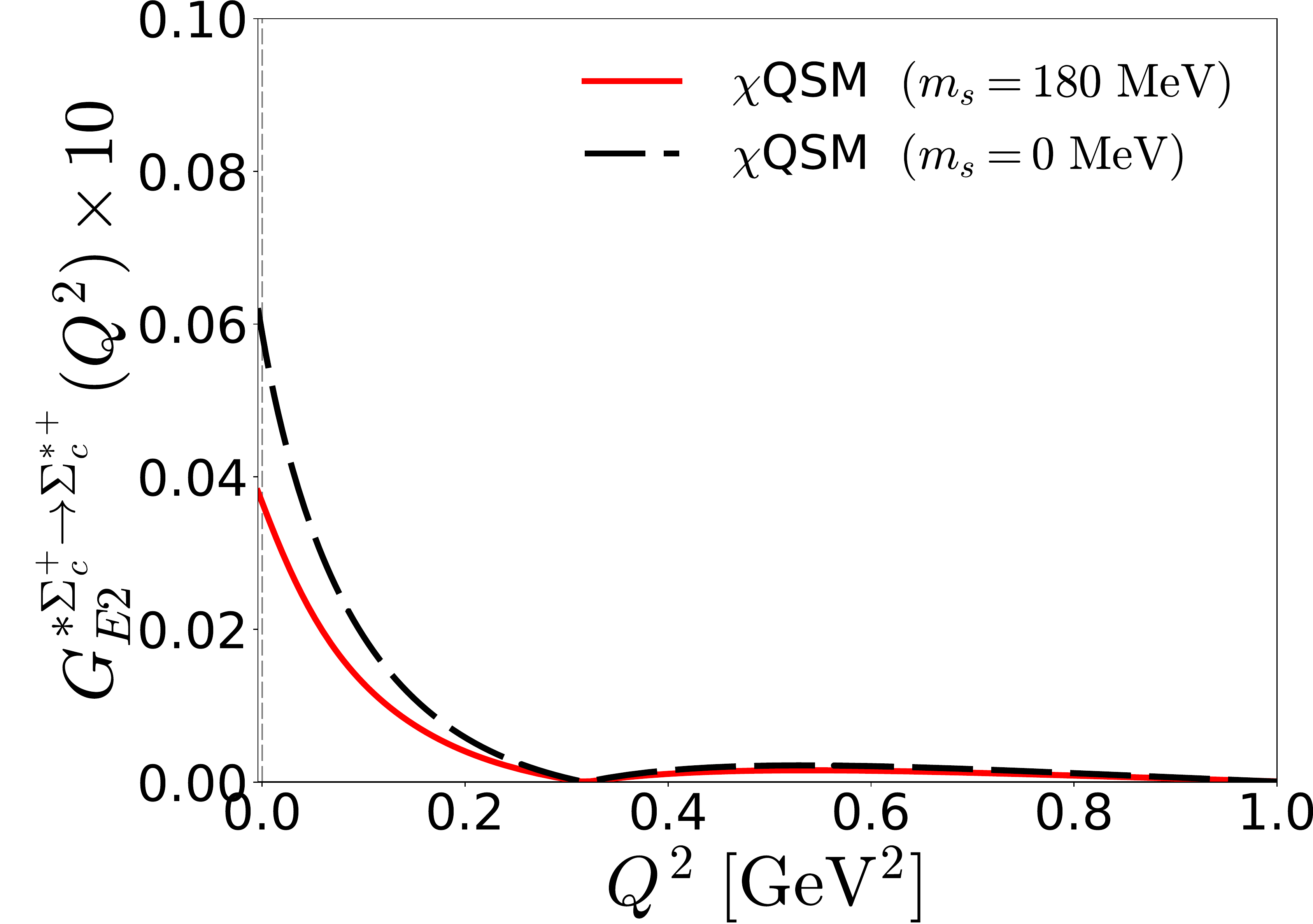}
\includegraphics[scale=0.25]{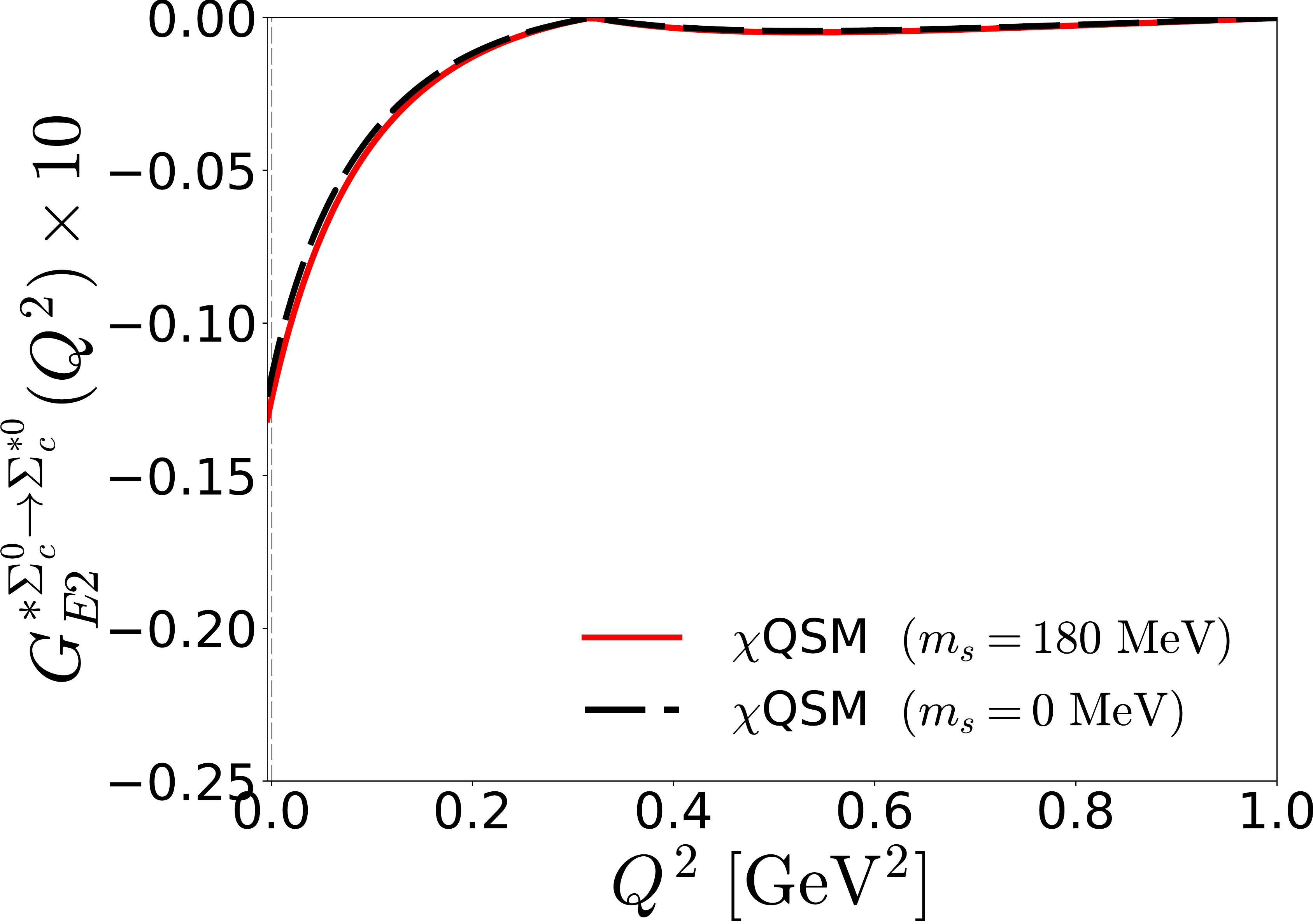}
\includegraphics[scale=0.25]{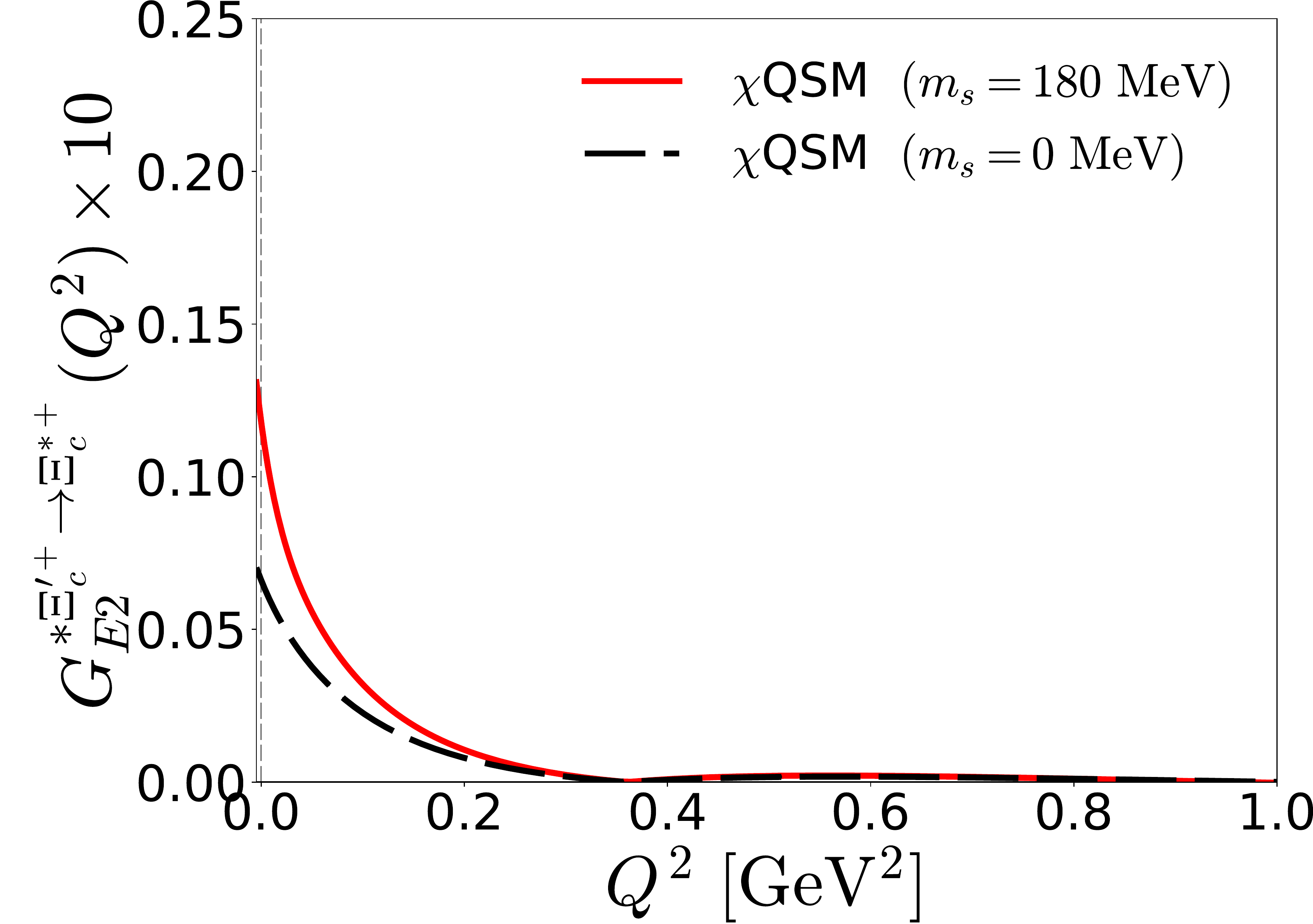}
\includegraphics[scale=0.25]{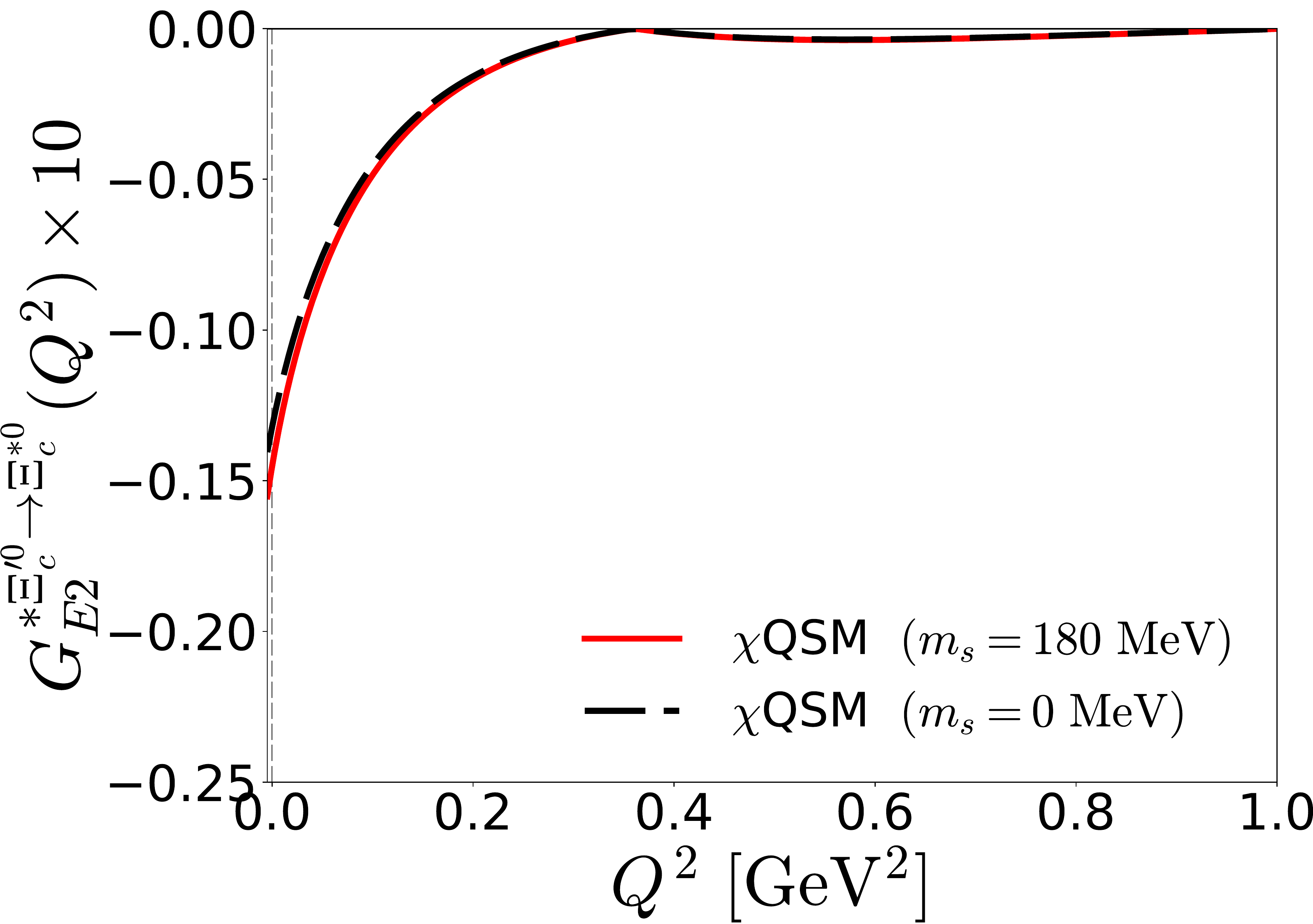}
\includegraphics[scale=0.25]{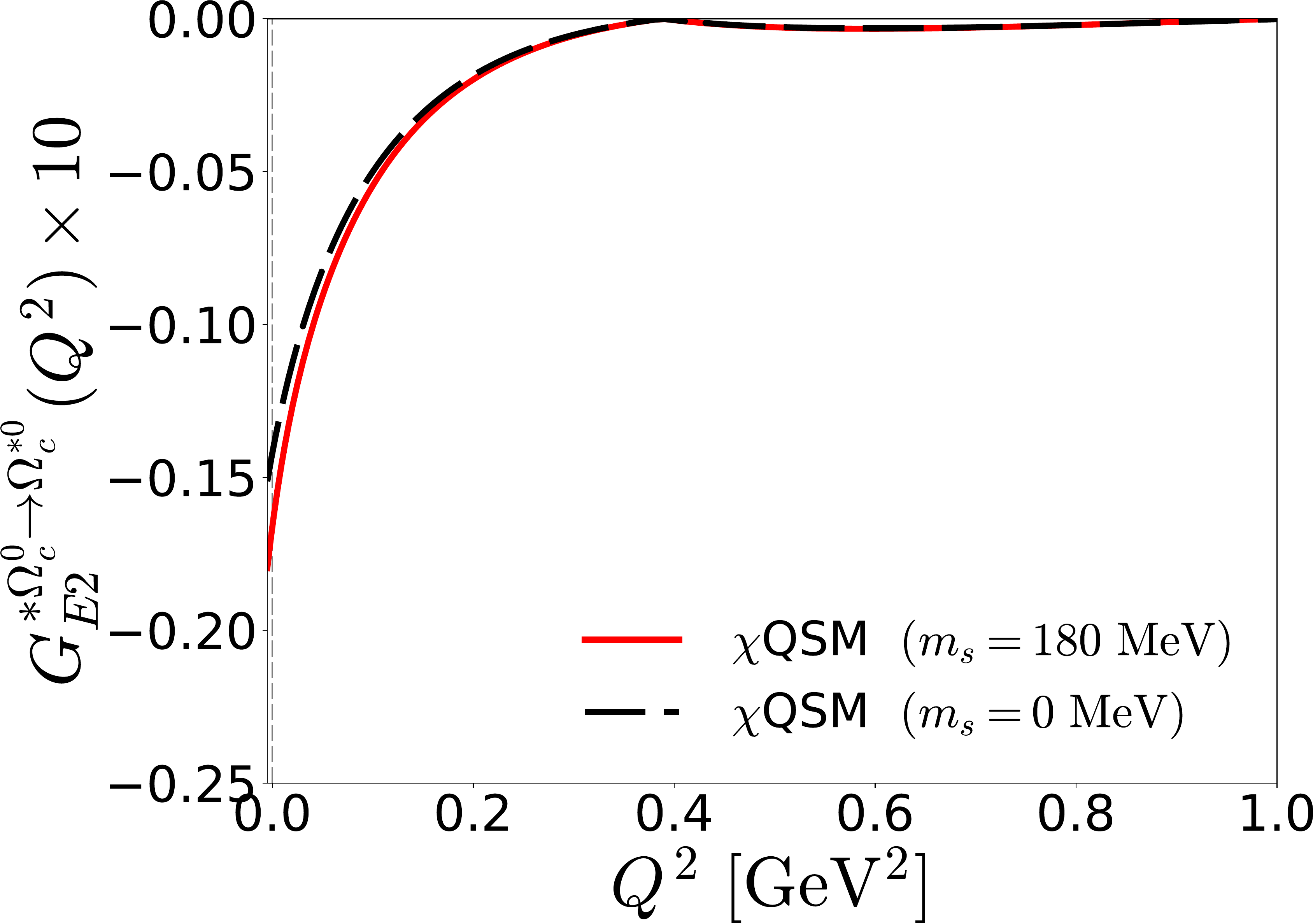}
\caption{Results for the electric qudrupole transition form
  factors from the baryon sextet with spin 1/2 to the baryon sextet
  with spin 3/2. The notations are the same as in Fig.~\ref{fig:6}.} 
\label{fig:8}
\end{figure}

\begin{figure}
\centering
\includegraphics[scale=0.25]{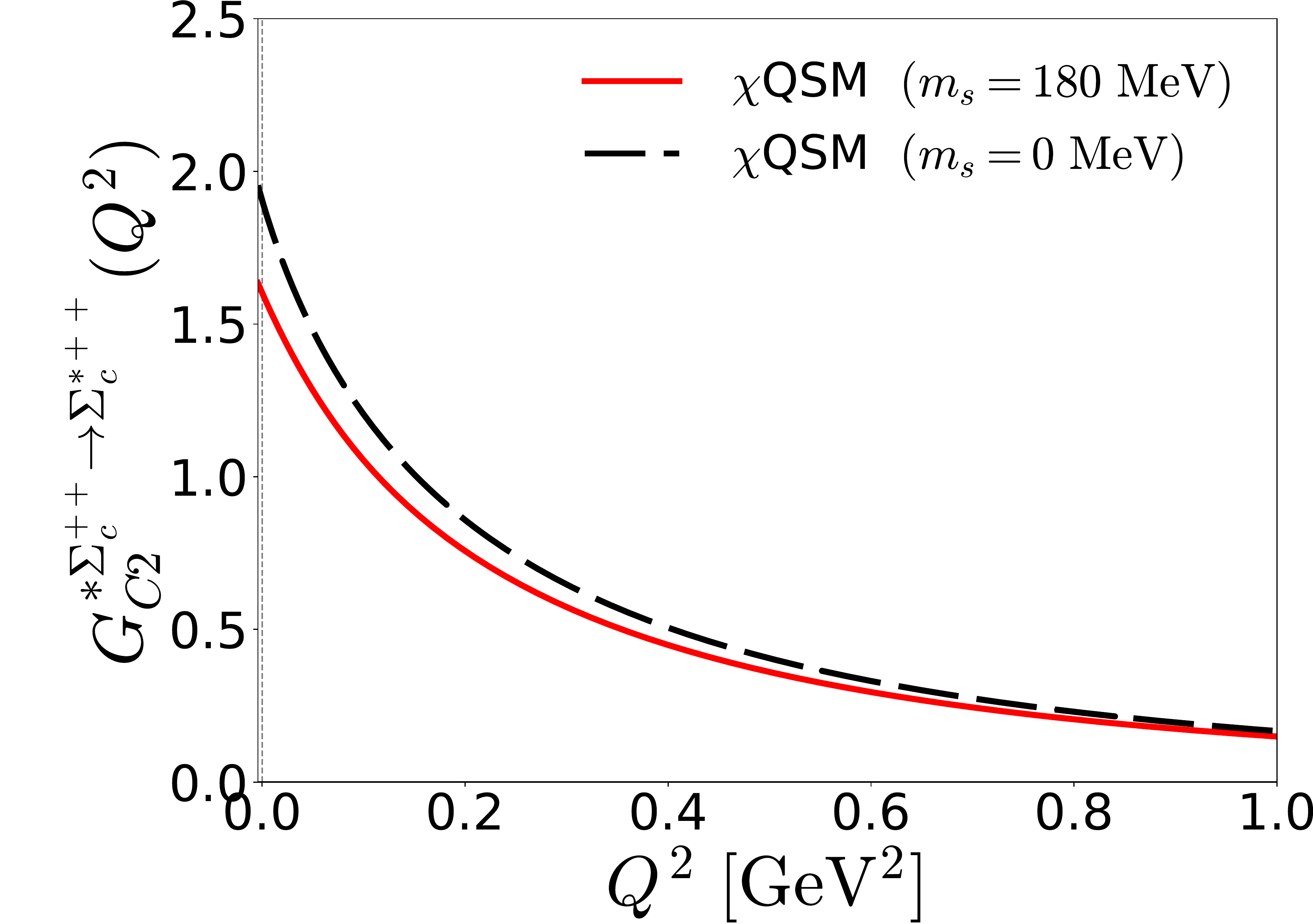}
\includegraphics[scale=0.25]{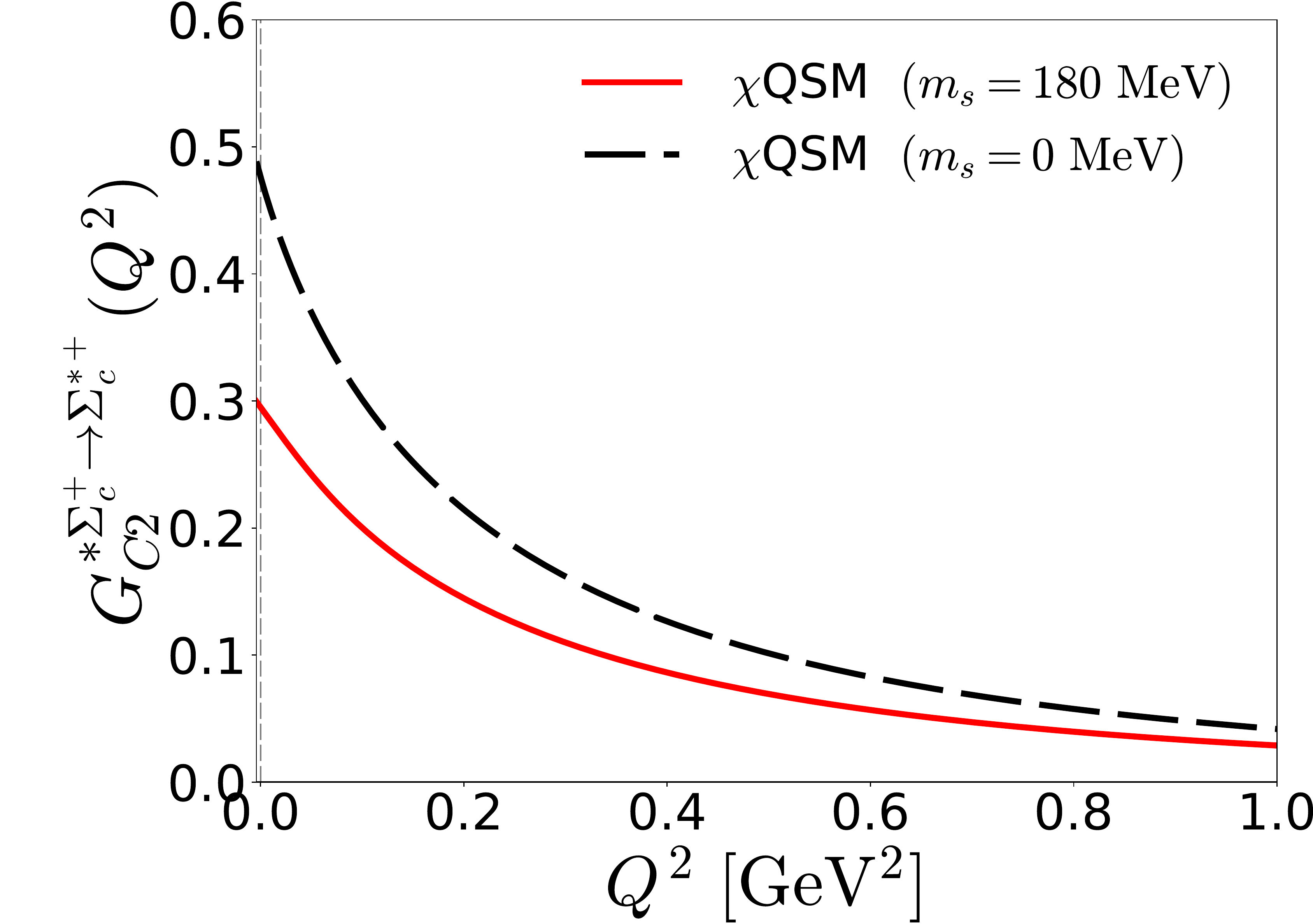}
\includegraphics[scale=0.25]{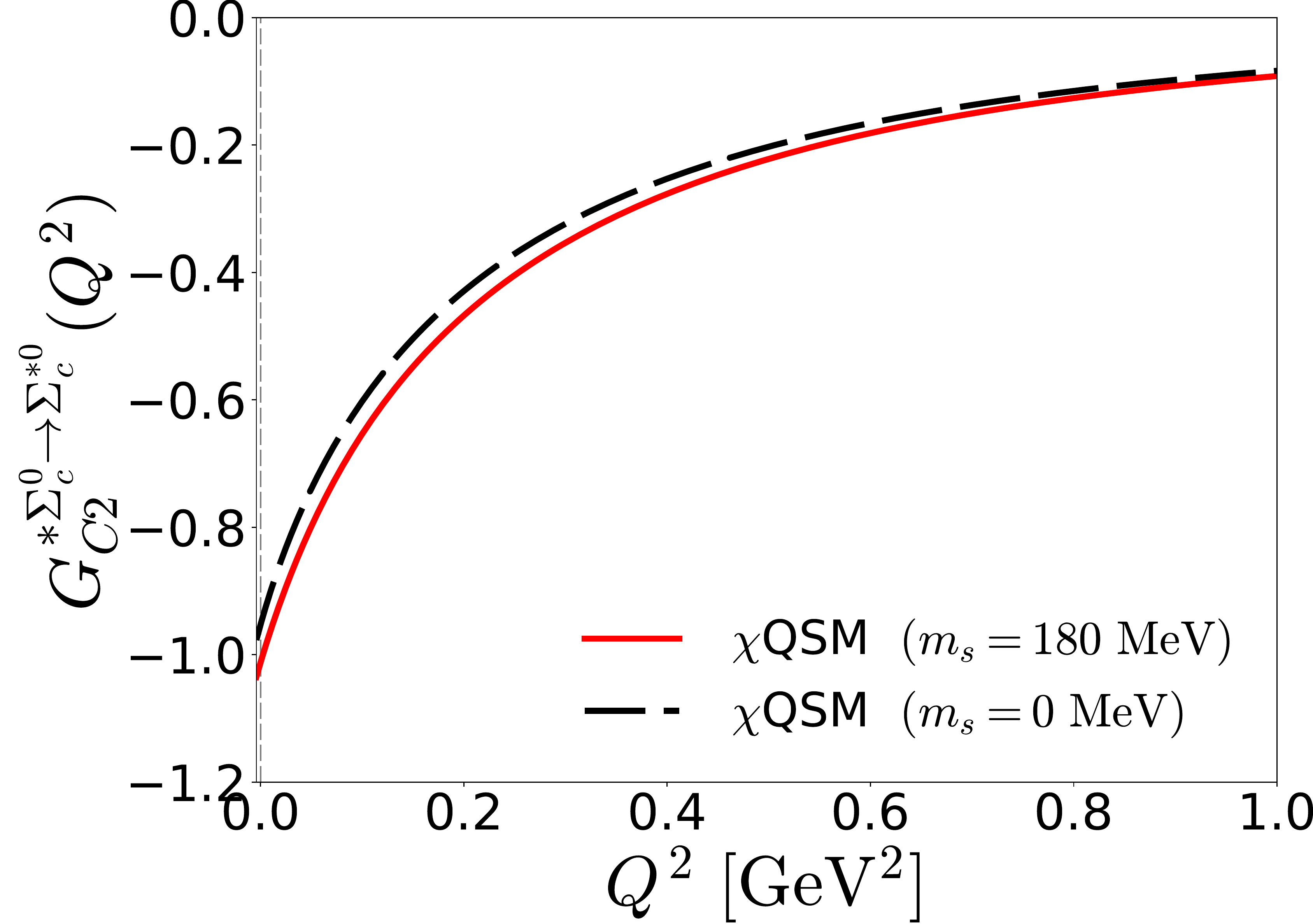}
\includegraphics[scale=0.25]{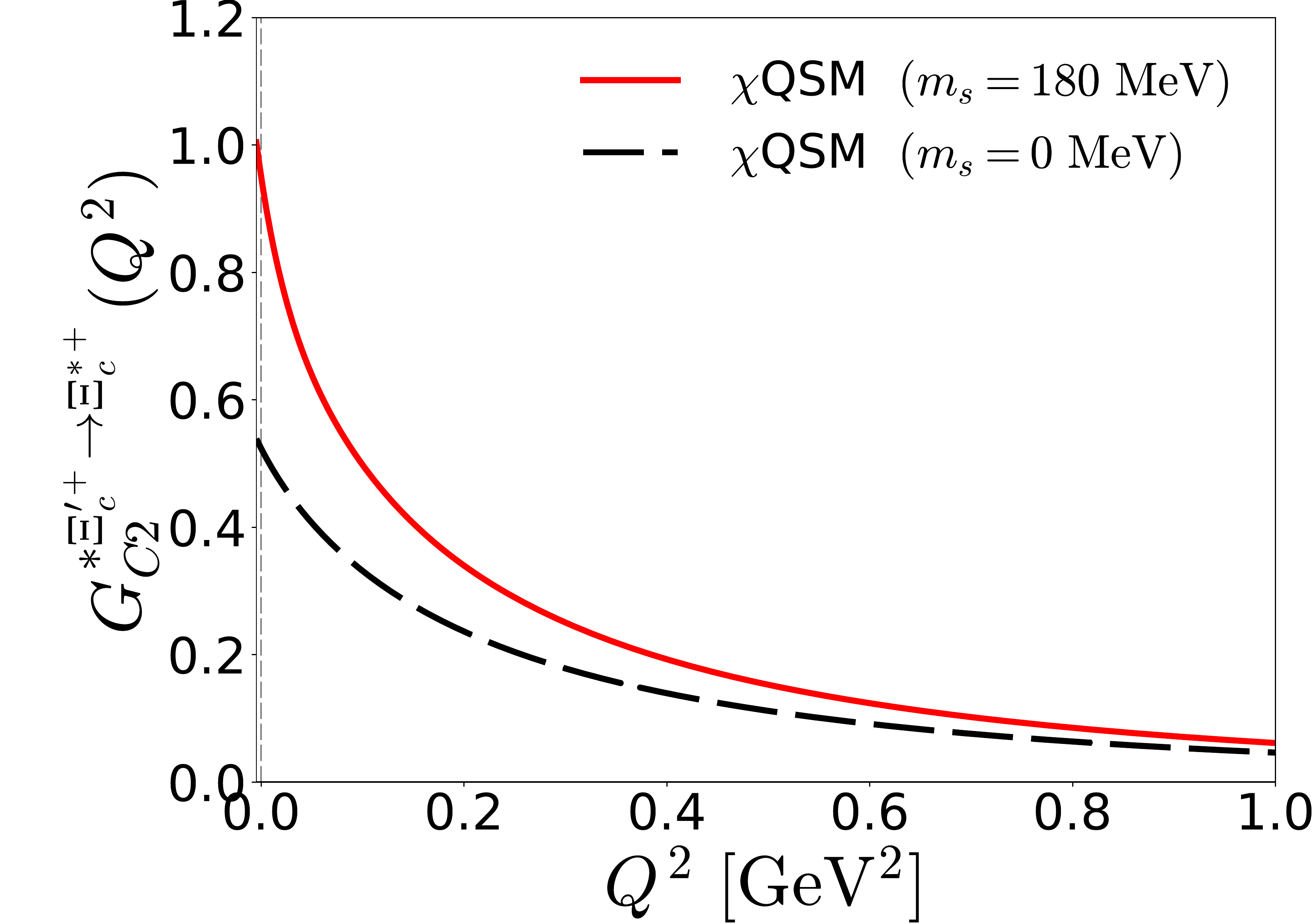}
\includegraphics[scale=0.25]{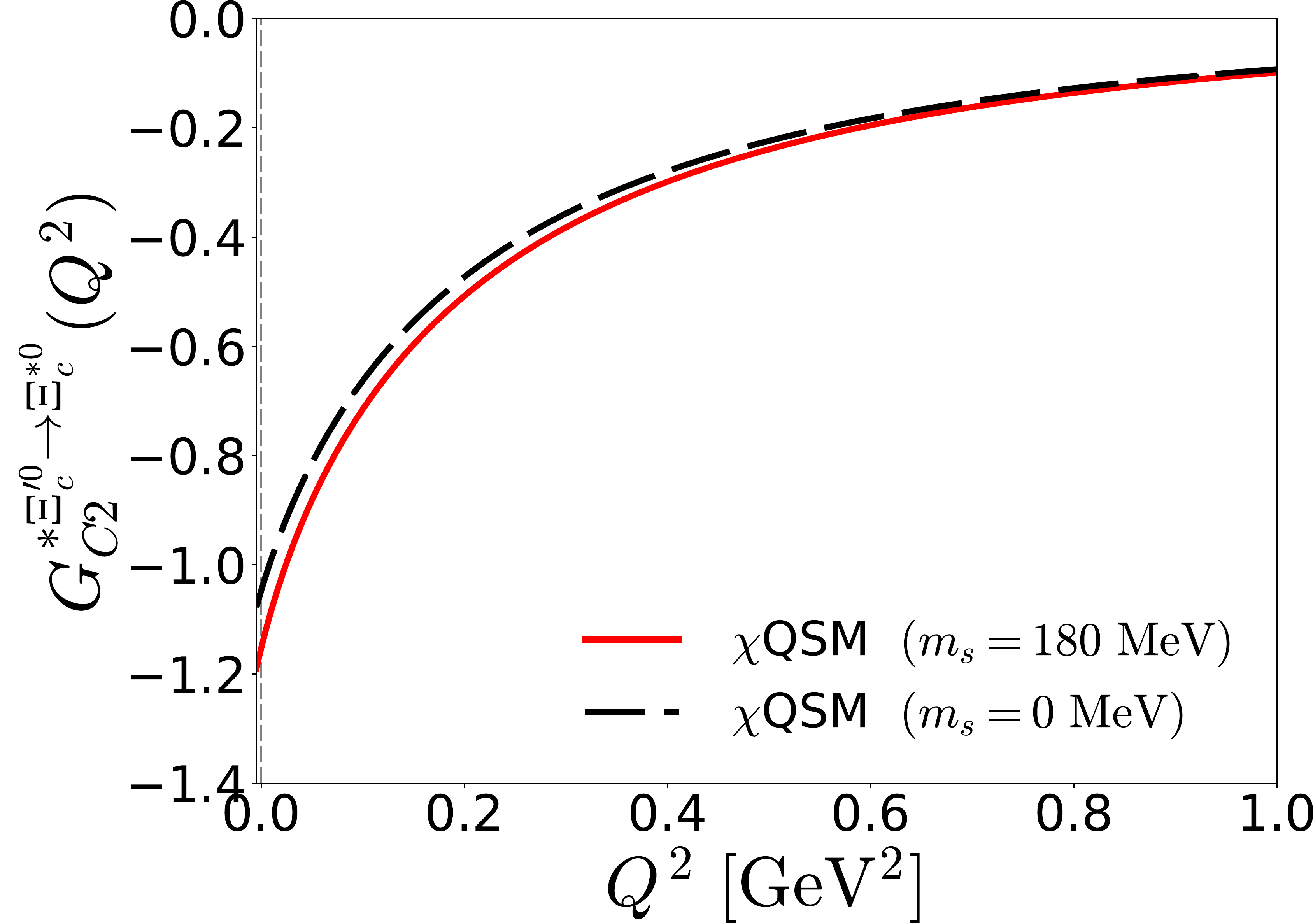}
\includegraphics[scale=0.25]{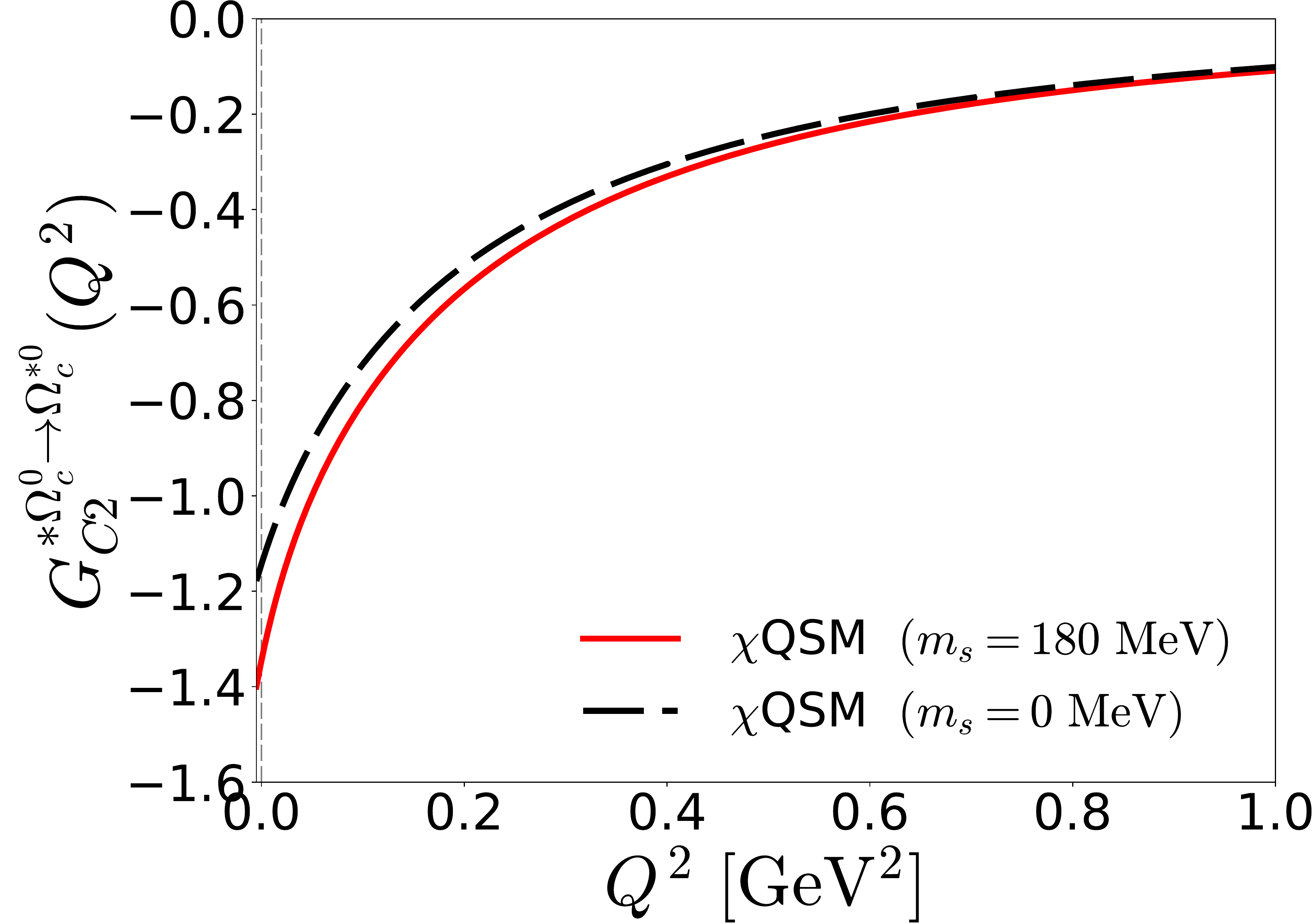}
\caption{Results for the Coulomb quadrupole transition form
  factors from the baryon sextet with spin 1/2 to the baryon sextet
  with spin 3/2. The notations are the same as in Fig.~\ref{fig:6}.} 
\label{fig:9}
\end{figure}
In Fig.~\ref{fig:6}, we examine the effects of flavor SU(3) symmetry
breaking on the M1 transition form factors. While the linear
$m_{\mathrm{s}}$ corrections are very small to the $\Lambda_c^+\to
\Sigma_c^{*+}$ and $\Xi_c^+\to \Xi_c^{*+}$ magnetic transitions, they
become the leading contributions to the $\Xi_c^0\to \Xi_c^{*0}$. The
reason is clear as mentioned previously. The $U$-spin symmetry forbids
the EM transition from $\Xi_c^0\to \Xi_c^{*0}$ as shown in
Eq.~\eqref{eq:M1leadingcon}. Note that $\Xi_c^0$ belongs to the
$U$-spin singlet, while $\Xi_c^{*0}$ is the $U$-spin triplet. 
Only when $m_{\mathrm{s}}\neq m_{\mathrm{u.\,d}}$, it allows the
$\Xi_c^0\to \Xi_c^{*0}$ transition mode. As a result, the magnitude of
the $\Xi_c^0\to \Xi_c^{*0}$ M1 form factor is tiny, compared to those
of the other M1 transition form 
factors. In Fig.~\ref{fig:7}, we depict the results for the M1
transition form factors from the baryon sextet with spin 1/2 to the
baryon sextet with spin 3/2 with the linear $m_{\mathrm{s}}$
considered. The effects of 
the flavor SU(3) symmetry breaking are again negligibly small. It is
interesting to compare these results with those for the M1 form
factors of the baryon decuplet presented in
Ref.~\cite{Kim:2020lgp}. While the linear $m_{\mathrm{s}}$ corrections
are also very small in the case of the M1 transition form factors for 
the baryon decuplet, they turn out to be much smaller for the baryon
sextet than for the decuplet. In the case of the E2 transition form
factors, the linear $m_{\mathrm{s}}$ corrections are sizable for the
$\Sigma_c^{+} \to \Sigma_c^{*+}$ and $\Xi_c^{\prime +} \to \Xi_c^{*+}$
E2 excitations as shown in Fig.~\ref{fig:8}. It is also
interesting to see that the linear $m_{\mathrm{s}}$ corrections
suppress the E2 transition form factors for the $\Sigma_c^{+} \to
\Sigma_c^{*+}$ excitation. These can be understood by examining 
Eqs.~\eqref{eq:E2op} and~\eqref{eq:E2wf}. In Fig.~\ref{fig:9}, we draw
the numerical results for the C2 transition form factors from the
baryon sextet with spin 1/2 to that with spin 3/2. Knowing that the
densities for them are the same as those for the E2 transition form
factors, we can understand the sizable effects of the linear
$m_{\mathrm{s}}$ on the $\Sigma_c^{+} \to \Sigma_c^{*+}$ and
$\Xi_c^{\prime +} \to \Xi_c^{*+}$ transitions.

\subsection{Decay widths of the radiative decay for the baryon sextet
  with spin 3/2}
\begin{table}[htp]
\setlength{\tabcolsep}{5pt} 
\global\long\def\arraystretch{1.2}%
 \caption{Results for the radiative decay widths of $B\gamma\to B^{*}$
   with and without flavor SU(3) symmetry breaking. } 
   {\footnotesize
\begin{tabular}{ccccccccc}
  \hline\hline
\multirow{2}{*}{$\Gamma(B_{c}\gamma\to B_{c}^{*})$} 
  & $\chi$QSM
  & $\chi$QSM
& \multirow{2}{*}{$\chi$SM~\cite{Yang:2019tst}} 
& \multirow{2}{*}{LQCD~\citep{Bahtiyar:2015sga}} 
& \multirow{2}{*}{Bag~\citep{Bernotas:2013eia}} 
& \multirow{2}{*}{$\chi$PT~\citep{Wang:2018cre}} 
& \multirow{2}{*}{QCDSR~\citep{Aliev:2009jt,Aliev:2014bma}} 
& \multirow{2}{*}{QM~\citep{Ivanov:1999bk}}
\tabularnewline
& ($m_{s}=0$ MeV) 
& ($m_{s}=180$ MeV) 
&  
&  
&  
&  
&  
& 
\tabularnewline
\hline 
$\Lambda_{c}^{+}\gamma\to\Sigma_{c}^{*+}$ 
& 63.37 
& 69.76 
& $191.13\pm15.15$ 
& -- 
& 126 
& 161.8 
& 130(45) 
& 151(4)
\tabularnewline
$\Xi_{c}^{+}\gamma\to\Xi_{c}^{*+}$ 
& 34.14 
& 31.97 
& $55.77\pm5.22$ 
& -- 
& 44.3 
& 21.6 
& 52(25) 
& 54(3)
\tabularnewline
$\Xi_{c}^{0}\gamma\to\Xi_{c}^{*0}$ 
& 0 
& 0.08 
& $1.61\pm0.42$ 
& -- 
& 0.908 
& 1.84 
& 0.66(32) 
& 0.68(4)
\tabularnewline
\hline 
$\Sigma_{c}^{++}\gamma\to\Sigma_{c}^{*++}$ 
& 1.12 
& 1.08 
& $2.41\pm0.22$ 
& -- 
& 0.826 
& 1.20 
& 2.65(1.20) 
& --
\tabularnewline
$\Sigma_{c}^{+}\gamma\to\Sigma_{c}^{*+}$ 
& 0.07 
& 0.06 
& $0.11\pm0.02$ 
& -- 
& 0.004 
& 0.04 
& 0.40(16) 
& 0.140(4)
\tabularnewline
$\Sigma_{c}^{0}\gamma\to\Sigma_{c}^{*0}$ 
& 0.28 
& 0.30 
& $0.80\pm0.06$ 
& -- 
& 1.08 
& 0.49 
& 0.08(3) 
& --
\tabularnewline
$\Xi_{c}^{\prime+}\gamma\rightarrow\Xi_{c}^{\ast+}$ 
& 0.09 
& 0.09 
& $0.21\pm0.02$ 
& -- 
& 0.011 
& 0.07 
& 0.274 
& --
\tabularnewline
$\Xi_{c}^{\prime0}\gamma\rightarrow\Xi_{c}^{\ast0}$ 
& 0.35 
& 0.34 
& $0.64\pm0.05$ 
& -- 
& 1.03 
& 0.42 
& 2.142 
& --
\tabularnewline
$\Omega_{c}^{0}\gamma\to\Omega_{c}^{*0}$ 
& 0.38 
& 0.34 
& $0.49\pm0.08$ 
& 0.074 
& 1.07 
& 0.32 
& 0.932 
  & --

\tabularnewline
\hline \hline
\end{tabular}
}
\label{tab:1}
\end{table}
In Table~\ref{tab:1} we list the results for the widths of the
radiative decays from the baryon sextet with spin 3/2 to the baryon
antitriplet in the first three lines  and to the baryon sextet with
spin 1/2 in the next lines. As written in Eq.~\eqref{eq:decay_width},
the decay width for the $B_{3/2}^*\to B_{1/2}\gamma$ is proportional
to $|G_{M1}^*|^2 + 3|G_{E2}^*|^2$. Since we have already shown that
the values of $G_{E2}^*$ are much smaller than those of $G_{M1}^*$, 
the decay widths are approximately proportional to
$|G_{M1}^*|^2$. Table~\ref{tab:1} indicates that the baryon sextet
with spin 3/2 decay more likely into the baryon antitriplet. As
explained previously, the spin state of the decaying baryon is flipped
in the M1 transition. This explains why the decay rates from the
baryon sextet with spin 3/2 to the baryon antitriplet are much larger
than the $\bm{6}_{3/2}\to \bm{6}_{1/2}\gamma$ decays. As we have
already seen from the results for the form factors, the effects of the
flavor SU(3) symmetry are rather small. The fourth column lists the
results from the chiral quark-soliton model in a model-independent
approach~\cite{Yang:2019tst} where all dynamical variables were
determined by experimental data on the light baryons\footnote{Note
  that in Ref.~\cite{Yang:2019tst} the formula for the decay width
  contains an error. The present results listed in the fourth column
  are the corrected ones.}. The present results seem overall
underestimated, compared with those from
Ref.~\cite{Yang:2019tst}. What is interesting is that the present
results are in agreement with those from chiral perturbation
theory~\cite{Wang:2018cre}. On the other hand, lattice QCD yields a
very small value of the decay width for $\Omega_c^{*0}\to \Omega_c^0
\gamma$, compared with the results from all other works.  

\begin{table}[htp]
  \setlength{\tabcolsep}{5pt}
\renewcommand{\arraystretch}{1.5}
\caption{Results for the $R_{EM}$ and $R_{SM}$ on $B\gamma\to
  B^{*}$ with and without flavor SU(3) symmetry breaking.} 
    \scalebox{0.94}{
\begin{tabular}{c c c c c c c c c c c c}
\hline
\hline
 & \multicolumn{2}{c}{$\chi$QSM($m_{s}=0$ MeV)}  &
\multicolumn{2}{c}{$\chi$QSM($m_{s}=180$ MeV)} \\  
 $[\%]$&$R_{EM}$ & $R_{SM}$ &$R_{EM}$ & $R_{SM}$ \\  
 \hline
 ${\Sigma_{c}^{++}\gamma\to\Sigma_{c}^{*++}}$  & -0.87 & -0.88 & -0.75 & -0.76 \\ 
 ${\Sigma_{c}^{+}\gamma\to\Sigma_{c}^{*+}}$    & -0.87 & -0.88 & -0.58 & -0.59 \\ 
 ${\Sigma_{c}^{0}\gamma\to\Sigma_{c}^{*0}}$    & -0.87 & -0.88 & -0.89 & -0.91 \\ 
 ${\Xi_{c}^{'+}\gamma\to\Xi_{c}^{*+}}$         & -0.93 & -0.95 & -1.63 & -1.69 \\ 
 ${\Xi_{c}^{'0}\gamma\to\Xi_{c}^{*0}}$         & -0.93 & -0.95 & -1.04 & -1.06 \\ 
 ${\Omega_{c}^{0}\gamma\to\Omega_{c}^{*0}}$    & -0.96 & -0.98 & -1.19 & -1.22 \\ 
 \hline \hline
\end{tabular}
\label{tab:2}}
\end{table}
We already observed that the E2 transition form factors are very
small. This means that the ratios $R_{EM}$ as defined in
Eq.~\eqref{eq:rem} will turn out to be 
also very small. As listed in Table~\ref{tab:2}, the values of
$R_{EM}$ for the radiative transitions of the baryon sextet with spin
3/2 are indeed very small. Except for the $\Xi_c^{\prime +}\gamma \to
\Xi_c^{*+}$ excitation, the values of the ratios $R_{EM}$ and $R_{SM}$ are
approximately two times smaller than those for the baryon decuplet. 

\section{Summary and conclusion}
We aimed in the present work at investigating the electromagnetic
transition form factors for the singly heavy baryons with spin 3/2,
based on the pion mean-field approach or the chiral quark-soliton
model. Having taken into account the rotational $1/N_c$ and linear
$m_{\mathrm{s}}$ corrections, we computed the magnetic dipole,
electric quadrupole and Coulomb quadrupole transition form factors
for the radiative excitations from the baryon antitriplet and sextet
with spin 1/2 to the baryon sextet with spin 3/2. 
Since the model parameters were already fixed in producing 
properties of the light baryons, we use exactly the same set of the
parameters for the present investigation. We compared the results for
the magnetic dipole (M1) and electric quadrupole (E2) transition form 
factors with those from a lattice calculation. However, the lattice
results for the form factor for the $\Omega_c^0\to \Omega_c^{*0}$ M1
transition seems to be underestimated in comparison with those from
other works. Since the lattice calculation suffers from large
uncertainties in the results for the corresponding E2 transition
form factor, we were not able to draw any conclusion from the
comparison of the present results with the lattice ones. We then
examined the valence- and sea-quark contributions to the M1, E2 and C2 
transition form factors. While the sea-quark contributions are
marginal to the M1 form factors, they dominate over the valence-quark
contributions in the smaller $Q^2$ region. On the other hand, the
sea-quark contributions fall off faster than the valence-quark ones as
$Q^2$ increases. The magnitudes of the M1 transitions form factors
from the baryon antitriplet to the baryon sextet with spin 3/2 are in
general larger than those from the baryon sextet with spin 1/2 to the
sextet with spin 3/2. This indicates that the M1 transitions occur
more naturally between the states with the total spin flipped. Since
the E2 and C2 transitions take place in the transitions without any spin
flip, we have null results for the transitions from the baryon
antitriplet to the sextet with spin 3/2.

We also examined the effects of flavor SU(3) symmetry breaking by
considering the linear $m_{\mathrm{s}}$ corrections. Except for the
$\Xi_c^0\to \Xi_c^{*0}$ transition that is forbidden by the $U$-spin
symmetry, we found that the effects of the flavor SU(3) symmetry
breaking are negligibly small. Since the $U$-spin symmetry is broken
by the finite value of the strange current quark mass, the M1
transition form factor for the $\Xi_c^0\to \Xi_c^{*0}$ radiative
excitation is finite but tiny, compared with those for other
transition modes. Similarly, we found that the linear $m_{\mathrm{s}}$
corrections are also very small to the E2 transition form factors
except for the $\Sigma_c^{+}\to \Sigma_c^{*+}$ and $\Xi_c^{\prime +}
\to \Xi_c^{*+}$ transitions. Similar features were seen in the results
for the Coulomb qudrupole form factors. We also computed the 
widths of the radiative decays for the baryon sextet with spin
3/2. The results for the transitions within the baryon sextet are in
agreement with those from chiral perturbation theory. Finally, we
presented the results for the ratios of the E2 over M1 and C2 over
M1. They turn out to be approximately two times smaller than those for
the baryon decuplet. As the electromagnetic (transition) form factors
provide the most valuable information on the structure of the
hadrons, it is extremely important to measure them, experimentally
and/or by lattice QCD computations. So far very limited information is
given for the charmed baryons, and further experimental and
computational efforts are highly desired.

\begin{acknowledgments}
The authors are grateful to K.~U. Can for valuable discussion related
to the lattice results. The present work was supported by Basic
Science Research Program through the National Research Foundation of
Korea funded by the Ministry of Education, Science and Technology 
(Grant-No. 2018R1A2B2001752 and 2018R1A5A1025563). J.-Y.K is supported
by the Deutscher Akademischer Austauschdienst(DAAD) doctoral
scholarship. The work of Gh.-S.-Y. is supported by
NRF-2019R1A2C1010443. The work of M.O. is supported in part by JSPS 
KAKENHI Grants No. JP19H05159 and JP20K03959. 
\end{acknowledgments}

\appendix

\section{Densities for the EM transition form factors
  \label{app:a}} 
The densities of the magnetic dipole transition form factors are
expressed explicitly as follows:
\begin{align}
 {\cal{Q}}_{0} (\bm{r})  =&    (N_{c}-1)\langle \mathrm{val}| \bm{r}
 \rangle  \gamma^5\{\hat{\bm{r}} \times \bm{\sigma} \}\cdot\bm{\tau} \langle
  \bm{r} | \mathrm{val}  \rangle 
  +N_{c}\sum_{n}   \mathcal{R}_{1}(E_{n}) \langle n| \bm{r} \rangle
  \gamma^5  \{ \hat{\bm{r}} \times \bm{\sigma} \}\cdot \bm{\tau} \langle
 \bm{r} |  n \rangle,  \cr 
 {\cal{Q}}_{1} (\bm{r})  =&  \frac{i}{2} (N_{c}-1) \sum_{n \ne
 \mathrm{val}} \frac{\mathrm{sign}(E_{n})}{E_{n}-E_{\mathrm{val}}}  
 \langle n| \bm{r} \rangle\gamma^5[  \{\hat{\bm{r}} \times \bm{\sigma}
 \}\times\bm{\tau} ] \langle  \bm{r} | \mathrm{val} \rangle \cdot
 \langle  {\mathrm{val}} | \bm{\tau}| n \rangle
 \cr 
& + \frac{i}{4}N_{c}\sum_{n,m}  {\mathcal R}_{4}(E_{n},E_{m})  \langle
  m | \bm{r} \rangle\gamma^5 [\{\hat{\bm{r}} \times
  \bm{\sigma}\}\times\bm{\tau}] \langle \bm{r} | n \rangle \cdot
  \langle m |\bm{\tau}| n \rangle ,  \cr  
 {\mathcal{X}}_{1} (\bm{r}) =& (N_{c}-1)  \sum_{n \ne
  \mathrm{val}}\frac{1}{E_{n}-E_{\mathrm{val}}}
   \langle \mathrm{val}| \bm{r} \rangle\gamma^5
  \{\hat{\bm{r}}\times\bm{\sigma}\} \langle \bm{r} | \mathrm{val}
 \rangle\cdot \langle n|  \bm{\tau} | \mathrm{val}  \rangle \cr  
&+\frac{1}{2}N_{c}\sum_{n, m} \mathcal{R}_{5}(E_{n},E_{m})\langle n |  
  \bm{r} \rangle \gamma^5  \{\hat{\bm{r}}\times\bm{\sigma}\}  \langle \bm{r}
  | m  \rangle \cdot  \langle m | \bm{\tau} | n \rangle , \cr 
{\cal{X}}_{2} (\bm{r})   =&    (N_{c}-1) \sum_{n^{0}}
  \frac{1}{E_{n^{0}}-E_{\mathrm{val}}}
  \langle \mathrm{val}| \bm{r} \rangle\gamma^5 \{\hat{\bm{r}} \times
  \bm{\sigma} \}\cdot\bm{\tau}  \langle \bm{r} | n^0 \rangle \langle
 n^0   |  \mathrm{val} \rangle  \cr 
& +N_{c}\sum_{n^0,m}\mathcal{R}_{5}(E_{m},E_{n^{0}}) \langle m |
  \bm{r} \rangle \gamma^5  \{\hat{\bm{r}} \times
  \bm{\sigma} \}\cdot\bm{\tau} \langle  \bm{r}| n^0 \rangle \langle
  n^0 | m \rangle ,  \cr 
 {\cal{M}}_{0} (\bm{r})   =&  (N_{c}-1) \sum_{n \ne
\mathrm{val}} \frac{1}{E_{n}-E_{\mathrm{val}}}  \langle
\mathrm{val}| \bm{r} \rangle \gamma^5  \{\hat{\bm{r}} \times
\bm{\sigma} \}\cdot\bm{\tau} \langle \bm{r} |  n \rangle \langle n|
\gamma^{0} |   \mathrm{val} \rangle  \cr 
& - \frac{1}{2}N_{c}\sum_{n,m} \mathcal{R}_{2}(E_{n},E_{m}) \langle m |
  \bm{r} \rangle \gamma^5  \{\hat{\bm{r}} \times
\bm{\sigma} \}\cdot\bm{\tau} \langle \bm{r} |  n \rangle \langle n|
\gamma^{0} | m \rangle  ,  \cr 
{\cal{M}}_{1} (\bm{r})=& (N_{c}-1)  \sum_{n \ne
 \mathrm{val}}\frac{1}{E_{n}-E_{\mathrm{val}}}
 \langle \mathrm{val}| \bm{r} \rangle  \gamma^5  \{\hat{\bm{r}} \times
  \bm{\sigma} \} \langle \bm{r} | n \rangle\cdot \langle n| \gamma^{0}
 \bm{\tau}  | \mathrm{val}  \rangle    \cr 
& -\frac{1}{2}N_{c}\sum_{n,m} {\cal R}_{2}(E_{n},E_{m})
  \langle m | \bm{r} \rangle \gamma^5  \{\hat{\bm{r}} \times
  \bm{\sigma} \} \langle \bm{r} | n \rangle\cdot \langle n| \gamma^{0} 
 \bm{\tau} | m \rangle,  \cr 
{\cal{M}}_{2} (\bm{r}) =& (N_{c}-1) \sum_{n^{0}}
  \frac{1}{E_{n^{0}}-E_{\mathrm{val}}}
  \langle \mathrm{val}| \bm{r} \rangle  \gamma^5  \{\hat{\bm{r}} \times
\bm{\sigma} \}\cdot\bm{\tau} \langle \bm{r} | n^{0} \rangle \langle
 n^{0}|  \gamma^{0} | \mathrm{val} \rangle  \cr 
& - N_{c}\sum_{n^{0},m} \mathcal{R}_{2}(E_{n^0},E_{m})
  \langle m | \bm{r} \rangle   \gamma^5  \{\hat{\bm{r}} \times
\bm{\sigma} \}\cdot\bm{\tau}
\langle  \bm{r}| n^0 \rangle\langle n^0 | \gamma^{0} | m \rangle.
\label{eq:M1den1}                                                   
\end{align}
The densities of the electric quadrupole transition form factors are
given as 
\begin{align}
-2\sqrt{10} \mathcal{I}_{1E2}(\bm{r})&= (N_{c}-1) \sum_{n \ne
\mathrm{val} }\frac{1}{E_{n}-E_{\mathrm{val}}}{\langle\mathrm{val}
| \bm{\tau} | n\rangle} \cdot{\langle n |\bm{r} \rangle
\{ \sqrt{4\pi}Y_{2}  \otimes\tau_{1}  \}_{1}\langle \bm{r} |
\mathrm{val}\rangle} \cr 
&  + \frac{1}{2} N_{c}\sum_{n,m} \mathcal{R}_{3}(E_n,E_m)   {\langle n |
  \bm{\tau} | m \rangle} \cdot{\langle m | \bm{r} \rangle \{
  \sqrt{4\pi} Y_{2}  \otimes \tau_{1}  \}_{1} \langle \bm{r} | n
  \rangle}  ,\cr 
-2\sqrt{10} \mathcal{K}_{1E2}(\bm{r})&= (N_{c}-1)  \sum_{n \ne
 \mathrm{val} } \frac{1}{E_{n}-E_{\mathrm{val}}} {\langle\mathrm{val}
 |  \gamma^{0} \bm{\tau} | n \rangle} \cdot {\langle n | \bm{r} \rangle
 \{ \sqrt{4\pi} Y_{2}  \otimes \tau_{1}  \}_{1} \langle \bm{r} |
 \mathrm{val} \rangle} \cr  
& + \frac{1}{2} N_{c} \sum_{n,m} \mathcal{R}_{5}(E_n,E_m)     {\langle n |
  \gamma^{0} \bm{\tau} | m \rangle} \cdot {\langle m | \bm{r} \rangle
  \{ \sqrt{4\pi} Y_{2}  \otimes \tau_{1}  \}_{1} \langle \bm{r} | n
                               \rangle}.
\label{eq:E2den1}                               
\end{align}
The regularization functions in Eqs.~\eqref{eq:M1den1}
and~\eqref{eq:E2den1} are defined by
\begin{align}
&\mathcal{R}_{1}(E_{n}) = -\frac{1}{2 \sqrt{\pi}} E_{n}
  \int^{\infty}_{0} \phi(u) \frac{du}{u} e^{-u E_{n}^{2}}, \cr 
&\mathcal{R}_{2}(E_{n},E_{m}) = \frac{1}{2 \sqrt{\pi}} \int^{\infty}_{0}
  \phi(u) \frac{du}{\sqrt{u}} \frac{E_{m} e^{-u E_{m}^{2}}-E_{n} e^{-u
  E_{n}^{2}}}{E_{n} - E_{m}}, \cr 
&\mathcal{R}_{3}(E_{n},E_{m}) = \frac{1}{2 \sqrt{\pi}} \int^{\infty}_{0}
  \phi(u) \frac{du}{\sqrt{u}} \left[ \frac{ e^{-u E_{m}^{2}}- e^{-u
  E_{n}^{2}}}{u(E^{2}_{n} - E^{2}_{m})} -\frac{E_{m} e^{-u
  E_{m}^{2}}+E_{n} e^{-u E_{n}^{2}}}{E_{n} + E_{m}}  \right ], \cr 
&\mathcal{R}_{4}(E_{n},E_{m}) = \frac{1}{2 {\pi}} \int^{\infty}_{0}
  \phi(u) {du} \int^{1}_{0} d\alpha e^{-u E_{n}^{2}(1-\alpha) - u
  E^{2}_{m}\alpha}  \frac{E_{n}(1-\alpha) - \alpha
  E_{m}}{\sqrt{\alpha(1-\alpha)}}, \cr 
&\mathcal{R}_{5}(E_{n},E_{m}) =
  \frac{\mathrm{sign}(E_{n})-\mathrm{sign}(E_{m})}{2(E_{n}-E_{m})},
\end{align}
with proper-time regulator $\phi(u)=c \theta(u-\Lambda^{-2}_{1}) +
(1-c) \theta(u-\Lambda^{-2}_{2})$. The cutoff parameters $c,
\Lambda_{1}$ and $\Lambda_{2}$ were determined in
Ref.~\cite{Christov:1995vm}. $|\mathrm{val}\rangle$ and $|n\rangle$
denote the states of the valence and sea quarks with the corresponding
eigenenergies $E_{\mathrm{val}}$ and $E_n$ of the single-quark
Hamiltonian $h(U_c)$, respectively~\cite{Christov:1995vm}.

\section{Collective matrix elements of electromagnetic form factor
  \label{app:b}} 
In Tables~\ref{tab:3}, \ref{tab:4}, \ref{tab:5} and \ref{tab:6}, we
present the results for all relevant matrix elements of the SU(3)
Wigner $D$ functions. 
   \begin{table}[htp]
\setlength{\tabcolsep}{4pt}
\renewcommand{\arraystretch}{1.75}
  \caption{The matrix elements of the collective operators for the
    leading-order contributions and the $1/N_c$ rotational corrections
    to the electrimagnetic transition form factors.}  
\begin{center}
\begin{tabular}{ c |  c c | c |  c  c c  } 
 \hline 
  \hline 
  $B_{\overline{3}}\gamma^{*} \to B_{6}$& $\Lambda_{c} \gamma^{*} \to \Sigma^{*}_{c}$ & $\Xi_{c} \gamma^{*} \to \Xi^{*}_{c}$ & $B_{6}\gamma^{*} \to B_{6}$ & $\Sigma_{c} \gamma^{*} \to \Sigma^{*}_{c}$ & $\Xi'_{c} \gamma^{*} \to \Xi^{*}_{c}$ & $\Omega_{c} \gamma^{*} \to \Omega^{*}_{c}$ \\  
 \hline
$\langle B_{{6}} |D^{(8)}_{33} | B_{\overline{3}}\rangle$  
& $\frac{1}{2\sqrt{3}} $& $-\frac{1}{2\sqrt{3}}T_{3}$&
$\langle B_{{6}} |D^{(8)}_{33} | B_{6}\rangle$ 
&$\frac{1}{5\sqrt{2}}T_{3}$ & $\frac{1}{5\sqrt{2}}T_{3}$ & 0 \\  

$\langle B_{{6}} |D^{(8)}_{83} | B_{\overline{3}}\rangle$  
& $0$ & $-\frac{1}{4}$ &
$\langle B_{{6}} |D^{(8)}_{83} | B_{6}\rangle$  
 & $\frac{1}{5\sqrt{6}}$& $-\frac{1}{10\sqrt{6}}$ & $-\frac{1}{5}\sqrt{\frac{2}{3}}$ \\  

$\langle B_{{6}}|D^{(8)}_{38}J_{3}  | B_{\overline{3}}\rangle$  
& $0$& $0$
&$\langle B_{{6}}|D^{(8)}_{38}J_{3}  | B_{6}\rangle$  & $-\frac{1}{5\sqrt{6}}T_{3}$ & $-\frac{1}{5\sqrt{6}}T_{3}$ & $0$ \\    

$\langle B_{{6}}|D^{(8)}_{88}J_{3}  | B_{\overline{3}}\rangle$  
& $0$  & $0$
& $\langle B_{{6}}|D^{(8)}_{88}J_{3}  | B_{6}\rangle$  & $-\frac{1}{15\sqrt{2}}$ & $\frac{1}{30\sqrt{2}}$ & $\frac{\sqrt{2}}{15}$ \\  

$\langle B_{{6}}|d_{ab3}D^{(8)}_{3a}J_{b}  | B_{\overline{3}}\rangle$  
&$-\frac{1}{4\sqrt{3}}$ &$\frac{1}{4\sqrt{3}}T_{3}$
&$\langle B_{{6}}|d_{ab3}D^{(8)}_{3a}J_{b}  | B_{6}\rangle$  &$-\frac{1}{10\sqrt{2}}T_{3}$ &$-\frac{1}{10\sqrt{2}}T_{3}$ & $0$  \\  

$\langle B_{{6}}|d_{ab3}D^{(8)}_{8a}J_{b} | B_{\overline{3}}\rangle$
& $0$ & $\frac{1}{8}$ 
&$\langle B_{{6}}|d_{ab3}D^{(8)}_{8a}J_{b} | B_{6}\rangle$ & $-\frac{1}{10\sqrt{6}}$ & $\frac{1}{20\sqrt{6}}$ & $\frac{1}{5\sqrt{6}}$  \\  

$\langle B_{{6}}|D^{(8)}_{3i}J_{i} | B_{\overline{3}}\rangle$  
& $0$& $0$ 
& $\langle B_{{6}}|D^{(8)}_{3i}J_{i} | B_{6}\rangle$  & $0$ & $0$ & $0$ \\  

$\langle B_{{6}}|D^{(8)}_{8i}J_{i} | B_{\overline{3}}\rangle$  
& $0$& $0$
&$\langle B_{{6}}|D^{(8)}_{8i}J_{i} | B_{6}\rangle$  &  $0$ & $0$ & $0$ \\  
 \hline 
 \hline
\end{tabular}
\end{center}
  \label{tab:3}
\end{table}

   \begin{table}[htp]
\setlength{\tabcolsep}{4pt}
\renewcommand{\arraystretch}{1.75}
  \caption{The matrix elements of the collective operators for the
    $m_s$ corrections to the electromagnetic transition form factors.} 
 \begin{center}
\begin{tabular}{ c |  c c | c |  c  c c  } 
 \hline 
  \hline 
  $B_{\overline{3}}\gamma^{*} \to B_{6}$& $\Lambda_{c} \gamma^{*} \to \Sigma^{*}_{c}$ & $\Xi_{c} \gamma^{*} \to \Xi^{*}_{c}$ & $B_{6}\gamma^{*} \to B_{6}$ & $\Sigma_{c} \gamma^{*} \to \Sigma^{*}_{c}$ & $\Xi'_{c} \gamma^{*} \to \Xi^{*}_{c}$ & $\Omega_{c} \gamma^{*} \to \Omega^{*}_{c}$ \\  
 \hline
$\langle B_{{6}} |D^{(8)}_{88}D^{(8)}_{33} | B_{\overline{3}}\rangle$  
& $\frac{\sqrt{3}}{20}$& $0$&
$\langle B_{{6}} |D^{(8)}_{88}D^{(8)}_{33} | B_{6}\rangle$ 
&$\frac{\sqrt{2}}{45}T_{3}$&$\frac{1}{45\sqrt{2}}T_{3}$ & $0$ \\  

$\langle B_{{6}} |D^{(8)}_{88}D^{(8)}_{83} | B_{\overline{3}}\rangle$  
& $0$ & $\frac{1}{20}$ &
$\langle B_{{6}} |D^{(8)}_{88}D^{(8)}_{83} | B_{6}\rangle$  
 & $-\frac{1}{30\sqrt{6}}$& $0$ & $\frac{1}{10\sqrt{6}}$ \\  

$\langle B_{{6}}|D^{(8)}_{83}D^{(8)}_{38}  | B_{\overline{3}}\rangle$  
& $\frac{1}{20\sqrt{3}}$& $-\frac{1}{5\sqrt{3}}T_{3}$&
$\langle B_{{6}}|D^{(8)}_{83}D^{(8)}_{38}  | B_{6}\rangle$
&$\frac{\sqrt{2}}{45}T_{3}$&$\frac{1}{45\sqrt{2}}T_{3}$ & $0$ \\    

$\langle B_{{6}}|D^{(8)}_{83}D^{(8)}_{88}  | B_{\overline{3}}\rangle$  
& $0$ & $\frac{1}{20}$ 
& $\langle B_{{6}}|D^{(8)}_{83}D^{(8)}_{88}  | B_{6}\rangle$  
& $-\frac{1}{30\sqrt{6}}$& $0$ & $\frac{1}{10\sqrt{6}}$ \\  

$\langle B_{{6}}|d_{ab3}D^{(8)}_{8a}D^{(8)}_{8b}  | B_{\overline{3}}\rangle$  
&$0$ &$\frac{\sqrt{3}}{20}$
&$\langle B_{{6}}|d_{ab3}D^{(8)}_{8a}D^{(8)}_{8b}  | B_{6}\rangle$  &$-\frac{\sqrt{2}}{45}$ &$\frac{1}{30\sqrt{2}}$ &  $\frac{1}{15\sqrt{2}}$   \\  

$\langle B_{{6}}|d_{ab3}D^{(8)}_{3a}D^{(8)}_{8b} | B_{\overline{3}}\rangle$
& $\frac{1}{10}$ & $-\frac{1}{10} T_{3}$ 
&$\langle B_{{6}}|d_{ab3}D^{(8)}_{3a}D^{(8)}_{8b} | B_{6}\rangle$ & $\frac{1}{9\sqrt{6}}T_{3}$ & $\frac{7}{45\sqrt{6}}T_{3}$ & $0$  \\  

$\langle B_{{6}}|D^{(8)}_{83}D^{(8)}_{33} | B_{\overline{3}}\rangle$  
& $0$& $0$ 
& $\langle B_{{6}}|D^{(8)}_{83}D^{(8)}_{33} | B_{6}\rangle$  & $-\frac{1}{45}\sqrt{\frac{2}{3}}T_{3}$ & $\frac{4}{45}\sqrt{\frac{2}{3}}T_{3}$  & $0$\\  

$\langle B_{{6}}|D^{(8)}_{83}D^{(8)}_{83} | B_{\overline{3}}\rangle$  
& $0$& $0$
&$\langle B_{{6}}|D^{(8)}_{83}D^{(8)}_{83} | B_{6}\rangle$  &  $-\frac{1}{45\sqrt{2}}$ & $\frac{1}{15\sqrt{2}}$ & $-\frac{1}{15\sqrt{2}}$   \\  

$\langle B_{{6}}|D^{(8)}_{8i}D^{(8)}_{3i} | B_{\overline{3}}\rangle$  
& $0$& $0$
&$\langle B_{{6}}|D^{(8)}_{8i}D^{(8)}_{3i} | B_{6}\rangle$  & $\frac{1}{45}\sqrt{\frac{2}{3}}T_{3}$ & $-\frac{4}{45}\sqrt{\frac{2}{3}}T_{3}$  & $0$\\  

$\langle B_{{6}}|D^{(8)}_{8i}D^{(8)}_{8i} | B_{\overline{3}}\rangle$  
& $0$& $0$
&$\langle B_{{6}}|D^{(8)}_{8i}D^{(8)}_{8i} | B_{6}\rangle$  &  $\frac{1}{45\sqrt{2}}$ & $-\frac{1}{15\sqrt{2}}$ & $\frac{1}{15\sqrt{2}}$   \\  
 \hline 
 \hline
\end{tabular}
\end{center}
 \label{tab:4}
\end{table}

   \begin{table}[htp]
\setlength{\tabcolsep}{4pt}
\renewcommand{\arraystretch}{1.75}
  \caption{The relevant transition matrix elements of the collective
    operators coming from the 15-plet component of the baryon wave 
    functions.}  
\begin{center}
\begin{tabular}{ c |  c c | c |  c  c c  } 
 \hline 
  \hline 
  $B_{\overline{15}}\gamma^{*} \to B_{6}$& $\Lambda_{c} \gamma^{*} \to \Sigma^{*}_{c}$ & $\Xi_{c} \gamma^{*} \to \Xi^{*}_{c}$ & $B_{\overline{15}}\gamma^{*} \to B_{6}$ & $\Sigma_{c} \gamma^{*} \to \Sigma^{*}_{c}$ & $\Xi'_{c} \gamma^{*} \to \Xi^{*}_{c}$ & $\Omega_{c} \gamma^{*} \to \Omega^{*}_{c}$ \\  
 \hline
$\langle B_{{6}} |D^{(8)}_{33} | B_{\overline{15}}\rangle$  
& $\frac{1}{6\sqrt{15}} $& $-\frac{\sqrt{5}}{18}T_{3}$&
$\langle B_{{6}} |D^{(8)}_{33} | B_{\overline{15}}\rangle$ 
&$\frac{1}{9\sqrt{5}}T_{3}$ & $\frac{1}{9}\sqrt{\frac{5}{6}}T_{3}$ & $0$ \\  

$\langle B_{{6}} |D^{(8)}_{83} | B_{\overline{15}}\rangle$  
& $0$ & $\frac{1}{4\sqrt{15}}$ &
$\langle B_{{6}} |D^{(8)}_{83} | B_{\overline{15}}\rangle$  
 & $-\frac{1}{3\sqrt{15}}$& $-\frac{1}{6\sqrt{10}}$ & $0$ \\  

$\langle B_{{6}}|D^{(8)}_{38}J_{3}  | B_{\overline{15}}\rangle$  
& $0$& $0$
&$\langle B_{{6}}|D^{(8)}_{38}J_{3}  | B_{\overline{15}}\rangle$  & $\frac{1}{3\sqrt{15}}T_{3}$ & $\frac{1}{9}\sqrt{\frac{5}{2}}T_{3}$ & $0$ \\    

$\langle B_{{6}}|D^{(8)}_{88}J_{3}  | B_{\overline{15}}\rangle$  
& $0$  & $0$
& $\langle B_{{6}}|D^{(8)}_{88}J_{3}  | B_{\overline{15}}\rangle$  & $-\frac{1}{3\sqrt{5}}$ & $-\frac{1}{2\sqrt{30}}$ & $0$ \\  

$\langle B_{{6}}|d_{ab3}D^{(8)}_{3a}J_{b}  | B_{\overline{15}}\rangle$  
&$\frac{1}{4\sqrt{15}}$ &$-\frac{\sqrt{5}}{12}T_{3}$
&$\langle B_{{6}}|d_{ab3}D^{(8)}_{3a}J_{b}  | B_{\overline{15}}\rangle$  &$\frac{1}{18\sqrt{5}}T_{3}$ &$\frac{1}{18}\sqrt{\frac{5}{6}}T_{3}$ & $0$  \\  

$\langle B_{{6}}|d_{ab3}D^{(8)}_{8a}J_{b} | B_{\overline{15}}\rangle$
& $0$ & $\frac{1}{8}\sqrt{\frac{3}{5}}$ 
&$\langle B_{{6}}|d_{ab3}D^{(8)}_{8a}J_{b} | B_{\overline{15}}\rangle$ & $-\frac{1}{6\sqrt{15}}$ & $-\frac{1}{12\sqrt{10}}$ & $0$  \\  

$\langle B_{{6}}|D^{(8)}_{3i}J_{i} | B_{\overline{15}}\rangle$  
& $0$& $0$ 
& $\langle B_{{6}}|D^{(8)}_{3i}J_{i} | B_{\overline{15}}\rangle$  & $0$ & $0$ & $0$ \\  

$\langle B_{{6}}|D^{(8)}_{8i}J_{i} | B_{\overline{15}}\rangle$  
& $0$& $0$
&$\langle B_{{6}}|D^{(8)}_{8i}J_{i} | B_{\overline{15}}\rangle$  &  $0$ & $0$ & $0$ \\  
 \hline 
 \hline
\end{tabular}
\end{center}
  \label{tab:5}
\end{table}

   \begin{table}[htp]
\setlength{\tabcolsep}{4pt}
\renewcommand{\arraystretch}{1.75}
  \caption{The relevant transition matrix elements of the collective
    operators coming from the 15- and 24-plet components of the baryon wave 
    functions.}  
\begin{center}
\begin{tabular}{ c |  c c | c |  c  c c  } 
 \hline 
  \hline 
  $B_{\overline{3}}\gamma^{*} \to B_{\overline{15}}$& $\Lambda_{c} \gamma^{*} \to \Sigma^{*}_{c}$ & $\Xi_{c} \gamma^{*} \to \Xi^{*}_{c}$ & $B_{6}\gamma^{*} \to B_{\overline{24}}$ & $\Sigma_{c} \gamma^{*} \to \Sigma^{*}_{c}$ & $\Xi'_{c} \gamma^{*} \to \Xi^{*}_{c}$ & $\Omega_{c} \gamma^{*} \to \Omega^{*}_{c}$ \\  
 \hline
$\langle B_{\overline{15}} |D^{(8)}_{33} | B_{\overline{3}}\rangle$  
& $\frac{1}{\sqrt{30}} $& $-\frac{\sqrt{5}}{6}T_{3}$&
$\langle B_{\overline{24}} |D^{(8)}_{33} | B_{6}\rangle$ 
&$\frac{1}{90\sqrt{2}}T_{3}$ & $\frac{1}{45\sqrt{3}}T_{3}$ & $0$ \\  

$\langle B_{\overline{15}} |D^{(8)}_{83} | B_{\overline{3}}\rangle$  
& $0$ & $\frac{1}{4}\sqrt{\frac{3}{5}}$ &
$\langle B_{\overline{24}} |D^{(8)}_{83} | B_{6}\rangle$  
 & $\frac{1}{15\sqrt{6}}$& $\frac{1}{30}$ & $\frac{1}{30}$ \\  

$\langle B_{\overline{15}}|D^{(8)}_{38}J_{3}  | B_{\overline{3}}\rangle$  
& $0$& $0$
&$\langle B_{\overline{24}}|D^{(8)}_{38}J_{3}  | B_{6}\rangle$  & $-\frac{1}{15\sqrt{6}}T_{3}$ & $-\frac{2}{45}T_{3}$ & $0$ \\    

$\langle B_{\overline{15}}|D^{(8)}_{88}J_{3}  | B_{\overline{3}}\rangle$  
& $0$  & $0$
& $\langle B_{\overline{24}}|D^{(8)}_{88}J_{3}  | B_{6}\rangle$  & $-\frac{\sqrt{2}}{15}$ & $-\frac{1}{5\sqrt{3}}$ & $-\frac{1}{5\sqrt{3}}$ \\  

$\langle B_{\overline{15}}|d_{ab3}D^{(8)}_{3a}J_{b}  | B_{\overline{3}}\rangle$  
&$\frac{1}{2\sqrt{30}}$ &$-\frac{1}{12\sqrt{5}}T_{3}$
&$\langle B_{\overline{24}}|d_{ab3}D^{(8)}_{3a}J_{b}  | B_{6}\rangle$  &$\frac{1}{45\sqrt{2}}T_{3}$ &$\frac{2}{45\sqrt{3}}T_{3}$ & $0$  \\  

$\langle B_{\overline{15}}|d_{ab3}D^{(8)}_{8a}J_{b} | B_{\overline{3}}\rangle$
& $0$ & $\frac{1}{8}\sqrt{\frac{3}{5}}$ 
&$\langle B_{\overline{24}}|d_{ab3}D^{(8)}_{8a}J_{b} | B_{6}\rangle$ & $\frac{1}{15}{\sqrt\frac{2}{3}}$ & $\frac{1}{15}$ & $\frac{1}{15}$  \\  

$\langle B_{\overline{15}}|D^{(8)}_{3i}J_{i} | B_{\overline{3}}\rangle$  
& $0$& $0$ 
& $\langle B_{\overline{24}}|D^{(8)}_{3i}J_{i} | B_{6}\rangle$  & $0$ & $0$ & $0$ \\  

$\langle B_{\overline{15}}|D^{(8)}_{8i}J_{i} | B_{\overline{3}}\rangle$  
& $0$& $0$
&$\langle B_{\overline{24}}|D^{(8)}_{8i}J_{i} | B_{6}\rangle$  &  $0$ & $0$ & $0$ \\  
 \hline 
 \hline
\end{tabular}
\end{center}
  \label{tab:6}
\end{table}


\end{document}